\documentclass[12pt]{article}
\usepackage{amsmath,amsthm}
\usepackage{eucal}
\usepackage{amsfonts}
\newcommand{\ar}{\renewcommand{\arraystretch}{1}} % 1.0 % 0.6
\DeclareMathAlphabet{\bb}{U}{msb}{m}{n}
\gdef\C{\Bbb C}
\gdef\dZ{\Bbb Z}

\gdef\dS{\Bbb S}
\gdef\R{\Bbb R}
\gdef\K{\Bbb K}
\gdef\BH{\Bbb H}
\gdef\F{\Bbb F}
\gdef\dO{\Bbb O}

\DeclareMathOperator{\spin}{{\bf Spin}}
\DeclareMathOperator{\pin}{{\bf Pin}}
\DeclareMathOperator{\fD}{\mathfrak{D}}
\DeclareMathOperator{\Id}{Id}

\DeclareMathOperator{\Aut}{Aut}
\DeclareMathOperator{\Ker}{Ker}

\DeclareMathOperator{\Sym}{Sym}

\newcommand{\s}{\!}

\newcommand{\re}{\mbox{\rm Re}\,}
\newcommand{\im}{\mbox{\rm Im}\,}

\newcommand{\cA}{\mathcal{A}}

\newcommand{\cE}{\mathcal{E}}
\newcommand{\cN}{\mathcal{N}}
\newcommand{\cL}{\mathcal{L}}

\newcommand{\M}{{\bf\sf M}}

\newcommand{\sA}{{\sf A}}

\newcommand{\sI}{{\sf I}}
\newcommand{\sW}{{\sf W}}
\newcommand{\sE}{{\sf E}}
\newcommand{\sC}{{\sf C}}
\newcommand{\sT}{{\sf T}}

\newcommand{\sP}{{\sf P}}
\newcommand{\sU}{{\sf U}}
\newcommand{\sQ}{{\sf Q}}
\newcommand{\sAut}{{\sf Aut}}

\newcommand{\bx}{{\bf x}}

\newcommand{\bF}{{\bf F}}
\newcommand{\bE}{{\bf E}}
\newcommand{\bH}{{\bf H}}
\newcommand{\fM}{\mathfrak{M}}
\newcommand{\fC}{\mathfrak{C}}
\newcommand{\fR}{\mathfrak{R}}
\newcommand{\fH}{\mathfrak{H}}
\newcommand{\fO}{\mathfrak{O}}
\newcommand{\fG}{\mathfrak{G}}

\newcommand{\cl}{C\kern -0.2em \ell}

\newcommand{\p}{\prime}
\newcommand{\e}{\mbox{\bf e}}

\newcommand{\ld}{\left[}
\newcommand{\rd}{\right]}
\newcommand{\lf}{\left\{}
\newcommand{\rf}{\right\}}
\newtheorem{theorem}{Theorem}
\begin{document}
\title{Clifford Algebras and Lorentz Group}
\author{V.~V. Varlamov\thanks{Department of Mathematics, Siberia State
University of Industry, Kirova 42, Novokuznetsk 654007, Russia}}
\date{}
\maketitle
%\begin{center}
%{\bf\LARGE Clifford Algebras and Lorentz Group}\\[0.5cm]
%{\large V.~V. Varlamov\footnote{Department of Mathematics, Siberia State
%University of Industry, Kirova 42, Novokuznetsk 654007, Russia}}
%\end{center}
%\medskip
\begin{abstract}
Finite--dimensional representations of the proper orthochronous Lorentz
group are studied in terms of spinor representations of the Clifford
algebras. The Clifford algebras are understood as an `algebraic covering'
of a full system of the finite--dimensional representations of the
Lorentz group. Space--time discrete symmetries $P$, $T$ and $PT$,
represented by fundamental automorphisms of the Clifford algebras, are
defined on all the representation spaces. Real, complex, quaternionic and
octonionic representations of the Lorentz group are considered. Physical
fields of the different types are formulated within such representations.
The Atiyah--Bott--Shapiro periodicity is defined on the Lorentz group.
It is shown that modulo 2 and modulo 8 periodicities of the Clifford
algebras allow to take a new look at the de Broglie--Jordan neutrino theory
of light and the Gell-Mann--Ne'emann eightfold way in particle physics.
On the representation spaces the charge conjugation $C$ is represented by
a pseudoautomorphism of the complex Clifford algebra. Quotient
representations of the Lorentz group are introduced. It is shown that
quotient representations are the most suitable for description of the
massless physical fields. By way of example, neutrino field is described
via the simplest quotient representation. Weyl--Hestenes equations for
neutrino field are given.
\end{abstract}
{\bf 1998 Physics and Astronomy Classification Scheme:} 02.10.Tq, 11.30.Er,
11.30.Cp
\section{Introduction}
Importance of discrete transformations is well--known, many textbooks
on quantum theory began with description of the discrete symmetries, and
famous L\"{u}ders--Pauli $CPT$--Theorem is a keystone of quantum field
theory. However, usual practice of definition of the discrete symmetries
from the analysis of relativistic wave equations does not give a full and
consistent theory of the discrete transformations. In the standard approach,
except a well studied case of the spin $j=1/2$ (Dirac equation), a situation
with the discrete symmetries remains vague for the fields of higher spin
$j>1/2$. It is obvious that a main reason of this is an absence of a fully
adequate formalism for description of higher--spin fields (all widely
accepted higher--spin formalisms such as Rarita--Schwinger approach \cite{RS41},
Bargmann--Wigner \cite{BW48} and Gel'fand--Yaglom \cite{GY48} multispinor
theories, and also Joos--Weinberg $2(2j+1)$--component formalism 
\cite{Joo62,Wein} have many intrinsic contradictions and difficulties).
The first attempt of going out from this situation was initiated by
Gel'fand, Minlos and Shapiro in 1958 \cite{GMS}. In the
Gel'fand--Minlos--Shapiro approach the discrete symmetries are represented
by outer involutory automorphisms of the Lorentz group (there are also other
realizations of the discrete symmetries via the outer automorphisms
\cite{Mic64,Kuo71,Sil92}).
At present the
Gel'fand--Minlos--Shapiro ideas have been found further development in the
works of Buchbinder, Gitman and Shelepin \cite{BGS00,GS00}, where the
discrete symmetries are represented by both outer and inner automorphisms of
the Poincar\'{e} group.

%Discrete symmetries $P$ and $T$ transform (reflect) space and time
%(two the most fundamental notions in physics), but in the Minkowski
%4--dimensional space--time continuum \cite{Min} space and time are not
%separate and independent. By this reason a transformation of one (space or
%time) induces a transformation of another. Therefore, discrete symmetries
%should be expressed by such trnsformations of the continuum, that transformed 
%all its structure totally with a full preservation of discrete nature. The
%only possible candidates on the role of such transformations are
%automorphisms. In such a way the idea of representation of the discrete
%symmetries via the automorphisms of the Lorentz group (`rotation' group of the
%4--dimensional continuum) is appearred in the Gel'fand--Minlos--Shapiro
%approach.

In 1957, Shirokov pointed out \cite{Shi57} that an universal covering of the
inhomogeneous Lorentz group has eight inequivalent realizations. Later on,
in the eighties this idea was applied to a general orthogonal group
$O(p,q)$ by D\c{a}browski \cite{Dab88}.
As known, the orthogonal
group $O(p,q)$ of the real space $\R^{p,q}$ is represented by the semidirect
product of a connected component $O_0(p,q)$ and a discrete subgroup
$\{1,P,T,PT\}$.
Further, a double covering of the orthogonal group $O(p,q)$ is a
Clifford--Lipschitz group $\pin(p,q)$ which is completely constructed within
a Clifford algebra $\cl_{p,q}$. In accordance with squares of elements of the
discrete subgroup ($a=P^2,\,b=T^2,\,c=(PT)^2$) there exist eight double
coverings (D\c{a}browski groups \cite{Dab88}) of the orthogonal group
defining by the signatures $(a,b,c)$, where $a,b,c\in\{-,+\}$. Such in brief is
a standard description scheme of the discrete transformations.
However, in this scheme there is one essential flaw. Namely, the
Clifford--Lipschitz group is an intrinsic notion of the algebra $\cl_{p,q}$
(a set of the all invertible elements of $\cl_{p,q}$), whereas the discrete
subgroup is introduced into the standard scheme in an external way, and the
choice of the signature $(a,b,c)$ of the discrete subgroup is not
determined by the signature of the space $\R^{p,q}$. Moreover, it is suggest
by default that for any signature $(p,q)$ of the vector space there exist
the all eight kinds of the discrete subgroups. It is obvious that a
consistent description of the double coverings of $O(p,q)$ in terms of
the Clifford--Lipschitz groups $\pin(p,q)\subset\cl_{p,q}$ can be obtained
only in the case when the discrete subgroup $\{1,P,T,PT\}$ is also defined
within the algebra $\cl_{p,q}$. Such a description has been given in the
works \cite{Var99,Var00,Var03}, where the discrete symmetries are
represented by fundamental automorphisms of the Clifford algebras.
Moreover, this description allows to incorporate the
Gel'fand--Minlos--Shapiro automorphism theory into 
Shirokov--D\c{a}browski scheme and further to unite them on the basis
of the Clifford algebras theory.

In the present paper such an unification is given. First of all,
Clifford algebras are understood as `algebraic coverings' of 
finite--dimensional representations of the proper Lorentz group $\fG_+$.
In the section 2 Clifford algebras $\C_n$ over the field $\F=\C$ are
associated with complex finite--dimensional representations $\fC$ of the
group $\fG_+$. It allows to define a new class of the finite--dimensional
representations of $\fG_+$ (quotient representations) corresponded to the
type $n\equiv 1\pmod{2}$ of the algebras $\C_n$. In its turn, representation
spaces of $\fC$ are the spinspaces $\dS_{2^{n/2}}$ or the minimal left
ideals of the algebras $\C_n$. In virtue of this in the section 3 the
discrete symmetries representing by spinor representations of the
fundamental automorphisms of $\C_n$ are defined for both complex and real
finite--dimensional representations of the group $\fG_+$.

A full system $\fM=\fM^+\oplus\fM^-$ of the finite--dimensional
representations of the group $\fG_+$ allows to define in the section 4
the Atiyah--Bott--Shapiro periodicity \cite{AtBSh} on the Lorentz group.
In case of the field $\F=\C$ we have modulo 2 periodicity on the
representations $\fC$, $\fC\cup\fC$, that allows to take a new look at the
de Broglie--Jordan neutrino theory of light \cite{Bro32,Jor35}. In its turn,
over the field $\F=\R$ we have on the system $\fM$ the modulo 8 periodicity
which relates with octonionic representations of the Lorentz group and
the G\"{u}naydin--G\"{u}rsey construction of the quark structure in terms
of an octonion algebra $\dO$ \cite{GG73}. In essence, the modulo 8
periodicity on the system $\fM$ gives an another realization of the
well--known Gell-Mann--Ne'emann eightfold way \cite{GN64}. It should be
noted here that a first attempt in this direction was initiated by
Coquereaux in 1982 \cite{Coq82}.

Other important discrete symmetry is the charge conjugation $C$. In contrast
with the transformations $P$, $T$, $PT$ the operation $C$ is not
space--time discrete symmetry. This transformation is firstly appearred
on the representation spaces of the Lorentz group and its nature is
strongly different from other discrete symmetries. By this reason in the
section 5 the charge conjugation $C$ is represented by a
pseudoautomorphism $\cA\rightarrow\overline{\cA}$ which is not fundamental
automorphism of the Clifford algebra. All spinor representations of the
pseudoautomorphism $\cA\rightarrow\overline{\cA}$ are given in
Theorem \ref{tpseudo}.

Quotient representations of the group $\fG_+$ compose the second half
$\fM^-$ of the full system $\fM$ and correspond to the types $n\equiv 1\pmod{2}$
($\F=\C$) and $p-q\equiv 1,5\pmod{8}$ ($\F=\R$). An explicit form
of the quotient representations is given in the section 6 
(Theorem \ref{tfactor}). In the section 7 the first simplest physical
field (neutrino field), corresponded to a fundamental representation
$\fC^{1,0}$ of the group $\fG_+$, is studied within a quotient
representation ${}^\chi\fC^{1,0}_c\cup{}^\chi\fC^{0,-1}_c$. Such a
description of the neutrino was firstly given in the work \cite{Var99},
but in \cite{Var99} this description looks like an exotic case, whereas
in the present paper it is a direct consequence of all mathematical
background developed in the previous sections. It is shown also that
the neutrino field $(1/2,0)\cup(0,1/2)$ can be defined in terms of a
Dirac--Hestenes spinor field \cite{Hest66,Hest67}, and the wave function
of this field satisfies the Weyl--Hestenes equations 
(massless Dirac--Hestenes equations).\section{Finite-dimensional representations of the Lorentz group and
complex Clifford algebras}
It is well--known \cite{RF,ED79,BR77} that representations of the Lorentz
group play a fundamental role in the quantum field theory. Physical fields
are defined in terms of finite--dimensional irreducible representations
of the Lorentz group $O(1,3)\simeq O(3,1)$
(correspondingly, Poincar\'{e} group $O(1,3)\odot T(4)$, where
$T(4)$ is a subgroup of four--dimensional translations). It should be
noted that in accordance with \cite{Nai58} any finite--dimensional 
irreducible representation of the proper Lorentz group 
$\fG_+=O_0(1,3)\simeq O_0(3,1)\simeq SL(2;\C)/\dZ_2$ is
equivalent to some spinor representation. Moreover, spinor representations
exhaust in essence all the finite--dimensional representations of the
group $\fG_+$. This fact we will widely use below.

Let us consider in brief the basic facts concerning the theory of spinor
representations of the Lorentz group. The initial point of this theory
is a correspondence between transformations of the proper Lorentz group
and complex matrices of the second order. Indeed, follows to
\cite{GMS} let us compare the Hermitian matrix of the second order
\begin{equation}\label{6.1}
X=\ar\begin{pmatrix}
x_0+x_3 & x_1-ix_2\\
x_1+ix_2 & x_0-x_3
\end{pmatrix}
\end{equation}
to the vector $v$ of the Minkowski space--time $\R^{1,3}$ with coordinates
$x_0,x_1,x_2,x_3$. At this point $\det X=x^2_0-x^2_1-x^2_2-x^2_3=S^2(x)$.
The correspondence between matrices $X$ and vectors $v$ is one--to--one and
linear. Any linear transformation $X^\prime=aXa^\ast$ in a space of the
matrices $X$ may be considered as a linear transformation $g_a$ in
$\R^{1,3}$, where $a$ is a complex matrix of the second order with
$\det a=1$. The correspondence $a\sim g_a$ possesses following properties:
1) $\ar\begin{pmatrix}
1 & 0 \\ 0 & 1\end{pmatrix}\sim e$ (identity element); 2)
$g_{a_1}g_{a_2}=g_{a_1a_2}$ (composition); 3) two different matrices
$a_1$ and $a_2$ correspond to one and the same transformation $g_{a_1}=g_{a_2}$
only in the case $a_1=-a_2$. Since every complex matrix is defined by
eight real numbers, then from the requirement $\det a=1$ it follow two
conditions $\re\det a=1$ and $\im\det a=0$. These conditions leave six
independent parameters, that coincides with parameter number of the
proper Lorentz group.

Further, a set of all complex matrices of the second order forms a full
matrix algebra $\M_2(\C)$ that is isomorphic to a biquaternion algebra
$\C_2$. In its turn, Pauli matrices
\begin{equation}\label{6.2}
\sigma_0=\ar\begin{pmatrix} 
1 & 0 \\
0 & 1
\end{pmatrix},\quad
\sigma_1=\begin{pmatrix}
0 & 1\\
1 & 0
\end{pmatrix},\quad
\sigma_2=\begin{pmatrix}
0 & -i\\
i & 0
\end{pmatrix},\quad
\sigma_3=\begin{pmatrix}
1 & 0\\
0 &-1
\end{pmatrix}.
\end{equation}
form one from a great number of isomorphic spinbasis of the algebra $\C_2$
(by this reason in physics the algebra $\C_2\simeq\cl^+_{1,3}\simeq\cl_{3,0}$ 
is called Pauli algebra). Using the basis (\ref{6.2}) we can write the
matrix (\ref{6.1}) in the form
\begin{equation}\label{6.3}
X=x^\mu\sigma_\mu.
\end{equation}
The Hermitian matrix (\ref{6.3}) is correspond to a spintensor
$(1,1)$ $X^{\lambda\dot{\nu}}$ with following coordinates
\begin{eqnarray}
x^0=+(1/\sqrt{2})(\xi^1\xi^{\dot{1}}+\xi^2\xi^{\dot{2}}),&&
x^1=+(1/\sqrt{2})(\xi^1\xi^{\dot{2}}+\xi^2\xi^{\dot{1}}),\nonumber\\
x^2=-(i/\sqrt{2})(\xi^1\xi^{\dot{2}}-\xi^2\xi^{\dot{1}}),&&
x^3=+(1/\sqrt{2})(\xi^1\xi^{\dot{1}}-\xi^2\xi^{\dot{2}}),\label{6.4}
\end{eqnarray}
where $\xi^\mu$ and $\xi^{\dot{\mu}}$ are correspondingly coordinates of
spinors and cospinors of spinspaces $\dS_2$ and $\dot{\dS}_2$. Linear
transformations of `vectors' (spinors and cospinors) of the spinspaces
$\dS_2$ and $\dot{\dS}_2$ have the form
\begin{equation}\label{6.5}\ar
\begin{array}{ccc}
\begin{array}{ccc}
{}^\prime\xi^1&=&\alpha\xi^1+\beta\xi^2,\\
{}^\prime\xi^2&=&\gamma\xi^1+\delta\xi^2,
\end{array} & \phantom{ccc} &
\begin{array}{ccc}
{}^\prime\xi^{\dot{1}}&=&\dot{\alpha}\xi^{\dot{1}}+\dot{\beta}\xi^{\dot{2}},\\
{}^\prime\xi^{\dot{2}}&=&\dot{\gamma}\xi^{\dot{1}}+\dot{\delta}\xi^{\dot{2}},
\end{array}\\
\sigma=\ar\begin{pmatrix}
\alpha & \beta\\
\gamma & \delta
\end{pmatrix} & \phantom{ccc} &
\dot{\sigma}=\begin{pmatrix}
\dot{\alpha} & \dot{\beta}\\
\dot{\gamma} & \dot{\delta}
\end{pmatrix}.
\end{array}
\end{equation}
Transformations (\ref{6.5}) form the group $SL(2;\C)$, since
$\sigma\in\M_2(\C)$ and
\[
SL(2;\C)=\left\{\ar\begin{pmatrix} \alpha & \beta\\ \gamma &\delta\end{pmatrix}
\in\C_2:\;\det\begin{pmatrix} \alpha & \beta \\ \gamma & \delta\end{pmatrix}=1
\right\}\simeq\spin_+(1,3).
\]
The expressions (\ref{6.4}) and (\ref{6.5}) compose a base of the 2--spinor
van der Waerden formalism \cite{Wa29,Rum36}, in which the spaces
$\dS_2$ and $\dot{\dS}_2$ are called correspondingly spaces of
{\it undotted and dotted spinors}. The each of the spaces
$\dS_2$ and $\dot{\dS}_2$ is homeomorphic to an extended complex plane
$\C\cup\infty$ representing an absolute (the set of infinitely distant points)
of a Lobatchevskii space $S^{1,2}$. At this point a group of fractional
linear transformations of the plane $\C\cup\infty$ is isomorphic to a motion
group of $S^{1,2}$ \cite{Roz55}. Besides, in accordance with
\cite{Kot27} the Lobatchevskii space $S^{1,2}$ is an absolute of the
Minkowski world $\R^{1,3}$ and, therefore, the group of fractional linear
transformations of the plane $\C\cup\infty$ (motion group of
$S^{1,2}$) twice covers a `rotation group' of the space--time $\R^{1,3}$,
that is the proper Lorentz group.
\begin{theorem}\label{tprod}
Let $\C_2$ be a biquaternion algebra and let $\sigma_i$ be a canonical spinor
representations (Pauli matrices) of the units of $\C_2$, then
$2k$ tensor products of the $k$ matrices $\sigma_i$ form a basis of the
full matrix algebra $\M_{2^k}(\C)$, which is a spinor representation of 
a complex Clifford algebra $\C_{2k}$. The set containing $2k+1$ tensor
products of the $k$ matrices $\sigma_i$ is homomorphically mapped onto a
set consisting of the same $2k$ tensor products and forming a basis
of the spinor representation of a quotient algebra 
${}^\epsilon\C_{2k}$.
\end{theorem} 
\begin{proof} 
As a basis of the spinor representation of the algebra
$\C_2$ we take the Pauli matrices (\ref{6.2}). This choice is explained by
physical applications only (from mathematical viewpoint the choice of the
spinbasis for $\C_2$ is not important).
Let us compose now $2k$ $2^k$--dimensional matrices:
\begin{equation}\label{6.6}
{\renewcommand{\arraystretch}{1.2}
\begin{array}{lcl}
\cE_{1}&=&\sigma_{1}\otimes\sigma_{0}\otimes\cdots\otimes\sigma_{0}\otimes 
\sigma_{0}\otimes\sigma_{0},\\
\cE_{2}&=&\sigma_{3}\otimes\sigma_{1}\otimes\sigma_{0}\otimes\cdots\otimes 
\sigma_{0}\otimes\sigma_{0},\\
\cE_{3}&=&\sigma_{3}\otimes\sigma_{3}\otimes\sigma_{1}\otimes\sigma_{0}
\otimes\cdots\otimes\sigma_{0},\\
\hdotsfor[2]{3}\\
\cE_{k}&=&\sigma_{3}\otimes\sigma_{3}\otimes\cdots\otimes\sigma_{3}
\otimes\sigma_{1},\\
\cE_{k+1}&=&\sigma_{2}\otimes\sigma_{0}\otimes\cdots\otimes\sigma_{0},\\
\cE_{k+2}&=&\sigma_{3}\otimes\sigma_{2}\otimes\sigma_{0}\otimes\cdots\otimes 
\sigma_{0},\\
\hdotsfor[2]{3}\\
\cE_{2k}&=&\sigma_{3}\otimes\sigma_{3}\otimes\cdots\otimes\sigma_{3}
\otimes\sigma_{2}.
\end{array}}\end{equation}
Since $\sigma^2_i=\sigma_0$, then for a square of any matrix from the set
(\ref{6.6}) we have
\begin{equation}\label{6.7}
\cE^2_i=\sigma_0\otimes\sigma_0\otimes\cdots\otimes\sigma_0,\quad
i=1,2,\ldots,2k,
\end{equation}
where the product $\sigma_0\otimes\sigma_0\otimes\cdots\otimes\sigma_0$ equals
to $2^k$--dimensional unit matrix. Further,
\begin{equation}\label{6.8}
\cE_{i}\cE_{j}=-\cE_{j}\cE_{i},\quad  i<j;\;\;i,j=1,2,\ldots 2k.
\end{equation}
Indeed, when $i=1$ and $j=3$ we obtain
\begin{eqnarray}
\cE_{1}\cE_{3}&=&\sigma_{1}\sigma_{3}\otimes\sigma_{3}\otimes\sigma_{1}
\otimes\sigma_{0}\otimes\cdots\otimes\sigma_{0},\nonumber \\
\cE_{3}\cE_{1}&=&\sigma_{3}\sigma_{1}\otimes\sigma_{3}\otimes\sigma_{1}
\otimes\sigma_{0}\otimes\cdots\otimes\sigma_{0}.\nonumber
\end{eqnarray}
Thus, the equalities (\ref{6.7}) and (\ref{6.8}) show that the matrices of
the set (\ref{6.6}) satisfy the multiplication law of the Clifford algebra.
Moreover, let us show that a set of the matrices
$\cE^{\alpha_1}_1\cE^{\alpha_2}_2\ldots\cE^{\alpha_{2k}}_{2k}$, where
each of the indices $\alpha_1,\alpha_2,\ldots,\alpha_{2k}$ takes either of
the two values 0 or 1, consists of $2^{2k}$ matrices. At this point
these matrices form a basis of the full
$2^k$--dimensional matrix algebra (spinor representation of $\C_{2k}$).
Indeed, in virtue of $i\sigma_1\sigma_2=\sigma_3$ from (\ref{6.6}) it follows
\begin{gather}
\cN_{j}=\cE_{j}\cE_{k+j}=\sigma_{0}\otimes\sigma_{0}\otimes\cdots\otimes 
\sigma_{3}\otimes\sigma_{0}\otimes\cdots\otimes\sigma_{0},\label{6.9}\\
j=1,2,\ldots k,\nonumber
\end{gather}
here the matrix $\sigma_3$ occurs in the $j$--th position. Further, since
the tensor product
$\sigma_{0}\otimes\cdots\otimes\sigma_{0}$ of the unit matrices of the
second order is also unit matrix $\cE_0$ of the $2^k$--order, then we can
write
\begin{eqnarray}
Z^{++}_{j}&=&\frac{1}{2}(\cE_{0}-\cN_{j})=\sigma_{0}\otimes\sigma_{0}
\otimes\cdots
\otimes Q^{++}\otimes\sigma_{0}\otimes\cdots\otimes\sigma_{0},\nonumber \\
Z^{--}_{j}&=&\frac{1}{2}(\cE_{0}+\cN_{j})=\sigma_{0}\otimes\sigma_{0}
\otimes\cdots
\otimes Q^{--}\otimes\sigma_{0}\otimes\cdots\otimes\sigma_{0},\nonumber
\end{eqnarray}
where the matrices $Q^{++}$ and $Q^{--}$ occur in the $j$--th position
and have the following form
\[
\ar\begin{pmatrix}
1 & 0 \\
0 & 0
\end{pmatrix},\quad
\begin{pmatrix}
0 & 0 \\
0 & 1
\end{pmatrix}.
\]
From (\ref{6.9}) it follows
\[
\cN_{1}\cN_{2}\ldots\cN_{j}=\sigma_{3}\otimes\sigma_{3}\otimes\cdots\otimes
\sigma_{3}\otimes\sigma_{0}\otimes\cdots\otimes\sigma_{0},
\]
where on the left side we have $j$ matrices $\sigma_3$. In virtue of this
equality we obtain
\begin{eqnarray}
\cL_{j}&=&\cN_{1}\ldots\cN_{j-1}\cE_{j}=
\sigma_{0}\otimes\sigma_{0}\otimes\cdots\otimes
\sigma_{1}\otimes\sigma_{0}\otimes\cdots\otimes\sigma_{0},\nonumber \\
\cL_{k+j}&=&\cN_{1}\ldots\cN_{j-1}\cE_{k+j}=
\sigma_{0}\otimes\sigma_{0}\otimes\cdots\otimes
\sigma_{2}\otimes\sigma_{0}\otimes\cdots\otimes\sigma_{0}.\nonumber
\end{eqnarray}
here the matrices $\sigma_1$ and $\sigma_2$ occur in the $j$--th position,
$j=1,2,\ldots,k$. Thus,
\begin{eqnarray}
Z^{+-}_{j}&=&\frac{1}{2}(\cL_{j}+i\cL_{k+j})=
\sigma_{0}\otimes\sigma_{0}\otimes\cdots
\otimes Q^{+-}\otimes\sigma_{0}\otimes\cdots\otimes\sigma_{0},\nonumber \\
Z^{-+}_{j}&=&\frac{1}{2}(\cL_{j}-i\cL_{k+j})=
\sigma_{0}\otimes\sigma_{0}\otimes\cdots
\otimes Q^{-+}\otimes\sigma_{0}\otimes\cdots\otimes\sigma_{0},\nonumber
\end{eqnarray}
where the matrices $Q^{+-}$ and $Q^{-+}$ also occur in the $j$--th position
and have correspondingly the following form
\[
\ar\begin{pmatrix}
0 & 1 \\
0 & 0
\end{pmatrix},\quad
\begin{pmatrix}
0 & 0 \\
1 & 0
\end{pmatrix}.
\]
Therefore, a matrix
\begin{gather}
\prod^{k}_{j}(Z^{s_{j}r_{j}}_{j})=Q^{s_{1}r_{1}}\otimes
Q^{s_{2}r_{2}}\otimes\cdots\otimes Q^{s_{k}r_{k}},\label{6.10}\\
s_{j}=\pm,\;r_{j}=\pm\nonumber
\end{gather}
has unit elements at the intersection of
$(s_1,s_2,\ldots,s_k)$--th row and $(r_1,r_2,\ldots,r_k)$--th column,
the other elements are equal to zero. In virtue of (\ref{6.7}) and (\ref{6.8}) 
each of the matrices $\cL_j,\,\cL_{k+j}$ and, therefore, each of the matrices
$Z^{s_jr_j}_j$, is represented by a linear combination of the matrices
$\cE^{\alpha_1}_1\cE^{\alpha_2}_2\cdots\cE^{\alpha_{2k}}_{2k}$. Hence it
follows that the matrices $\prod^k_j(Z^{s_jr_j}_j)$ and, therefore, all the
$2^k$--dimensional matrices, are represented by such the linear combinations.
Thus, $2k$ matrices $\cE_1,\ldots,\cE_{2k}$ generate a group consisting of
the products $\pm\cE^{\alpha_1}_1\cE^{\alpha_2}_2\cdots
\cE^{\alpha_{2k}}_{2k}$, and an enveloped algebra of this group is a full
$2^k$--dimensional matrix algebra.

The following part of this Theorem tells that the full matrix algebra, forming
by the tensor products (\ref{6.6}), is a spinor representation of the algebra
$\C_{2k}$. Let us prove this part on the several examples. First of all,
in accordance with (\ref{6.6}) tensor products of the $k=2$ order are
\begin{gather}
\ar\cE_{1}=\sigma_{1}\otimes\sigma_{0}=
\begin{pmatrix}
0 & \phantom{-}1 \\
1 & 0
\end{pmatrix}\otimes
\begin{pmatrix}
1 & 0 \\
0 & \phantom{-}1
\end{pmatrix}=
\begin{pmatrix}
0 & 0 & 1 & 0 \\
0 & 0 & 0 & 1 \\
1 & 0 & 0 & 0 \\
0 & 1 & 0 & 0
\end{pmatrix},\nonumber\\
\cE_{2}=\sigma_{3}\otimes\sigma_{1}=\ar
\begin{pmatrix}
1 & 0 \\
0 & -1
\end{pmatrix}\otimes
\begin{pmatrix}
0 & 1 \\
1 & 0
\end{pmatrix}=
\begin{pmatrix}
0 & 1 & 0 & 0 \\
1 & 0 & 0 & 0 \\
0 & 0 & 0 & -1 \\
0 & 0 & -1 & 0
\end{pmatrix},\nonumber\\
\cE_{3}=\sigma_{2}\otimes\sigma_{0}=\ar
\begin{pmatrix}
0 & -i \\
i & 0
\end{pmatrix}\otimes
\begin{pmatrix}
1 & 0 \\
0 & 1
\end{pmatrix}=
\begin{pmatrix}
0 & 0 & -i & 0 \\
0 & 0 & 0 & -i \\
i & 0 & 0 & 0 \\
0 & i & 0 & 0
\end{pmatrix},\nonumber\\
\cE_{4}=\sigma_{3}\otimes\sigma_{2}=\ar
\begin{pmatrix}
1 & 0 \\
0 & -1
\end{pmatrix}\otimes
\begin{pmatrix}
0 & -i \\
i & 0
\end{pmatrix}=
\begin{pmatrix}
0 & -i & 0 & 0 \\
i & 0 & 0 & 0 \\
0 & 0 & 0 & i \\
0 & 0 & -i & 0
\end{pmatrix}.\label{6.11}
\end{gather}
It is easy to see that $\cE^2_i=\cE_0$ and $\cE_i\cE_j=-\cE_j\cE_i$
$(i,j=1,\ldots,4)$. The set of the matrices $\cE^{\alpha_1}_1\cE^{\alpha_2}_2
\cdots\cE^{\alpha_4}_4$, where each of the indices $\alpha_1,\alpha_2,
\alpha_3,\alpha_4$ takes either of the two values 0 or 1, consists of
$2^4=16$ matrices. At this point these matrices form a basis of the full
4--dimensional matrix algebra. Thus, we can define a one--to--one
correspondence between sixteen matrices $\cE^{\alpha_1}_1\cE^{\alpha_2}_2\cdots
\cE^{\alpha_4}_4$ and sixteen basis elements $\e_{i_1}\e_{i_2}\ldots\e_{i_k}$
of the Dirac algebra $\C_4$. Therefore, the matrices (\ref{6.11}) form a
basis of the spinor representation of $\C_4$. Moreover, from (\ref{6.11}) it
follows that $\C_4\simeq\C_2\otimes\C_2$, that is, {\it the Dirac algebra
is a tensor product of the two Pauli algebras}.

Analogously, when $k=3$ from (\ref{6.6}) we obtain following products
(matrices of the eighth order):
\begin{gather}
\cE_1=\sigma_1\otimes\sigma_0\otimes\sigma_0,\quad
\cE_2=\sigma_3\otimes\sigma_1\otimes\sigma_0,\quad
\cE_3=\sigma_3\otimes\sigma_3\otimes\sigma_1,\nonumber\\
\cE_4=\sigma_2\otimes\sigma_0\otimes\sigma_0,\quad
\cE_5=\sigma_3\otimes\sigma_2\otimes\sigma_0,\quad
\cE_6=\sigma_3\otimes\sigma_3\otimes\sigma_2,\nonumber
\end{gather}
The set of the matrices
$\cE^{\alpha_1}_1\cE^{\alpha_2}_2\cdots\cE^{\alpha_6}_6$, consisting of
$2^6=64$ matrices, forms a basis of the full 8--dimensional matrix algebra,
which is isomorphic to a spinor representation of the algebra
$\C_6\simeq\C_2\otimes\C_2\otimes\C_2$. Generalizing we obtain that $2k$
tensor products of the $k$ Pauli matrices form a basis of the spinor
representation of the complex Clifford algebra
\begin{equation}\label{6.11'}
\C_{2k}\simeq
\underbrace{\C_2\otimes\C_2\otimes\cdots\otimes\C_2}_{k\;\text{times}}.
\end{equation}

Let us consider now a case of odd dimensions. When $n=2k+1$ we add to the
set of the tensor products (\ref{6.6}) a matrix
\begin{equation}\label{6.12}
\cE_{2k+1}=\underbrace{\sigma_{3}\otimes\sigma_{3}\otimes\cdots\otimes\sigma_{3}
}_{k\;\text{times}},
\end{equation}
which satisfying the conditions
\begin{gather}
\cE^{2}_{2k+1}=\cE_{0},\quad \cE_{2k+1}\cE_{i}=-\cE_{i}\cE_{2k+1},\nonumber\\
i=1,2,\ldots, k.\nonumber
\end{gather}
It is obvious that a product
$\cE_{1}\cE_{2}\cdots\cE_{2k}\cE_{2k+1}$ commutes with all the products of the
form $\cE^{\alpha_1}_1\cE^{\alpha_2}_2\cdots\cE^{\alpha_{2k}}_{2k}$.
Further, let
\[
\varepsilon=\begin{cases}
1&, \text{if $k\equiv 0\!\!\!\pmod{2}$},\\
i&, \text{if $k\equiv 1\!\!\!\pmod{2}$}.
\end{cases}
\]
Then a product
\[
\sU=\varepsilon\cE_{1}\cE_{2}\cdots\cE_{2k}\cE_{2k+1}
\]
satisfies the condition $\sU^{2}=\cE_{0}$. Let $\sP$ be a set of all
$2^{2k+1}$ matrices $\cE^{\alpha_1}_1\cE^{\alpha_2}_2\cdots
\cE^{\alpha_{2k}}_{2k}\cE^{\alpha_{2k+1}}_{2k+1}$, where $\alpha_j$ equals to 0
or 1, $j=1,2,\ldots,2k+1$. Let us divide the set $\sP$ into two subsets
by the following manner:
\begin{equation}\label{6.13}
\sP=\sP^{1}+\sP^{0},
\end{equation}
where the subset $\sP^0$ contains products with the matrix $\cE_{2k+1}$,
and $\sP^1$ contains products without the matrix $\cE_{2k+1}$. Therefore,
products
$\cE^{\alpha_1}\cE^{\alpha_2}_2\cdots\cE^{\alpha_{2k}}_{2k}\subset\sP^1$
form a full $2^k$--dimensional matrix algebra. Further, when we multiply
the matrices from the subset $\sP^0$ by the matrix $\sU$ the
factors $\cE_{2k+1}$ are mutually annihilate. Thus, matrices of the set
$\sU\sP^0$ are also belong to the $2^k$--dimensional matrix algebra.
Let us denote $\sU\sP^0$ via $\sP^2$. Taking into account that
$\sU^2=\cE_0$ we obtain $\sP^0=\sU\sP^2$ and
\[
\sP=\sP^1+\sP^2,
\]
where $\sP^1,\sP^2\in\M_{2^k}$. Let
\begin{equation}\label{6.14}
\chi\;:\;\;\sP^{1}+\sU\sP^{2}\longrightarrow
\sP^{1}+\sP^{2}
\end{equation}
be an homomorphic mapping of the set (\ref{6.13}) containing the matrix
$\cE_{2k+1}$ onto the set $\sP^1+\sP^2$ which does not contain this matrix.
The mapping $\chi$ preserves addition and multiplication operations.
Indeed, let $\sP=\sP^1+\sU\sP^2$ and $\sQ=\sQ^1+\sU\sQ^2$, then
\[
\sP+\sQ=\sP^1+\sU\sP^2+\sQ^1+\sU\sQ^2\longrightarrow
\sP^1+\sP^2+\sQ^1+\sQ^2,
\]
\begin{multline}
\sP\sQ=(\sP^1+\sU\sP^2)(\sQ^1+\sU\sQ^2)=(\sP^1\sQ^1+\sP^2\sQ^2)+
\sU(\sP^1\sQ^2+\sP^2\sQ^1)\longrightarrow\\
(\sP^1\sQ^1+\sP^2\sQ^2)+(\sP^1\sQ^2+\sP^2\sQ^1)=(\sP^1+\sP^2)(\sQ^1+\sQ^2),
\end{multline}
that is, an image of the products equals to the product of factor images in
the same order. In particular, when $\sP=\sU$ we obtain
$\sP^1=0,\,\sP^2=\cE_0$ and
\begin{equation}\label{6.15}
\sU\longrightarrow\cE_0.
\end{equation}\begin{sloppypar}\noindent
In such a way, at the mapping $\chi$ all the matrices of the form
$\cE^{\alpha_1}_1\cE^{\alpha_2}_2\cdots\cE^{\alpha_{2k}}_{2k}-
\sU\cE^{\alpha_1}_1\cE^{\alpha_2}_2\cdots\cE^{\alpha_{2k}}_{2k}$
are mapped into zero. Therefore, a kernel of the homomorphism $\chi$ is
defined by an expression $\Ker\chi=\left\{\sP^1-\sU\sP^1\right\}$. We obtain
the homomorphic mapping of the set of all the $2^{2k+1}$ matrices
$\cE^{\alpha_1}_1\cE^{\alpha_2}_2\cdots\cE^{\alpha_{2k}}_{2k}
\cE^{\alpha_{2k+1}}_{2k+1}$ onto the full matrix algebra $\M_{2^k}$.
In the result of this mapping we have a quotient algebra
${}^\chi\M_{2^k}\simeq\sP/\Ker\chi$. As noted previously,
$2k$ tensor products of the $k$ Pauli matrices (or other $k$ matrices
defining spinor representation of the biquaternion algebra $\C_2$)
form a basis of the spinor representation of the algebra $\C_{2k}$. 
It is easy to see that there exists one--to--one correspondence 
between $2^{2k+1}$ matrices of the set $\sP$ and 
basis elements of the odd--dimensional
Clifford algebra $\C_{2k+1}$. It is well--known \cite{Rash,Port} that
$\C_{2k+1}$ is isomorphic to a direct sum of two even--dimensional subalgebras:
$\C_{2k+1}\simeq\C_{2k}\oplus\C_{2k}$. Moreover, there exists an
homomorphic mapping $\epsilon:\;\C_{2k+1}\rightarrow\C_{2k}$, in the result
of which we have a quotient algebra
${}^\epsilon\C_{2k}\simeq\C_{2k+1}/\Ker\epsilon$, where
$\Ker\epsilon=\left\{\cA^1-\varepsilon\omega\cA^1\right\}$ is a kernel of the
homomorphism $\epsilon$, $\cA^1$ is an arbitrary element of the subalgebra
$\C_{2k}$, $\omega$ is a volume element of the algebra $\C_{2k+1}$.
It is easy to see that the homomorphisms $\epsilon$ and $\chi$ have a
similar structure. Thus, hence it immediately follows an isomorphism
${}^\epsilon\C_{2k}\simeq{}^\chi\M_{2^k}$. Therefore, a basis of the matrix
quotient algebra ${}^\chi\M_{2^k}$ is also a basis of the spinor
representation of the Clifford quotient algebra ${}^\epsilon\C_{2k}$, that
proves the latter assertion of the theorem.\end{sloppypar}
\end{proof}
Let us consider now spintensor representations of the proper
Lorentz group $O_0(1,3)\simeq SL(2;\C)/\dZ_2\simeq\spin_+(1,3)/\dZ_2$ and their
relations with the complex Clifford algebras. From each complex Clifford
algebra $\C_n=\C\otimes\cl_{p,q}\;
(n=p+q)$ we obtain a spinspace $\dS_{2^{n/2}}$, which is a complexification
of the minimal left ideal of the algebra
$\cl_{p,q}$: $\dS_{2^{n/2}}=\C\otimes I_{p,q}=\C\otimes\cl_{p,q}
e_{pq}$, where $e_{pq}$ is a primitive idempotent of the algebra $\cl_{p,q}$.
Further, a spinspace corresponding the Pauli algebra $\C_2$ has a form
$\dS_2=\C\otimes I_{2,0}=\C\otimes\cl_{2,0}e_{20}$ or
$\dS_2=\C\otimes I_{1,1}=\C\otimes\cl_{1,1}e_{11}(\C\otimes I_{0,2}=
\C\otimes\cl_{0,2}e_{02})$. Therefore, the tensor product
(\ref{6.11'}) of the $k$ algebras $\C_2$ induces a tensor product of the $k$
spinspaces $\dS_2$:
\[
\dS_2\otimes\dS_2\otimes\cdots\otimes\dS_2=\dS_{2^k}.
\]
Vectors of the spinspace $\dS_{2^k}$ (or elements of the minimal left
ideal of $\C_{2k}$) are spintensors of the following form
\begin{equation}\label{6.16}
\xi^{\alpha_1\alpha_2\cdots\alpha_k}=\sum
\xi^{\alpha_1}\otimes\xi^{\alpha_2}\otimes\cdots\otimes\xi^{\alpha_k},
\end{equation}
where summation is produced on all the index collections
$(\alpha_1\ldots\alpha_k)$, $\alpha_i=1,2$. In virtue of (\ref{6.5}) for
each spinor $\xi^{\alpha_i}$ from (\ref{6.16}) we have a transformation rule
${}^\prime\xi^{\alpha^\prime_i}=
\sigma^{\alpha^\prime_i}_{\alpha_i}\xi^{\alpha_i}$. Therefore, in general
case we obtain
\begin{equation}\label{6.17}
{}^\prime\xi^{\alpha^\prime_1\alpha^\prime_2\cdots\alpha^\prime_k}=\sum
\sigma^{\alpha^\prime_1}_{\alpha_1}\sigma^{\alpha^\prime_2}_{\alpha_2}\cdots
\sigma^{\alpha^\prime_k}_{\alpha_k}\xi^{\alpha_1\alpha_2\cdots\alpha_k}.
\end{equation}
A representation (\ref{6.17}) is called
{\it undotted spintensor representation of the proper Lorentz group of the
rank $k$}.

Further, let $\overset{\ast}{\C}_2$ be a biquaternion algebra, the
coefficients of which are complex conjugate. Let us show that the algebra
$\overset{\ast}{\C}_2$ is obtained from $\C_2$ under action of the
automorphism $\cA\rightarrow\cA^\star$ or antiautomorphism
$\cA\rightarrow\widetilde{\cA}$. Indeed, in virtue of an isomorphism
$\C_2\simeq\cl_{3,0}$ a general element
\[
\cA=a^0\e_0+\sum^3_{i=1}a^i\e_i+\sum^3_{i=1}\sum^3_{j=1}a^{ij}\e_{ij}+
a^{123}\e_{123}
\]
of the algebra $\cl_{3,0}$ can be written in the form
\begin{equation}\label{6.17'}
\cA=(a^0+\omega a^{123})\e_0+(a^1+\omega a^{23})\e_1+(a^2+\omega a^{31})\e_2
+(a^3+\omega a^{12})\e_3,
\end{equation}
where $\omega=\e_{123}$. Since $\omega$ belongs to a center of the algebra
$\cl_{3,0}$ (commutes with all the basis elements) and $\omega^2=-1$, then
we can to suppose $\omega\equiv i$. The action of the automorphism $\star$
on the homogeneous element $\cA$ of a degree $k$ is defined by a formula
$\cA^\star=(-1)^k\cA$. In accordance with this the action of the
automorphism $\cA\rightarrow\cA^\star$, where $\cA$ is the element
(\ref{6.17'}), has a form
\begin{equation}\label{In1}
\cA\longrightarrow\cA^\star=-(a^0-\omega a^{123})\e_0-(a^1-\omega a^{23})\e_1
-(a^2-\omega a^{31})\e_2-(a^3-\omega a^{12})\e_3.
\end{equation}
Therefore, $\star:\,\C_2\rightarrow -\overset{\ast}{\C}_2$. Correspondingly,
the action of the antiautomorphism $\cA\rightarrow\widetilde{\cA}$ on the
homogeneous element $\cA$ of a degree $k$ is defined by a formula
$\widetilde{\cA}=(-1)^{\frac{k(k-1)}{2}}\cA$. Thus, for the element
(\ref{6.17'}) we obtain
\begin{equation}\label{In2}
\cA\longrightarrow\widetilde{\cA}=(a^0-\omega a^{123})\e_0+
(a^1-\omega a^{23})\e_1+(a^2-\omega a^{31})\e_2+(a^3-\omega a^{12})\e_3,
\end{equation}
that is, $\widetilde{\phantom{cc}}:\,\C_2\rightarrow
\overset{\ast}{\C}_2$.
This allows to define an algebraic analog of the Wigner's representation
doubling: $\C_2\oplus\overset{\ast}{\C}_2$. 
Further, from (\ref{6.17'})
it follows that $\cA=\cA_1+\omega\cA_2=(a^0\e_0+a^1\e_1+a^2\e_2+a^3\e_3)+
\omega(a^{123}\e_0+a^{23}\e_1+a^{31}\e_2+a^{12}\e_3)$. In general case,
by virtue of an isomorphism $\C_{2k}\simeq\cl_{p,q}$, where $\cl_{p,q}$ is a
real Clifford algebra with a division ring $\K\simeq\C$, $p-q\equiv 3,7
\pmod{8}$, we have for a general element of $\cl_{p,q}$ an expression
$\cA=\cA_1+\omega\cA_2$, here $\omega^2=\e^2_{12\ldots p+q}=-1$ and, therefore,
$\omega\equiv i$. Thus, from $\C_{2k}$ under action of the automorphism
$\cA\rightarrow\cA^\star$ we obtain a general algebraic doubling
\begin{equation}\label{D}
\C_{2k}\oplus\overset{\ast}{\C}_{2k}.
\end{equation}

Correspondingly, a tensor product
$\overset{\ast}{\C}_2\otimes\overset{\ast}{\C}_2\otimes\cdots\otimes
\overset{\ast}{\C}_2\simeq\overset{\ast}{\C}_{2r}$ of $r$ algebras
$\overset{\ast}{\C}_2$ induces a tensor product of $r$ spinspaces
$\dot{\dS}_2$:
\[
\dot{\dS}_2\otimes\dot{\dS}_2\otimes\cdots\otimes\dot{\dS}_2=\dot{\dS}_{2^r}.
\]
Te vectors of the spinspace $\dot{\dS}_{2^r}$ have the form
\begin{equation}\label{6.18}
\xi^{\dot{\alpha}_1\dot{\alpha}_2\cdots\dot{\alpha}_r}=\sum
\xi^{\dot{\alpha}_1}\otimes\xi^{\dot{\alpha}_2}\otimes\cdots\otimes
\xi^{\dot{\alpha}_r},
\end{equation}
where the each cospinor $\xi^{\dot{\alpha}_i}$ from (\ref{6.18}) in virtue of
(\ref{6.5}) is transformed by the rule ${}^\prime\xi^{\dot{\alpha}^\prime_i}=
\sigma^{\dot{\alpha}^\prime_i}_{\dot{\alpha}_i}\xi^{\dot{\alpha}_i}$.
Therefore,
\begin{equation}\label{6.19}
{}^\prime\xi^{\dot{\alpha}^\prime_1\dot{\alpha}^\prime_2\cdots
\dot{\alpha}^\prime_r}=\sum\sigma^{\dot{\alpha}^\prime_1}_{\dot{\alpha}_1}
\sigma^{\dot{\alpha}^\prime_2}_{\dot{\alpha}_2}\cdots
\sigma^{\dot{\alpha}^\prime_r}_{\dot{\alpha}_r}
\xi^{\dot{\alpha}_1\dot{\alpha}_2\cdots\dot{\alpha}_r}.
\end{equation}\begin{sloppypar}\noindent
A representation (\ref{6.19}) is called
{\it a dotted spintensor representation of the proper Lorentz group of the
rank $r$}.\end{sloppypar}

In general case we have a tensor product of $k$ algebras $\C_2$ and
$r$ algebras $\overset{\ast}{\C}_2$:
\[
\C_2\otimes\C_2\otimes\cdots\otimes\C_2\otimes
\overset{\ast}{\C}_2\otimes
\overset{\ast}{\C}_2\otimes\cdots\otimes\overset{\ast}{\C}_2\simeq
\C_{2k}\otimes\overset{\ast}{\C}_{2r},
\]
which induces a spinspace
\[
\dS_2\otimes\dS_2\otimes\cdots\otimes\dS_2\otimes\dot{\dS}_2\otimes
\dot{\dS}_2\otimes\cdots\otimes\dot{\dS}_2=\dS_{2^{k+r}}
\]
with the vectors
\begin{equation}\label{6.20'}
\xi^{\alpha_1\alpha_2\cdots\alpha_k\dot{\alpha}_1\dot{\alpha}_2\cdots
\dot{\alpha}_r}=\sum
\xi^{\alpha_1}\otimes\xi^{\alpha_2}\otimes\cdots\otimes\xi^{\alpha_k}\otimes
\xi^{\dot{\alpha}_1}\otimes\xi^{\dot{\alpha}_2}\otimes\cdots\otimes
\xi^{\dot{\alpha}_r}.
\end{equation}
In this case we have a natural unification of the representations
(\ref{6.17}) and (\ref{6.19}):
\begin{equation}\label{6.20}
{}^\prime\xi^{\alpha^\prime_1\alpha^\prime_2\cdots\alpha^\prime_k
\dot{\alpha}^\prime_1\dot{\alpha}^\prime_2\cdots\dot{\alpha}^\prime_r}=\sum
\sigma^{\alpha^\prime_1}_{\alpha_1}\sigma^{\alpha^\prime_2}_{\alpha_2}\cdots
\sigma^{\alpha^\prime_k}_{\alpha_k}\sigma^{\dot{\alpha}^\prime_1}_{
\dot{\alpha}_1}\sigma^{\dot{\alpha}^\prime_2}_{\dot{\alpha}_2}\cdots
\sigma^{\dot{\alpha}^\prime_r}_{\dot{\alpha}_r}
\xi^{\alpha_1\alpha_2\cdots\alpha_k\dot{\alpha}_1\dot{\alpha}_2\cdots
\dot{\alpha}_r}.
\end{equation}
So, a representation (\ref{6.20}) is called
{\it a spintensor representation of the proper Lorentz group of the
rank $(k,r)$}.

In general case, the representations defining by the formulas
(\ref{6.17}), (\ref{6.19}) and
(\ref{6.20}), are reducible, that is there exists possibility of decomposition 
of the initial spinspace $\dS_{2^{k+r}}$ (correspondingly, spinspaces
$\dS_{2^k}$ and $\dS_{2^r}$) into a direct sum of invariant (with respect to
transformations of the group $\fG_+$) spinspaces
$\dS_{2^{\nu_1}}\oplus\dS_{2^{\nu_2}}\oplus\cdots\oplus\dS_{2^{\nu_s}}$,
where $\nu_1+\nu_2+\ldots+\nu_s=k+r$.
\begin{sloppypar}
Further, an important notion of the {\it physical field} is closely related
with finite--dimensional representations of the proper Lorentz
group $\fG_+$.
In accordance with Wigner interpretation
\cite{Wig39}, an elementary particle is described by some irreducible
finite--dimensional representation of the Poincar\'{e} group. The double
covering of the proper Poincar\'{e} group is isomorphic
to a semidirect product $SL(2;\C)\odot T(4)$, or
$\spin_+(1,3)\odot T(4)$, where $T(4)$ is the subgroup of four--dimensional
translations. Let $\psi(x)$ be a physical field, then at the transformations
$(a,\Lambda)$ of the proper Poincar\'{e} group the field $\psi(x)$ 
is transformed by a following rule\end{sloppypar}
\begin{equation}\label{6.21}
\psi^\prime_mu(x)=\sum_\nu\fC_{\mu\nu}(\sigma)\psi_\nu
(\Lambda^{-1}(x-a)),
\end{equation}\begin{sloppypar}\noindent
where $a\in T(4)$, $\sigma\in\fG_+$, $\Lambda$ is a Lorentz transformation,
and $\fC_{\mu\nu}$ is a representation of the group $\fG_+$ in the space
$\dS_{2^{k+r}}$. Since the group $T(4)$ is Abelian, then all its
representations are one--dimensional. Thus, all the finite--dimensional
representations of the proper Poincar\'{e} group in essence
are equivalent to the representations $\fC$ of the group
$\fG_+$. If the representation $\fC$ is reducible, then the space
$\dS_{2^{k+r}}$ is decomposed into a direct sum of irreducible subspaces, that is,
it is possible to choose in $\dS_{2^{k+r}}$ such a basis, in which all the
matrices $\fC_{\mu\nu}$ take a block--diagonal form. Then the field $\psi(x)$ 
is reduced to some number of the fields corresponding to obtained irreducible
representations of the group $\fG_+$, each of which is transformed
independently from the other, and the field $\psi(x)$ in this case is a
collection of the fields with more simple structure. It is obvious that these
more simple fields correspond to irreducible representations $\fC$. 
As known \cite{Nai58,GMS,RF}, a system of irreducible finite--dimensional
representations of the group $\fG_+$ is realized in the space
$\Sym_{(k,r)}\subset\dS_{2^{k+r}}$ of symmetric spintensors. The
dimensionality of $\Sym_{(k,r)}$ is equal to $(k+1)(r+1)$. 
A representation of the group
$\fG_+$ by such spintensors is irreducible and denoted by the symbol
$\fC^{j,j^\prime}$, where $2j=k,\;2j^\prime=r$, numbers $j$ and
$j^\prime$ defining the spin are integer or half--integer. Then the field
$\psi(x)$ transforming by the formula (\ref{6.21}) is, in general case, a field
of the type $(j,j^\prime)$. In such a way, all the physical fields are
reduced to the fields of this type, the mathematical structure of which
requires a knowledge of representation matrices $\fC^{j,j^\prime}_{\mu\nu}$. 
As a rule, in physics there are two basic types of the fields:\\[0.3cm]
1) The field of type $(j,0)$. The structure of this field 
(or the field $(0,j)$) is described by the representation
$\fC^{j,0}$ ($\fC^{0,j^\prime}$), which is realized in the
space $\Sym_{(k,0)}\subset\dS_{2^k}$
($\Sym_{(0,r)}\subset\dS_{2^r}$). At this point in accordance with
Theorem \ref{tprod} the algebra
$\C_{2k}\simeq\C_2\otimes\C_2\otimes\cdots\otimes\C_2$
(correspondingly, $\overset{\ast}{\C}_2\simeq\overset{\ast}{\C}_2\otimes
\overset{\ast}{\C}_2\otimes\cdots\otimes\overset{\ast}{\C}_2$)
is associated with the field of type $(j,0)$ (correspondingly, $(0,j^\prime)$)
The trivial case $j=0$ corresponds to {\it a Pauli--Weisskopf
field} describing the scalar particles. In other particular case, when
$j=j^\prime=1/2$ we have {\it a Weyl field} describing the neutrino.
At this point the antineutrino is described by a fundamental representation
$\fC^{1/2,0}=\sigma$ of the group $\fG_+$ and the algebra
$\C_2$ related with this representation (Theorem \ref{tprod}). Correspondingly,
the neutrino is described by a conjugated representation $\fC^{0,1/2}$
and the algebra $\overset{\ast}{\C}_2$. In relation with this, it is hardly
too much to say that the neutrino field is a more fundamental physical field,
that is a kind of the basic building block, from which other physical fields
built by means of direct sum or tensor product.\\[0.3cm]
2) The field of type $(j,0)\oplus(0,j)$. The structure of this field admits
a space inversion and, therefore, in accordance with a Wigner's doubling
\cite{Wig64} is described by a representation $\fC^{j,0}\oplus\fC^{0,j}$
of the group $\fG_+$. This representation is realized in the space
$\Sym_{(k,k)}\subset
\dS_{2^{2k}}$. In accordance with (\ref{D})
the Clifford algebra related with this representation is a direct sum
$\C_{2k}\oplus\overset{\ast}{\C}_{2k}\simeq
\C_2\otimes\C_2\otimes\cdots\otimes\C_2\oplus
\overset{\ast}{\C}_2\otimes
\overset{\ast}{\C}_2\otimes\cdots\otimes\overset{\ast}{\C}_2$.
In the simplest case $j=1/2$ we have {\it bispinor (electron--positron)
Dirac field} $(1/2,0)\oplus(0,1/2)$ with the algebra $\C_2\oplus$
\raisebox{0.21ex}
{$\overset{\ast}{\C}_2$}. It should be noted that the Dirac algebra
$\C_4$ considered as a tensor product $\C_2\otimes\C_2$ 
(or
$\C_2\otimes\overset{\ast}{\C}_2$) 
in accordance with (\ref{6.16})
(or (\ref{6.20'})) gives rise to spintensors $\xi^{\alpha_1\alpha_2}$
(or $\xi^{\alpha_1\dot{\alpha}_1}$), but it contradicts with the usual
definition of the Dirac bispinor as a pair 
$(\xi^{\alpha_1},\xi^{\dot{\alpha}_1})$. Therefore, the Clifford algebra
associated with the Dirac field is 
$\C_2\oplus\overset{\ast}{\C}_2$, and
a spinspace of this sum in virtue of unique decomposition
$\dS_2\oplus\dot{\dS}_2=\dS_4$ ($\dS_4$ is a spinspace of $\C_4$) allows to
define $\gamma$--matrices in the Weyl basis.
The case $j=1$ corresponds to 
{\it Maxwell fields} $(1,0)$ and $(0,1)$ with the algebras
$\C_2\otimes\C_2$ and $\overset{\ast}{\C}_2\otimes\overset{\ast}{\C}_2$.
At this point the electromagnetic field is defined by complex linear
combinations $\bF=\bE-i\bH,\;\overset{\ast}{\bF}=\bE+i\bH$ (Helmholtz
representation). Besides, the algebra related with Maxwell field is a tensor
product of the two algebras $\C_2$ describing the neutrino fields.
In this connection it is of interest to recall {\it a neutrino theory of
light} was proposed by de Broglie and Jordan \cite{Bro32,Jor35}.
In the de Broglie--Jordan neutrino theory of light electromagnetic field
is constructed from the two neutrino fields (for more details and related
papers see \cite{Dvo99}). Traditionally, physicists attempt to describe
electromagnetic field in the framework of $(1,0)\oplus(0,1)$ representation
(see old works \cite{LU31,Rum36,RF} and recent developments based on the
Joos--Weinberg formalism \cite{Joo62,Wein} and its relation with a
Bargmann--Wightman--Wigner type quantum field theory \cite{AJG93,Dvo97}).
However, Weinberg's equations (or Weinberg--like equations) for electromagnetic
field obtained within the subspace $\Sym_{(k,r)}$ with dimension
$2(2j+1)$ have acausal (tachionic) solutions \cite{AE92}. 
Electromagnetic field in terms of a quotient representation
$(1,0)\cup (0,1)$ in the full representation space $\dS_{2^{k+r}}$ will be
considered in separate paper.
\end{sloppypar}

In this connection it should be noted two important circumstances related
with irreducible representations of the group $\fG_+$ and complex Clifford
algebras associated with these representations. The first circumstance relates
with the Wigner interpretation of an elementary particle. Namely, a relation
between finite--dimensional representations of the proper Lorentz group
and comlex Clifford algebras (Theorem
\ref{tprod}) allows to essentially extend the Wigner interpretation by means
of the use of an extraordinary rich and universal structure of the Clifford
algebras at the study of space--time (and also intrinsic) symmetries of
elementary particles. The second circumstance relates with the spin.
Usually, the Clifford algebra is associated with a half--integer spin
corresponding to fermionic fields, so--called `matter fields', whilst the
fields with an integer spin (bosonic fields) are eliminated from an algebraic
description. However, such a non--symmetric situation is invalid, since
the fields with integer spin have a natural description in terms of 
spintensor representations of the proper Lorentz group with even rank and
algebras $\C_{2k}$ and $\overset{\ast}{\C}_{2k}$ associated with these
representations, where $k$ is even (for example, Maxwell field). In this
connection it should be noted that generalized statistics in terms of
Clifford algebras have been recently proposed by Finkelstein and
collaborators \cite{FG00,BFGS}.

As known, complex Clifford algebras $\C_n$ are modulo 2 periodic
\cite{AtBSh} and, therefore, there exist two types of $\C_n$:
$n\equiv 0\s\pmod{2}$ and $n\equiv 1\s\pmod{2}$. Let us consider these two
types in the form of following series:
\[
\begin{array}{cccccccccc}
\C_2 && \C_4 && \cdots && \C_{2k} && \cdots \\
& \C_3 && \C_5 && \cdots && \C_{2k+1} && \cdots
\end{array}
\]
Let us consider the decomposition $\C_{2k+1}\simeq\C_{2k}\oplus\C_{2k}$
in more details. This decomposition may be represented by a following scheme
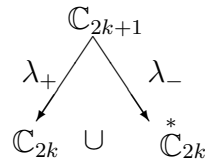
\[
\unitlength=0.5mm
\begin{picture}(70,50)
\put(35,40){\vector(2,-3){15}}
\put(35,40){\vector(-2,-3){15}}
\put(28.25,42){$\C_{2k+1}$}
\put(16,28){$\lambda_{+}$}
\put(49.5,28){$\lambda_{-}$}
\put(13.5,9.20){$\C_{2k}$}
\put(52.75,9){$\stackrel{\ast}{\C}_{2k}$}
\put(32.5,10){$\cup$}
\end{picture}
\]
Here central idempotents
\[
\lambda^+=\frac{1+\varepsilon\e_1\e_2\cdots\e_{2k+1}}{2},\quad
\lambda^-=\frac{1-\varepsilon\e_1\e_2\cdots\e_{2k+1}}{2},
\]
where
\[
\varepsilon=\begin{cases}
1,& \text{if $k\equiv 0\pmod{2}$},\\
i,& \text{if $k\equiv 1\pmod{2}$}
\end{cases}
\]
satisfy the relations $(\lambda^+)^2=\lambda^+$, $(\lambda^-)^2=\lambda^-$,
$\lambda^+\lambda^-=0$. Thus, we have a decomposition of the initial
algebra $\C_{2k+1}$ into a direct sum of two mutually annihilating simple
ideals: $\C_{2k+1}\simeq\frac{1}{2}(1+\varepsilon\omega)\C_{2k+1}\oplus
\frac{1}{2}(1-\varepsilon\omega)\C_{2k+1}$. Each of the ideals
$\lambda^{\pm}\C_{2k+1}$ is isomorphic to the subalgebra 
$\C_{2k}\subset\C_{2k+1}$. In accordance with Chisholm and Farwell \cite{CF97}
the idempotents $\lambda^+$ and $\lambda^-$ can be identified with
helicity projection operators which distinguish left and right handed
spinors. The Chisholm--Farwell notation for $\lambda^\pm$ we will widely
use below.Therefore,
in virtue of the isomorphism $\C_{2k+1}\simeq\C_{2k}\cup\C_{2k}$ and the
homomorphic mapping $\epsilon:\,\C_{2k+1}\rightarrow\C_{2k}$ the second series
(type $n\equiv 1\s\pmod{2}$) is replaced by a sequence of the quotient
algebras ${}^\epsilon\C_{2k}$, that is,
\[
\begin{array}{cccccccccc}
\C_2 && \C_4 && \cdots && \C_{2k} && \cdots \\
& {}^\epsilon\C_2 && {}^\epsilon\C_4 && \cdots && {}^\epsilon\C_{2k} && \cdots
\end{array}
\]
Representations corresponded these two series of $\C_n$ ($n\equiv 0,1\pmod{2}$)
form a full system $\fM=\fM^0\oplus\fM^1$ of finite--dimensional 
representations of the proper Lorentz group $\fG_+$.

All the physical fields used in quantum field theory and related 
representations of the group
$\fG_+$ $\fC^{j,0}\;(\C_{2k})$, $\fC^{j,0}\oplus\fC^{0,j}\;
(\C_{2k}\otimes\overset{\ast}{\C}_{2k}$) are constructed from the upper
series (type $n\equiv 0\s\pmod{2}$). Whilst the lower series
(type $n\equiv 1\s\pmod{2}$) is not considered in physics as yet.
In accordance with Theorem \ref{tprod} we have an isomorphism
${}^\epsilon\C_{2k}\simeq{}^\chi\M_{2^k}$. Therefore, the quotient algebra
${}^\epsilon\C_{2k}$ induces a spinspace
${}^\epsilon\dS_{2^k}$ that is a space of a quotient representation
${}^\chi\fC^{j,0}$ of the group
$\fG_+$. Analogously, a quotient representation ${}^\chi\fC^{0,j^\prime}$ 
is realised in the space ${}^\epsilon\dot{\dS}_{2^r}$ which
induced by the quotient algebra ${}^\epsilon\overset{\ast}{\C}_{2r}$.
In general case, we have a quotient representation ${}^\chi\fC^{j,j^\prime}$ 
defined by a tensor product ${}^\epsilon\C_{2k}\otimes
{}^\epsilon\overset{\ast}{\C}_{2r}$. Thus, {\it the complex type
$n\equiv 1\s\pmod{2}$ corresponds to a full system of irreducible
finite--dimensional quotient representations ${}^\chi\fC^{j,j^\prime}$
of the proper Lorentz group}. Therefore, until now in physics only one half
($n\equiv 0\s\pmod{2}$) of all possible finite--dimensional representations
of the Lorentz group has been used.

Let us consider now a full system of physical fields with different types.
First of all, the field
\begin{equation}\label{F1}
(j,0)=(1/2,0)\otimes(1/2,0)\otimes\cdots\otimes(1/2,0)
\end{equation}
in accordance with Theorem \ref{tprod} is a tensor product of the $k$ fields 
of type $(1/2,0)$, each of which corresponds to the fundamental representation
$\fC^{1/2,0}=\sigma$ of the group $\fG_+$ and the biquaternion
algebra $\C_2$ related with fundamental representation. In its turn, the
field
\begin{equation}\label{F2}
(0,j^\prime)=(0,1/2)\otimes(0,1/2)\otimes\cdots\otimes(0,1/2)
\end{equation}
is a tensor product of the $r$ fundamental fields of the type $(0,1/2)$,
each of which corresponds to a conjugated representation 
$\fC^{0,1/2}=\dot{\sigma}$ and the conjugated algebra
$\overset{\ast}{\C}_2$ obtained in accordance with (\ref{In1})--(\ref{In2})
under action of the automorphism $\cA\rightarrow\cA^\star$ (space inversion),
or under action of the antiautomorphism $\cA\rightarrow\widetilde{\cA}$
(time reversal). The numbers $j$ and $j^\prime$ are integer (bosonic fields) if
in the products (\ref{F1}) and (\ref{F2}) there are $k,r\equiv 0\pmod{2}$
$(1/2,0)$ (or $(0,1/2)$) factors, and the numbers $j$ and $j^\prime$ are
half--integer (fermionic fields) if in the products (\ref{F1})--(\ref{F2})
there are $k,r\equiv 1\pmod{2}$ factors. Further, the field
\begin{equation}\label{F3}
(j,j^\prime)=(1/2,0)\otimes(1/2,0)\otimes\cdots\otimes(1/2,0)\bigotimes
(0,1/2)\otimes(0,1/2)\otimes\cdots\otimes(0,1/2)
\end{equation}
is a tensor product of the fields (\ref{F1}) and (\ref{F2}). As consequence
of the doubling (\ref{D}) we have the field of type $(j,0)\oplus(0,j)$: 
\[
(j,0)\oplus(0,j)=(1/2,0)\otimes(1/2,0)\otimes\cdots\otimes(1/2,0)\bigoplus
(0,1/2)\otimes(0,1/2)\otimes\cdots\otimes(0,1/2)
\]
In general, all the fields (\ref{F1})--(\ref{F3}) describe multiparticle
states. The decompositions of these multiparticle states into single states
provided in the full representation space $\dS_{2^{k+r}}$, where
$\Sym_{(k,r)}$ and $\Sym_{(k,k)}$ with dimensions
$(2j+1)(2j^\prime+1)$ and $2(2j+1)$ are subspaces of $\dS_{2^{k+r}}$
(for example, the Clebsh--Gordan decomposition of two spin 1/2 particles
into singlet and triplet: $(1/2,1/2)=(1/2,0)\otimes(0,1/2)=(0,0)\oplus(1,0)$).
In the papers \cite{Hol88,DLG93,SLD99} a multiparticle state is described in 
the framework of a tensor product 
$\cl_{3,0}\otimes\cl_{3,0}\otimes\cdots\otimes\cl_{3,0}$. It is easy to see
that in virtue of the isomorphism $\cl_{3,0}\simeq\C_2$ the tensor product
of the algebras $\cl_{3,0}$ is isomorphic to the product (\ref{6.11'}).
Therefore, the Holland approach naturally incorporates into a more general
scheme considered here. Finally, for the type $n\equiv 1\pmod{2}$ we have
quotient representations ${}^\chi\fC^{j,j^\prime}$ of the group
$\fG_+$. The physical fields corresponding to the
quotient representations are constructed like the fields (\ref{F1})-(\ref{F3}).
%However, a conjugated field $(0,j)$ there exists 
%if and only if the automorphism $\cA\rightarrow\cA^\star$
%(or the antiautomorphism $\cA\rightarrow\widetilde{\cA}$) transfers from
%$\C_n$ to $\C_{n-1}$ under action of the homomorphism 
%$\epsilon:\C_n\rightarrow\C_{n-1}$. 
Due to the decomposition
$\C_n\simeq\C_{n-1}\cup\C_{n-1}$ ($n\equiv 1\pmod{2}$) we have a field
\begin{equation}\label{F4}
(j,0)\cup(j,0)=(1/2,0)\otimes(1/2,0)\otimes\cdots\otimes(1/2,0)\bigcup
(1/2,0)\otimes(1/2,0)\otimes\cdots\otimes(1/2,0),
\end{equation}
and also we have fields $(0,j)\cup(0,j)$ and $(j,0)\cup(0,j)$ if the quotient
algebras ${}^\epsilon\C_{n-1}$ admit space inversion or time reversal.
The field
\begin{equation}\label{F5}
(j,0)\cup(0,j)=(1/2,0)\otimes(1/2,0)\otimes\cdots\otimes(1/2,0)\bigcup
(0,1/2)\otimes(0,1/2)\otimes\cdots\otimes(0,1/2)
\end{equation}
is analogous to the field $(j,0)\oplus(0,j)$, but, in general, the field
$(j,0)\cup(0,j)$ has a quantity of violated discrete symmetries.
An explicit form of the quotient representations and their relations with
discrete symmetries will be explored in the following sections.
\section{Discrete symmetries on the representation spaces of the
Lorentz group}
Since all the physical fields are defined in terms of finite--dimensional
representations of the group $\fG_+$, then a construction of the
discrete symmetries 
(space inversion $P$, time reversal $T$ and combination 
$PT$) on the representation spaces of the Lorentz group has a primary
importance.

In the recent paper \cite{Var99} it has been shown that the space inversion
$P$, time reversal $T$ and full reflection $PT$ correspond to
fundamental automorphisms
$\cA\rightarrow\cA^\star$ (involution),
$\cA\rightarrow\widetilde{\cA}$ (reversion) and
$\cA\rightarrow\widetilde{\cA^{\star}}$ (conjugation) of the Clifford
algebra $\cl$. 
Moreover, there exists an isomorphism between a discrete subgroup
$\{1,P,T,PT\}\simeq\dZ_2\otimes\dZ_2$
($P^2=T^2=(PT)^2=1,\;PT=TP$) of the orthogonal group $O(p,q)$ 
and an automorphism group
$\Aut(\cl)=\{\Id,\star,\widetilde{\phantom{cc}},\widetilde{\star}\}$:
\begin{equation}\label{6.22}
{\renewcommand{\arraystretch}{1.4}
%{\renewcommand{\arraystretch}{0.85}
\begin{tabular}{|c||c|c|c|c|}\hline
        & $\Id$ & $\star$ & $\widetilde{\phantom{cc}}$ & $\widetilde{\star}$\\ \hline\hline
$\Id$   & $\Id$ & $\star$ & $\widetilde{\phantom{cc}}$ & $\widetilde{\star}$\\ \hline
$\star$ & $\star$ & $\Id$ & $\widetilde{\star}$ & $\widetilde{\phantom{cc}}$\\ \hline
$\widetilde{\phantom{cc}}$ & $\widetilde{\phantom{cc}}$ &$\widetilde{\star}$
& $\Id$ & $\star$ \\ \hline
$\widetilde{\star}$ & $\widetilde{\star}$ & $\widetilde{\phantom{cc}}$ &
$\star$ & $\Id$\\ \hline
\end{tabular}
\;\sim\;
\begin{tabular}{|c||c|c|c|c|}\hline
    & $1$ & $P$ & $T$ & $PT$\\ \hline\hline
$1$ & $1$ & $P$ & $T$ & $PT$\\ \hline
$P$ & $P$ & $1$ & $PT$& $T$\\ \hline
$T$ & $T$ & $PT$& $1$ & $P$\\ \hline
$PT$& $PT$& $T$ & $P$ & $1$\\ \hline
\end{tabular}
}
\end{equation}
Further, in the case $P^2=T^2=(PT)^2=\pm 1$ and $PT=-TP$ there is an isomorphism
between the group $\{1,P,T,PT\}$ and automorphism group
$\sAut(\cl)=\{\sI,\sW,\sE,\sC\}$.
Spinor representations of the fundamental automorphisms of the algebras
$\C_n$ was first obtained by Rashevskii in 1955
\cite{Rash}: 1) Involution: $\sA^\star=\sW\sA\sW^{-1}$, where $\sW$ is a matrix
of the automorphism $\star$ (matrix representation of the volume element
$\omega$); 2) Reversion: $\widetilde{\sA}=\sE\sA^{\sT}\sE^{-1}$, where
$\sE$ is a matrix of the antiautomorphism $\widetilde{\phantom{cc}}$
satisfying the conditions $\cE_i\sE-\sE\cE^{\sT}_i=0$ and
$\sE^{\sT}=(-1)^{\frac{m(m-1)}{2}}\sE$, here $\cE_i=\gamma(\e_i)$ are
matrix representations of the units of the algebra $\cl$; 3) Conjugation:
$\widetilde{\sA^\star}=\sC\sA^{\sT}\sC^{-1}$, where $\sC=\sE\sW^{\sT}$ 
is a matrix of the antiautomorphism
$\widetilde{\star}$ satisfying the conditions
$\sC\cE^{\sT}_i+\cE_i\sC=0$ and
$\sC^{\sT}=(-1)^{\frac{m(m+1)}{2}}\sC$.
So, for the Dirac algebra $\C_4$ in the canonical
$\gamma$--basis there exists a standard (Wigner) representation
$P=\gamma_0$ and $T=\gamma_1\gamma_3$ \cite{BLP89}, therefore,
$\{1,P,T,PT\}=\{1,\gamma_0,\gamma_1\gamma_3,\gamma_0\gamma_1\gamma_3\}$.
On the other hand, the automorphism group of the algebra $\C_4$ 
for $\gamma$--basis has a form
$\sAut(\C_4)=\{\sI,\sW,\sE,\sC\}=
\{\sI,\gamma_0\gamma_1\gamma_2\gamma_3,\gamma_1\gamma_3,\gamma_0\gamma_2\}$.
In \cite{Var99} it is shown that  $\{1,P,T,PT\}=
\{1,\gamma_0,\gamma_1\gamma_3,\gamma_0\gamma_1\gamma_3\}\simeq\sAut(\C_4)
\simeq\dZ_4$, where $\dZ_4$ is a complex group with the signature
$(+,-,-)$. 

In general case, according to Theorem \ref{tprod} a space of the
finite--dimensional representation of the group $SL(2;\C)$ is a spinspace
$\dS_{2^{k+r}}$, or a minimal left ideal of the algebra 
$\C_{2k}\otimes\overset{\ast}{\C}_{2r}$. 
Therefore, in the spinor representation the
fundamental automorphisms of the algebra
$\C_{2k}\otimes\overset{\ast}{\C}_{2r}$ 
(all spinor representations of the
fundamental automorphisms have been found in \cite{Var00}) by virtue of the
isomorphism (\ref{6.22}) induce discrete transformations
on the representation spaces (spinspaces) of the Lorentz group.
\begin{theorem}\label{tinf}
1) The field $\F=\C$. The tensor products 
$\C_2\otimes\C_2\otimes\cdots\otimes\C_2$, 
$\overset{\ast}{\C}_2\otimes\overset{\ast}{\C}_2\otimes\cdots\otimes
\overset{\ast}{\C}_2$, 
$\C_2\otimes\C_2\otimes\cdots\otimes\C_2\otimes
\overset{\ast}{\C}_2\otimes\overset{\ast}{\C}_2\otimes\cdots
\overset{\ast}{\C}_2$ of the Pauli algebra $\C_2$ correspond to
finite--dimensional representations $\fC^{l_0+l_1-1,0}$, $\fC^{0,l_0-l_1+1}$,
$\fC^{l_0+l_1-1,l_0-l_1+1}$ of the proper Lorentz group $\fG_+$, where
$(l_0,l_1)=\left(\frac{k}{2},\frac{k}{2}+1\right)$, $(l_0,l_1)=
\left(-\frac{r}{2},\frac{r}{2}+1\right)$, $(l_0,l_1)=\left(\frac{k-r}{2},
\frac{k+r}{2}+1\right)$, and spinspaces $\dS_{2^k}$, $\dS_{2^r}$,
$\dS_{2^{k+r}}$ are representation spaces of the group $\fG_+$,
$\C_2\leftrightarrow\fC^{1,0}$ 
($\overset{\ast}{\C}_2\leftrightarrow\fC^{0,-1}$) 
is a fundamental
representation of $\fG_+$. Then in a spinor representation of the
fundamental automorphisms of the algebra $\C_n$ for the matrix $\sW$ of the
automorphism $\cA\rightarrow\cA^\star$ (space inversion) and also for the
matrices $\sE$ and $\sC$ of the antiautomorphisms $\cA\rightarrow
\widetilde{\cA}$ (time reversal) and $\cA\rightarrow\widetilde{\cA^\star}$
(full reflection) the following permutation relations with infinitesimal
operators of the group $\fG_+$ take place:
\begin{alignat}{7}
\ld\sW,A_{23}\rd &=\ld\sW,A_{13}\rd &=\ld\sW,A_{12}\rd &=0,\quad
\lf\sW,B_1\rf &=\lf\sW,B_2\rf &=\lf\sW,B_3\rf &=0\label{T0}\\
\ld\sW,H_+\rd &=\ld\sW,H_-\rd &=\ld\sW,H_3\rd &=0,\quad
\lf\sW,F_+\rf &=\lf\sW,F_-\rf &=\lf\sW,F_3\rf &=0.\label{T0'}
\end{alignat}
\begin{alignat}{7}
\ld\sE,A_{23}\rd &=\ld\sE,A_{13}\rd &=\ld\sE,A_{12}\rd &=0,\quad
\lf\sE,B_1\rf &=\lf\sE,B_2\rf &=\lf\sE,B_3\rf &=0,\label{T1}\\
\ld\sC,A_{23}\rd &=\ld\sC,A_{13}\rd &=\ld\sC,A_{12}\rd &=0,\quad
\ld\sC,B_1\rd &=\ld\sC,B_2\rd &=\ld\sC,B_3\rd &=0,\label{T2}\\
\ld\sE,H_+\rd &=\ld\sE,H_-\rd &=\ld\sE,H_3\rd &=0,\quad
\lf\sE,F_+\rf &=\lf\sE,F_-\rf &=\lf\sE,F_3\rf &=0,\label{T3}\\
\ld\sC,H_+\rd &=\ld\sC,H_-\rd &=\ld\sC,H_3\rd &=0,\quad
\ld\sC,F_+\rd &=\ld\sC,F_-\rd &=\ld\sC,F_3\rd &=0.\label{T4}
\end{alignat}
\begin{alignat}{7}
\ld\sE,A_{23}\rd &=\ld\sE,A_{13}\rd &=\ld\sE,A_{12}\rd &=0,\quad
\ld\sE,B_1\rd &=\ld\sE,B_2\rd &=\ld\sE,B_3\rd &=0,\label{T5}\\
\ld\sC,A_{23}\rd &=\ld\sC,A_{13}\rd &=\ld\sC,A_{12}\rd &=0,\quad
\lf\sC,B_1\rf &=\lf\sC,B_2\rd &=\lf\sC,B_3\rd &=0,\label{T6}\\
\ld\sE,H_+\rd &=\ld\sE,H_-\rd &=\ld\sE,H_3\rd &=0,\quad
\ld\sE,F_+\rd &=\ld\sE,F_-\rd &=\ld\sE,F_3\rd &=0,\label{T7}\\
\ld\sC,H_+\rd &=\ld\sC,H_-\rd &=\ld\sC,H_3\rd &=0,\quad
\lf\sC,F_+\rf &=\lf\sC,F_-\rf &=\lf\sC,F_3\rf &=0.\label{T8}
\end{alignat}
\begin{alignat}{7}
\ld\sE,A_{23}\rd &=0,\quad\lf\sE,A_{13}\rf &=\lf\sE,A_{12}\rf &=0,\quad
\ld\sE,B_1\rd &=0,\quad\lf\sE,B_2\rf &=\lf\sE,B_3\rf &=0,\label{T9}\\
\ld\sC,A_{23}\rd &=0,\quad\lf\sC,A_{13}\rf &=\lf\sC,A_{12}\rf &=0,\quad
\lf\sC,B_1\rf &=0,\quad\ld\sC,B_2\rd &=\ld\sC,B_3\rd &=0.\label{T10}
\end{alignat}
\begin{alignat}{7}
\ld\sE,A_{23}\rd &=0,\quad\lf\sE,A_{13}\rf &=\lf\sE,A_{12}\rf &=0,\quad
\lf\sE,B_1\rf &=0,\quad\ld\sE,B_2\rd &=\ld\sE,B_3\rd &=0,\label{T11}\\
\ld\sC,A_{23}\rd &=0,\quad\lf\sC,A_{13}\rf &=\lf\sC,A_{12}\rf &=0,\quad
\ld\sC,B_1\rd &=0,\quad\lf\sC,B_2\rf &=\ld\sC,B_3\rf &=0.\label{T12}
\end{alignat}
\begin{alignat}{7}
\lf\sE,A_{23}\rf &=\lf\sE,A_{13}\rf &=0,\quad\ld\sE,A_{12}\rd &=0,\quad
\ld\sE,B_1\rd &=\ld\sE,B_2\rd &=0,\quad\lf\sE,B_3\rf &=0,\label{T13}\\
\lf\sC,A_{23}\rf &=\lf\sC,A_{13}\rf &=0,\quad\ld\sC,A_{12}\rd &=0,\quad
\lf\sC,B_1\rf &=\lf\sC,B_2\rf &=0,\quad\ld\sC,B_3\rd &=0,\label{T14}\\
\lf\sE,H_+\rf &=\lf\sE,H_-\rf &=0,\quad\ld\sE,H_3\rd &=0,\quad
\ld\sE,F_+\rd &=\ld\sE,F_-\rd &=0,\quad\lf\sE,F_3\rf &=0,\label{T15}\\
\lf\sC,H_+\rf &=\lf\sC,H_-\rf &=0,\quad\ld\sC,H_3\rd &=0,\quad
\lf\sC,F_+\rf &=\lf\sC,F_-\rf &=0,\quad\ld\sC,F_3\rd &=0.\label{T16}
\end{alignat}
\begin{alignat}{7}
\lf\sE,A_{23}\rf &=0,\;\;\ld\sE,A_{13}\rd &=0,\;\;\lf\sE,A_{12}\rf &=0,\;\;
\ld\sE,B_1\rd &=0,\;\;\lf\sE,B_2\rf &=0,\;\;\ld\sE,B_3\rd &=0,\label{T17}\\
\lf\sC,A_{23}\rf &=0,\;\;\ld\sC,A_{13}\rd &=0,\;\;\lf\sC,A_{12}\rf &=0,\;\;
\lf\sC,B_1\rf &=0,\;\;\ld\sC,B_2\rd &=0,\;\;\lf\sC,B_3\rf &=0.\label{T18}
\end{alignat}
\begin{alignat}{7}
\lf\sE,A_{23}\rf &=0,\;\;\ld\sE,A_{13}\rd &=0,\;\;\lf\sE,A_{12}\rf &=0,\;\;
\lf\sE,B_1\rf &=0,\;\;\ld\sE,B_2\rd &=0,\;\;\lf\sE,B_3\rf &=0,\label{T19}\\
\lf\sC,A_{23}\rf &=0,\;\;\ld\sC,A_{13}\rd &=0,\;\;\lf\sC,A_{12}\rf &=0,\;\;
\ld\sC,B_1\rd &=0,\;\;\lf\sC,B_2\rf &=0,\;\;\ld\sC,B_3\rd &=0.\label{T20}
\end{alignat}
\begin{alignat}{7}
\lf\sE,A_{23}\rf &=\lf\sE,A_{13}\rf &=0,\quad\ld\sE,A_{12}\rd &=0,\quad
\lf\sE,B_1\rf &=\lf\sE,B_2\rf &=0,\quad\ld\sE,B_3\rd &=0,\label{T21}\\
\lf\sC,A_{23}\rf &=\lf\sC,A_{13}\rf &=0,\quad\ld\sC,A_{12}\rd &=0,\quad
\ld\sC,B_1\rd &=\ld\sC,B_2\rd &=0,\quad\lf\sC,B_3\rf &=0,\label{T22}\\
\lf\sE,H_+\rf &=\lf\sE,H_-\rf &=0,\quad\ld\sE,H_3\rd &=0,\quad
\lf\sE,F_+\rf &=\lf\sE,F_-\rf &=0,\quad\ld\sE,F_3\rd &=0,\label{T23}\\
\lf\sC,H_+\rf &=\lf\sC,H_-\rf &=0,\quad\ld\sC,H_3\rd &=0,\quad
\ld\sC,F_+\rd &=\ld\sC,F_-\rd &=0,\quad\lf\sC,F_3\rf &=0.\label{T24}
\end{alignat}
where $A_{23},\,A_{13},\,A_{12}$ are infinitesimal operators of a subgroup of
three--dimensional rotations, $B_1,\,B_2,\,B_3$ are infinitesimal operators of
hyperbolic rotations.\\[0.2cm]
2) The field $\F=\R$. The factorization $\cl_{s_i,t_j}\otimes\cl_{s_i,t_j}
\otimes\cdots\otimes\cl_{s_i,t_j}$ of the real Clifford algebra $\cl_{p,q}$
corresponds to a real finite--dimensional representation of the group $\fG_+$
with a pair $(l_0,l_1)=\left(\frac{p+q}{4},0\right)$, that is equivalent to a
representation of the subgroup $SO(3)$ of three--dimensional rotations
($B_1=B_2=B_3=0$). Then there exist two classes of real representations
$\fR^{l_0}_{0,2}$ of the group $\fG_+$ corresponding to the algebras
$\cl_{p,q}$ with a division ring $\K\simeq\R$, $p-q\equiv 0,2\pmod{8}$,
and also there exist two classes of quaternionic representations 
$\fH^{l_0}_{4,6}$ of $\fG_+$ corresponding to the algebras $\cl_{p,q}$
with a ring $\K\simeq\BH$, $p-q\equiv 4,6\pmod{8}$. For the real
representations $\fR^{l_0}_{0,2}$ operators of the discrete subgroup of 
$\fG_+$ defining by the matrices $\sW,\,\sE,\,\sC$ of the fundamental
automorphisms of $\cl_{p,q}$ with $p-q\equiv 0,2\pmod{8}$ are always
commute with all the infinitesimal operators of the representation.
In turn, for the quaternionic representations $\fH^{l_0}_{4,6}$ following
relations hold:
\begin{equation}
\ld\sW,A_{23}\rd=\ld\sW,A_{13}\rd=\ld\sW,A_{12}\rd=0,\quad
\ld\sW,H_+\rd=\ld\sW,H_-\rd=\ld\sW,H_3\rd=0.
\end{equation}
\begin{alignat}{7}
\ld\sE,A_{23}\rd &=\ld\sE,A_{13}\rd &=\ld\sE,A_{12}\rd &=0,\quad
\ld\sC,A_{23}\rd &=\ld\sC,A_{13}\rd &=\ld\sC,A_{12}\rd &=0,\label{TR11}\\
\ld\sE,H_+\rd &=\ld\sE,H_-\rd &=\ld\sE,H_3\rd &=0,\quad
\ld\sC,H_+\rd &=\ld\sC,H_-\rd &=\ld\sC,H_3\rd &=0.
\end{alignat}
\begin{equation}
\ld\sE,A_{23}\rd=0,\quad\lf\sE,A_{13}\rf=\lf\sE,A_{12}\rf=0,\quad
\ld\sC,A_{23}\rd=0,\quad\lf\sC,A_{13}\rf=\lf\sC,A_{12}\rf=0.\label{TR12}
\end{equation}
\begin{alignat}{7}
\lf\sE,A_{23}\rf &=\lf\sE,A_{13}\rf &=0,\quad\ld\sE,A_{12}\rd=0,\quad
\lf\sC,A_{23}\rf &=\lf\sC,A_{13}\rf &=0,\quad\ld\sC,A_{12}\rd=0,\\
\lf\sE,H_+\rf &=\lf\sE,H_-\rf &=0,\quad\ld\sE,H_3\rd=0,\quad
\lf\sC,H_+\rf &=\lf\sC,H_-\rf &=0,\quad\ld\sC,H_3\rd=0.
\end{alignat}
\begin{equation}
\lf\sE,A_{23}\rf=0,\;\;\ld\sE,A_{13}\rd=0,\;\;\lf\sE,A_{12}\rf=0,\;\;
\lf\sC,A_{23}\rf=0,\;\;\ld\sC,A_{13}\rd=0,\;\;\lf\sC,A_{12}\rf=0.
\end{equation}
\end{theorem}
\begin{proof}
1) Complex representations.\\
As noted previously, a full representation space of the finite--dimensional
representation of the proper Lorentz group $\fG_+$ is defined in terms of
the minimal left ideal of the algebra $\C_2\otimes\C_2\otimes\cdots\C_2
\simeq\C_{2k}$. Indeed, in virtue of an isomorphism
\begin{equation}\label{Iso}
\C_{2k}\simeq\cl_{p,q},
\end{equation}
where $\cl_{p,q}$ is a Clifford algebra over the field $\F=\R$ with a
division ring $\K\simeq\C$, $p-q\equiv 3,7\pmod{8}$, we have for the minimal
left ideal of $\C_{2k}$ an expression $\dS=\cl_{p,q}f$, here
\[
f=\frac{1}{2}(1\pm\e_{\alpha_1})\frac{1}{2}(1\pm\e_{\alpha_2})\cdots
\frac{1}{2}(1\pm\e_{\alpha_t})
\]
is a primitive idempotent of the algebra $\cl_{p,q}$ \cite{Lou81}, and
$\e_{\alpha_1},\e_{\alpha_2},\ldots,\e_{\alpha_t}$ are commuting elements
with square 1 of the canonical basis of $\cl_{p,q}$ generating a group of
order $2^t$. The values of $t$ are defined by a formula $t=q-r_{q-p}$,
where $r_i$ are the Radon--Hurwitz numbers \cite{Rad22,Hur23}, values of
which form a cycle of the period 8: $r_{i+8}=r_i+4$. The values of all $r_i$
are
\begin{center}
\begin{tabular}{lcccccccc}
$i$  & 0 & 1 & 2 & 3 & 4 & 5 & 6 & 7\\ \hline
$r_i$& 0 & 1 & 2 & 2 & 3 & 3 & 3 & 3
\end{tabular}
\end{center}
The dimension of the minimal left ideal $\dS$ is equal to 
$2^k=2^{\frac{p+q-1}{2}}$. Therefore, for the each finite--dimensional
representation of the group $\fG_+$ we have $2^t$ copies of the spinspace
$\dS_{2^k}$ (full representation space). It should be noted that not all
these copies are equivalent to each other, some of them give rise to
different reflection groups (see \cite{Var00}).

In general, all the finite--dimensional representations of 
group $\fG_+$ in the spinspace $\dS_{2^k}$
are reducible.
Therefore, there exists a decomposition of the spinspace
$\dS_{2^k}\simeq\dS_2\otimes\dS_2\otimes\cdots\dS_2$ into a direct sum of
invariant subspaces $\Sym_{(k_j,0)}$ of symmetric spintensors with
dimensions $(k_j+1)$:
\[
\dS_{2^k}=\Sym_{(k_1,0)}\oplus\Sym_{(k_2,0)}\oplus\cdots\oplus\Sym_{k_s,0)},
\]
where $k_1+k_2+\ldots+k_s=k$, $k_j\in\dZ$. At this point there exists an
orthonormal basis with matrices of the form
\[
\ar\begin{pmatrix}
A^0_t & & & & & & &\\
& A^{1/2}_t & & & & & &\\
& & \ddots & & & & & \\
& & & A^s_t & & & &\\
& & & & B^0_t & & &\\
& & & & & B^{1/2}_t & &\\
& & & & & & \ddots &\\
& & & & & & & B^s_t
\end{pmatrix},
\]
where for the matrices $A^i_1,\,A^j_2,\,A^j_3,\,B^j_1,\,B^j_2,\,B^j_3$
(matrices of the infinitesimal operators of $\fG_+$) in accordance with
Gel'fand--Naimark formulas \cite{GMS,Nai58}
\begin{equation}\label{I1}
A_{23}\xi_{l,m}=-\frac{i}{2}\sqrt{(l+m+1)(l-m)}\xi_{l,m+1}-
\frac{i}{2}\sqrt{(l+m)(l-m+1)}\xi_{l,m-1},
\end{equation}
\begin{equation}\label{I2}
A_{13}\xi_{l,m}=\frac{1}{2}\sqrt{(l+m)(l-m+1)}\xi_{l,m-1}-
\frac{1}{2}\sqrt{(l+m+1)(l-m)}\xi_{l,m+1},
\end{equation}
\begin{equation}\label{I3}
A_{12}\xi_{l,m}=-im\xi_{l,m},
\end{equation}
\begin{multline}\label{I4}
B_1\xi_{l,m}=-\frac{i}{2}C_l\sqrt{(l-m)(l-m-1)}\xi_{l-1,m+1}+
\frac{i}{2}A_l\sqrt{(l-m)(l+m+1)}\xi_{l,m+1}-\\
\frac{i}{2}C_{l+1}\sqrt{(l+m+1)(l+m+2)}\xi_{l+1,m+1}+
\frac{i}{2}C_l\sqrt{(l+m)(l+m-1)}\xi_{l-1,m-1}+\\
\frac{i}{2}A_l\sqrt{(l+m)(l-m+1)}\xi_{l,m-1}+
\frac{i}{2}C_{l+1}\sqrt{(l-m+1)(l-m+2)}\xi_{l+1,m-1},
\end{multline}
\begin{multline}\label{I5}
B_2\xi_{l,m}=-\frac{1}{2}C_l\sqrt{(l+m)(l+m-1)}\xi_{l-1,m-1}-
\frac{1}{2}A_l\sqrt{(l+m)(l-m+1)}\xi_{l,m-1}-\\
\frac{1}{2}C_{l+1}\sqrt{(l-m+1)(l-m+2)}\xi_{l+1,m-1}-
\frac{1}{2}C_l\sqrt{(l-m)(l-m-1)}\xi_{l-1,m+1}+\\
\frac{1}{2}A_l\sqrt{(l-m)(l+m+1)}\xi_{l,m+1}-
\frac{1}{2}C_{l+1}\sqrt{(l+m+1)(l+m+2)}\xi_{l+1,m+1},
\end{multline}
\begin{equation}\label{I6}
B_3\xi_{l,m}=-iC_l\sqrt{l^2-m^2}\xi_{l-1,m}+iA_lm\xi_{l,m}+
iC_{l+1}\sqrt{(l+1)^2-m^2}\xi_{l+1,m},
\end{equation}
\begin{equation}\label{I6'}
A_l=\frac{il_0l_1}{l(l+1)},\quad
C_l=\frac{i}{l}\sqrt{\frac{(l^2-l^2_0)(l^2-l^2_1)}{4l^2-1}}
\end{equation}
\begin{gather}
m=-l,-l+1,\ldots,l-1,l\nonumber\\
l=l_0,l_0+1,\ldots\nonumber
\end{gather}
we have
\begin{equation}\label{A1}
A^j_{23}=-\frac{i}{2}
\ar\begin{bmatrix}
0 & \alpha_{-l_j+1} & 0 & \dots & 0 & 0\\
\alpha_{-l_j+1} & 0 & \alpha_{-l_j+2} & \dots & 0 & 0\\
0 & \alpha_{-l_j+2} & 0 & \dots & 0 & 0\\
\hdotsfor[2]{6}\\
\hdotsfor[2]{6}\\
0 & 0 & 0 & \dots & 0 & \alpha_{l_j}\\
0 & 0 & 0 & \dots & \alpha_{l_j} & 0
\end{bmatrix}
\end{equation}
\begin{equation}\label{A2}
A^j_{13}=\frac{1}{2}
\ar\begin{bmatrix}
0 & \alpha_{-l_j+1} & 0 & \dots & 0 & 0\\
-\alpha_{-l_j+1} & 0 & \alpha_{-l_j+2} & \dots & 0 & 0\\
0 & -\alpha_{-l_j+2} & 0 & \dots & 0 & 0\\
\hdotsfor[2]{6}\\
\hdotsfor[2]{6}\\
0 & 0 & 0 & \dots & 0 & \alpha_{l_j}\\
0 & 0 & 0 & \dots & -\alpha_{l_j} & 0
\end{bmatrix}
\end{equation}
\begin{equation}\label{A3}
A^j_{12}=
\ar\begin{bmatrix}
il_j & 0 & 0 & \dots & 0 & 0\\
0 & i(l_j-1) & 0 & \dots & 0 & 0\\
0 & 0 & i(l_j-2) & \dots & 0 & 0\\
\hdotsfor[2]{6}\\
\hdotsfor[2]{6}\\
0 & 0 & 0 & \dots & -i(l_j-1) & 0\\
0 & 0 & 0 & \dots & 0 & -il_j
\end{bmatrix}
\end{equation}
\begin{equation}\label{B1}
B^j_1=\frac{i}{2}A_j
\ar\begin{bmatrix}
0 & \alpha_{-l_j+1} & 0 & \dots & 0 & 0\\
\alpha_{-l_j+1} & 0 & \alpha_{-l_j+2} & \dots & 0 & 0\\
0 & \alpha_{-l_j+2} & 0 & \dots & 0 & 0\\
\hdotsfor[2]{6}\\
\hdotsfor[2]{6}\\
0 & 0 & 0 & \dots & 0 & \alpha_{l_j}\\
0 & 0 & 0 & \dots & \alpha_{l_j} & 0
\end{bmatrix}
\end{equation}
\begin{equation}\label{B2}
B^j_2=\frac{1}{2}A_j
\ar\begin{bmatrix}
0 & -\alpha_{-l_j+1} & 0 & \dots & 0 & 0\\
\alpha_{-l_j+1} & 0 & -\alpha_{-l_j+2} & \dots & 0 & 0\\
0 & \alpha_{-l_j+2} & 0 & \dots & 0 & 0\\
\hdotsfor[2]{6}\\
\hdotsfor[2]{6}\\
0 & 0 & 0 & \dots & 0 & -\alpha_{l_j}\\
0 & 0 & 0 & \dots & \alpha_{l_j} & 0
\end{bmatrix}
\end{equation}
\begin{equation}\label{B3}
B^j_3=\frac{1}{2}A_j
\ar\begin{bmatrix}
il_j & 0 & 0 & \dots & 0 & 0\\
0 & i(l_j-1) & 0 & \dots & 0 & 0\\
0 & 0 & i(l_j-2) & \dots & 0 & 0\\
\hdotsfor[2]{6}\\
\hdotsfor[2]{6}\\
0 & 0 & 0 & \dots & -i(l_j-1) & 0\\
0 & 0 & 0 & \dots & 0 & -il_j
\end{bmatrix}
\end{equation}
where $\alpha_m=\sqrt{(l_j+m)(l_j-m+1)}$. The formulas (\ref{I1})--(\ref{I6'})
define a finite--dimensional representation of the group $\fG_+$ when
$l^2_1=(l_0+p)^2$, $p$ is some natural number,
$l_0$ is an integer or half--integer number, $l_1$ is an arbitrary
complex number. In the case $l^2_1\neq(l_0+p)^2$ we have an
infinite--dimensional representation of $\fG_+$. We will deal below only
with the finite--dimensional representations, because these representations
are most useful in physics.

The relation between the numbers $l_0$, $l_1$ and the number $k$ of the
factors $\C_2$ in the product $\C_2\otimes\C_2\otimes\cdots\otimes\C_2$ is
given by a following formula
\[
(l_0,l_1)=\left(\frac{k}{2},\frac{k}{2}+1\right),
\]
whence it immediately follows that $k=l_0+l_1-1$. Thus, we have
{\it a complex representation $\fC^{l_0+l_1-1,0}$ of the proper Lorentz
group $\fG_+$ in the spinspace $\dS_{2^k}$}.

Let us calculate now infinitesimal operators of the fundamental representation
$\fC^{1,0}$ of $\fG_+$. The representation $\fC^{1,0}$ is defined by a pair
$(l_0,l_1)=\left(\frac{1}{2},\frac{3}{2}\right)$. In accordance with
(\ref{I1})--(\ref{I6'}) from (\ref{A1})--(\ref{B3}) we obtain
\begin{align}
A^{1/2}_{23} &=-\frac{i}{2}\ar\begin{bmatrix}
0 & \alpha_{1/2}\\
\alpha_{1/2} & 0
\end{bmatrix}=-\frac{i}{2}\ar\begin{bmatrix}
0 & 1\\
1 & 0
\end{bmatrix},\label{18}\\[0.2cm]
A^{1/2}_{13} &=\frac{1}{2}\ar\begin{bmatrix}
0 & \alpha_{1/2}\\
-\alpha_{1/2} & 0
\end{bmatrix}=\frac{1}{2}\ar\begin{bmatrix}
0 & 1\\
-1 & 0
\end{bmatrix},\label{19}\\[0.2cm]
A^{1/2}_{12} &=\frac{1}{2}\ar\begin{bmatrix}
i & 0\\
0 & -i
\end{bmatrix},\label{20}\\[0.2cm]
B^{1/2}_1 &=\frac{i}{2}A_{1/2}\ar\begin{bmatrix}
0 & \alpha_{1/2}\\
\alpha_{1/2} & 0
\end{bmatrix}=-\frac{1}{2}\ar\begin{bmatrix}
0 & 1\\
1 & 0
\end{bmatrix},\label{21}\\[0.2cm]
B^{1/2}_2 &=\frac{1}{2}A_{1/2}\ar\begin{bmatrix}
0 & -\alpha_{1/2}\\
\alpha_{1/2} & 0
\end{bmatrix}=\frac{1}{2}\ar\begin{bmatrix}
0 & -i\\
i & 0
\end{bmatrix},\label{22}\\[0.2cm]
B^{1/2}_3 &=\frac{i}{2}A_{1/2}\ar\begin{bmatrix}
1 & 0\\
0 & -1
\end{bmatrix}=\frac{1}{2}\ar\begin{bmatrix}
-1 & 0\\
0 & 1
\end{bmatrix}.\label{23}
\end{align}
The operators (\ref{18})--(\ref{23}) satisfy the relations
\begin{alignat}{3}
\ld A_{23},A_{13}\rd &=A_{12}, & \quad \ld A_{13},A_{12}\rd &=A_{23}, \quad
\ld A_{12},A_{23}\rd &=A_{13},\nonumber\\
\ld B_1,B_2\rd &=-A_{12}, & \quad \ld B_2,B_3\rd &=A_{23}, \quad
\ld B_3,B_1\rd &=A_{13},\nonumber\\
\ld A_{23},B_1\rd &=0, & \quad \ld A_{13},B_2\rd &=0, \quad
\ld A_{12},B_3\rd &=0,\nonumber\\
\ld A_{23},B_2\rd &=-B_3, & \quad \ld A_{23},B_3\rd &=B_2,&&\nonumber\\
\ld A_{13},B_3\rd &=-B_1, & \quad \ld A_{13},B_1\rd &=B_3,&&\nonumber\\
\ld A_{12},B_1\rd &=B_2, & \quad \ld A_{12},B_2\rd &=-B_1.&&\label{commut}
\end{alignat}
From (\ref{21})--(\ref{23}) it is easy to see that there is an equivalence
between infinitesimal operators $B^{1/2}_i$ and Pauli matrices:
\begin{equation}\label{IF1}
B^{1/2}_1=-\frac{1}{2}\sigma_1,\quad B^{1/2}_2=\frac{1}{2}\sigma_2,\quad
B^{1/2}_3=-\frac{1}{2}\sigma_3.
\end{equation}
In its turn, from (\ref{18})--(\ref{20}) it follows
\begin{equation}\label{IF2}
A^{1/2}_{23}=-\frac{1}{2}\sigma_2\sigma_3,\quad
A^{1/2}_{13}=-\frac{1}{2}\sigma_1\sigma_2,\quad
A^{1/2}_{12}=\frac{1}{2}\sigma_1\sigma_2.
\end{equation}
It is obvious that this equivalence takes place also for high dimensions,
that is, there exists an equivalence between infinitesimal operators
(\ref{A1})--(\ref{B3}) and tensor products of the Pauli matrices.
In such a way, let us suppose that 
\begin{alignat}{3}
A^j_{23} &\sim-\frac{1}{2}\cE_a\cE_b, & \quad
A^j_{13} &\sim-\frac{1}{2}\cE_c\cE_b, & \quad
A^j_{12} &\sim\frac{1}{2}\cE_c\cE_a,\label{O1}\\
B^j_1 &\sim-\frac{1}{2}\cE_c, & \quad
B^j_2 &\sim\frac{1}{2}\cE_a, & \quad
B^j_3 &\sim-\frac{1}{2}\cE_b,\label{O2}
\end{alignat}
where $\cE_i$ ($i=a,b,c$) are $k$--dimensional matrices (the tensor products
(\ref{6.6})) and $c<a<b$. It is easy to verify that the operators
(\ref{O1})--(\ref{O2}) satisfy the relations (\ref{commut}). Indeed,
for the commutator $\ld A_{23},A_{13}\rd$ we obtain
\begin{multline}
\ld A_{23},A_{13}\rd=A_{23}A_{13}-A_{13}A_{23}=
\frac{1}{4}\cE_a\cE_b\cE_c\cE_b-\\
-\frac{1}{4}\cE_c\cE_b\cE_a\cE_b=-\frac{1}{4}\cE_a\cE_c+
\frac{1}{4}\cE_c\cE_a=\frac{1}{2}\cE_c\cE_a=A_{12}\nonumber
\end{multline}
and so on. Therefore, the operator set (\ref{O1})--(\ref{O2}) isomorphically
defines the set of infinitesimal operators of the group $\fG_+$.

In accordance with Gel'fand--Yaglom approach \cite{GY48} (see also
\cite{GMS,Nai58}) an operation of space inversion $P$ commutes with all the
operators $A_{ik}$ and anticommutes with all the operators $B_i$:
\begin{alignat}{3}
PA_{23}P^{-1} &=A_{23}, & \quad
PA_{13}P^{-1} &=A_{13}, & \quad
PA_{12}P^{-1} &=A_{12}, \nonumber\\
PB_1P^{-1} &=-B_1, & \quad
PB_2P^{-1} &=-B_2, & \quad
PB_3P^{-1} &=-B_3.\label{GY}
\end{alignat}

Let us consider permutation conditions of the operators (\ref{O1})--(\ref{O2})
with the matrix $\sW$ of the automorphism $\cA\rightarrow\cA^\star$
(space inversion). Since $\sW=\cE_1\cE_2\cdots\cE_n$ is a volume element of
$\C_n$, then $\cE_a,\,\cE_b,\,\cE_c\in\sW$, $\cE^2_i=\sI$,
$\sW^2=\sI$ at $n\equiv 0\pmod{4}$ and $\sW^2=-\sI$ at $n\equiv 2\pmod{4}$.
Therefore, for the operator $A_{23}\sim-\frac{1}{2}\cE_a\cE_b$ we obtain
\begin{eqnarray}
A_{23}\sW&=&-(-1)^{a+b-2}\frac{1}{2}\cE_1\cE_2\cdots\cE_{a-1}\cE_{a+1}\cdots
\cE_{b-1}\cE_{b+1}\cdots\cE_n,\nonumber\\
\sW A_{23}&=&-(-1)^{2n-a-b}\frac{1}{2}\cE_1\cE_2\cdots\cE_{a-1}\cE_{a+1}\cdots
\cE_{b-1}\cE_{b+1}\cdots\cE_n,\nonumber
\end{eqnarray}
whence it immediately follows a comparison $a+b-2\equiv 2n-a-b\pmod{2}$ or
$2(a+b)\equiv 2(n+1)\pmod{2}$. Thus, $\sW$ and $A_{23}$ are always commute.
It is easy to verify that analogous conditions take place for the operators
$A_{13},\,A_{12}$ (except the case $n=2$). Further, for 
$B_1\sim-\frac{1}{2}\cE_c$ we obtain
\begin{eqnarray}
B_1\sW&=&-(-1)^{c-1}\frac{1}{2}\cE_1\cE_2\cdots\cE_{c-1}\cE_{c+1}\cdots\cE_n,
\nonumber\\
\sW B_1&=&-(-1)^{n-c}\frac{1}{2}\cE_1\cE_2\cdots\cE_{c-1}\cE_{c+1}\cdots\cE_n,
\nonumber
\end{eqnarray}
that is, $c-1\equiv n-c\pmod{2}$ or $n\equiv 2c-1\pmod{2}$. Therefore, the
matrix $\sW$ always anticommute with $B_1$ (correspondingly with
$B_1,\,B_2$), since $n\equiv 0\pmod{2}$. Thus, in full accordance with
Gel'fand--Yaglom relations (\ref{GY}) we have\footnote{Except the case of the
fundamental representation $\fC^{1,0}$ for which the automorphism group is
$\sAut_+(\C_2)=\{\sI,\sW,\sE,\sC\}=\{\sigma_0,-\sigma_3,\sigma_1,-i\sigma_2\}
\simeq D_4/\dZ_2$. It is easy to verify that the matrix $\sW\sim\sigma_3$
does not satisfy the relations (\ref{C1}). Therefore, in case of $\fC^{1,0}$
we have an anomalous behaviour of the parity transformation $\sW\sim P$.
This fact will be explained further within quotient representations.}
\begin{alignat}{3}
\sW A_{23}\sW^{-1} &=A_{23}, &\quad
\sW A_{13}\sW^{-1} &=A_{13}, &\quad
\sW A_{12}\sW^{-1} &=A_{12}, \nonumber\\
\sW B_1\sW^{-1} &=-B_1, &\quad
\sW B_2\sW^{-1} &=-B_2, &\quad
\sW B_3\sW^{-1} &=-B_3,\label{C1}
\end{alignat}
where the matrix $\sW$ of the automorphism $\cA\rightarrow\cA^\star$ is an
element of an Abelian automorphism group $\sAut_-(\C_n)\simeq\dZ_2\otimes\dZ_2$
with the signature $(+,+,+)$ at $n\equiv 0\pmod{4}$ and also a 
non--Abelian automorphism group $\sAut_+(\C_n)\simeq Q_4/\dZ_2$ with the
signature $(-,-,-)$ at $n\equiv 2\pmod{4}$, here $\dZ_2\otimes\dZ_2$ is a
Gauss--Klein group and $Q_4$ is a quaternionic group (see Theorem 9 in
\cite{Var99}).

Let us consider now permutation conditions of the operators
(\ref{O1})--(\ref{O2}) with the matrix $\sE$ of the antiautomorphism
$\cA\rightarrow\widetilde{\cA}$ (time reversal). Over the field $\F=\C$
the matrix $\sE$ has two forms (Theorem 9 in \cite{Var99}):
1) $\sE=\cE_1\cE_2\cdots\cE_m$ at $m\equiv 1\pmod{2}$, the group
$\sAut_+(\C_n)\simeq Q_4/\dZ_2$; 2) $\sE=\cE_{m+1}\cE_{m+2}\cdots\cE_n$ at
$m\equiv 0\pmod{2}$, the group $\sAut_+(\C_n)\simeq\dZ_2\otimes\dZ_2$.
Obviously, in both cases $\sW=\cE_1\cE_2\cdots\cE_m\cE_{m+1}\cE_{m+2}\cdots
\cE_n$, where the matrices $\cE_i$ are symmetric for $1<i\leq m$ and
skewsymmetric for $m<i\leq n$. 

So, let $\sE=\cE_{m+1}\cE_{m+2}\cdots\cE_n$ be a matrix of 
$\cA\rightarrow\widetilde{\cA}$, $m\equiv 0\pmod{2}$. Let us assume that
$\cE_a,\,\cE_b,\,\cE_c\in\sE$, then for the operator $A_{23}\sim
-\frac{1}{2}\cE_a\cE_b$ we obtain
\begin{eqnarray}
A_{23}\sE&=&-(-1)^{b+a-2}\frac{1}{2}\cE_{m+1}\cE_{m+2}\cdots\cE_{a-1}\cE_{a+1}
\cdots\cE_{b-1}\cE_{b+1}\cdots\cE_n,\nonumber\\
\sE A_{23}&=&-(-1)^{2m-a-b}\frac{1}{2}\cE_{m+1}\cE_{m+2}\cdots\cE_{a-1}\cE_{a+1}
\cdots\cE_{b-1}\cE_{b+1}\cdots\cE_n,\label{R3}
\end{eqnarray}
that is, $b+a-2\equiv 2m-a-b\pmod{2}$ and in this case the matrix $\sE$
commutes with $A_{23}$ (correspondingly with $A_{13},\,A_{12}$). For the
operator $B_1$ we have
(analogously for $B_2$ and $B_3$):
\begin{eqnarray}
B_1\sE&=&-(-1)^{c-1}\frac{1}{2}\cE_{m+1}\cE_{m+2}\cdots\cE_{c-1}\cE_{c+1}
\cdots\cE_n,\nonumber\\
\sE B_1&=&-(-1)^{m-c}\frac{1}{2}\cE_{m+1}\cE_{m+2}\cdots\cE_{c-1}\cE_{c+1}
\cdots\cE_n,\label{R4}
\end{eqnarray}
that is, $m\equiv 2c-1\pmod{2}$ and, therefore, the matrix $\sE$ in this case
always anticommutes with $B_i$, since $m\equiv 0\pmod{2}$. Thus,
\begin{alignat}{3}
\sE A_{23}\sE^{-1} &=A_{23}, &\quad
\sE A_{13}\sE^{-1} &=A_{13}, &\quad
\sE A_{12}\sE^{-1} &=A_{12}, \nonumber\\
\sE B_1\sE^{-1} &=-B_1, &\quad
\sE B_2\sE^{-1} &=-B_2, &\quad
\sE B_3\sE^{-1} &=-B_3.\label{C2}
\end{alignat}
Let us assume now that $\cE_a,\,\cE_b,\,\cE_c\not\in\sE$, then
\begin{equation}\label{R5}
A_{23}\sE=(-1)^{2m}\sE A_{23},
\end{equation}
that is, in this case $\sE$ and $A_{ik}$ are always commute. For the
hyperbolic operators we obtain
\begin{equation}\label{R6}
B_i\sE=(-1)^m\sE B_i\quad(i=1,2,3)
\end{equation}
and since $m\equiv 0\pmod{2}$, then $\sE$ and $B_i$ are also commute.
Therefore,
\begin{equation}\label{C3}
\ld\sE,A_{23}\rd=0,\;\;\ld\sE,A_{13}\rd=0,\;\;\ld\sE,A_{12}\rd=0,\;\;
\ld\sE,B_1\rd=0,\;\;\ld\sE,B_2\rd=0,\;\;\ld\sE,B_3\rd=0.
\end{equation}
Assume now that $\cE_a,\,\cE_b\in\sE$ and $\cE_c\not\in\sE$. Then in
accordance with (\ref{R3}) and (\ref{R6}) the matrix $\sE$ commutes with
$A_{23}$ and $B_1$, and according to (\ref{R4}) anticommutes with $B_2$ and
$B_3$. For the operator $A_{13}$ we find
\begin{eqnarray}
A_{13}\sE&=&-(-1)^{b-1}\frac{1}{2}\cE_c\cE_{m+1}\cE_{m+2}\cdots\cE_{b-1}
\cE_{b+1}\cdots\cE_n,\nonumber\\
\sE A_{23}&=&-(-1)^{2m-b}\frac{1}{2}\cE_c\cE_{m+1}\cE_{m+2}\cdots\cE_{b-1}
\cE_{b+1}\cdots\cE_n,\label{R7}
\end{eqnarray}
that is, $2m\equiv 2b-1\pmod{2}$ and, therefore, in this case $\sE$ always
anticommutes with $A_{13}$ and correspondingly with the operator $A_{12}$
which has the analogous structure. Thus,
\begin{equation}\label{C4}
\ld\sE,A_{23}\rd=0,\;\;\lf\sE,A_{13}\rf=0,\;\;\lf\sE,A_{12}\rf=0,\;\;
\ld\sE,B_1\rd=0,\;\;\lf\sE,B_2\rf=0,\;\;\lf\sE,B_3\rf=0.
\end{equation}
If we take $\cE_a\in\sE$, $\cE_b,\,\cE_c\not\in\sE$, then for the operator
$A_{23}$ it follows that
\begin{eqnarray}
A_{23}\sE&=&-(-1)^{a-2}\frac{1}{2}\cE_b\cE_{m+1}\cE_{m+2}\cdots\cE_{a-1}
\cE_{a+1}\cdots\cE_n,\nonumber\\
\sE A_{23}&=&-(-1)^{2m-a-1}\frac{1}{2}\cE_b\cE_{m+1}\cE_{m+2}\cdots\cE_{a-1}
\cE_{a+1}\cdots\cE_n,\label{R8}
\end{eqnarray}
that is, $a-2\equiv 2m-a-1\pmod{2}$ or $2m\equiv 2a-1\pmod{2}$. Therefore,
$\sE$ anticommutes with $A_{23}$. Further, in virtue of (\ref{R5}) $\sE$
commutes with $A_{13}$ and anticommutes with $A_{12}$ in virtue of (\ref{R7}).
Correspondingly, from (\ref{R6}) and (\ref{R4}) it follows that $\sE$
commutes with $B_1,\,B_3$ and anticommutes with $B_2$. Thus,
\begin{equation}\label{C5}
\lf\sE,A_{23}\rf=0,\;\;\ld\sE,A_{13}\rd=0,\;\;\lf\sE,A_{12}\rf=0,\;\;
\ld\sE,B_1\rd=0,\;\;\lf\sE,B_2\rf=0,\;\;\ld\sE,B_3\rd=0.
\end{equation}
Cyclic permutations of the indices in $\cE_i,\,\cE_j\in\sE$, 
$\cE_k\not\in\sE$ and $\cE_i\in\sE$, $\cE_j,\,\cE_k\not\in\sE$,
$i,j,k=\{a,b,c\}$, give the following relations
\begin{equation}\label{C6}
\begin{array}{lll}
\lf\sE,A_{23}\rf=0,&\lf\sE,A_{13}\rf=0,&\ld\sE,A_{12}\rd=0,\\
\lf\sE,B_1\rf=0,&\lf\sE,B_2\rf=0,&\ld\sE,B_3\rd=0,
\end{array}\quad\cE_a,\,\cE_c\in\sE,\;\cE_b\not\in\sE.
\end{equation}
\begin{equation}\label{C7}
\begin{array}{lll}
\lf\sE,A_{23}\rf=0,&\ld\sE,A_{13}\rd=0,&\lf\sE,A_{12}\rf=0,\\
\lf\sE,B_1\rf=0,&\ld\sE,B_2\rd=0,&\lf\sE,B_3\rf=0,
\end{array}\quad\cE_b,\,\cE_c\in\sE,\;\cE_a\not\in\sE.
\end{equation}
\begin{equation}\label{C8}
\begin{array}{lll}
\lf\sE,A_{23}\rf=0,&\lf\sE,A_{13}\rf=0,&\ld\sE,A_{12}\rd=0,\\
\ld\sE,B_1\rd=0,&\ld\sE,B_2\rd=0,&\lf\sE,B_3\rf=0,
\end{array}\quad\cE_b\in\sE,\;\cE_a,\,\cE_c\not\in\sE.
\end{equation}
\begin{equation}\label{C9}
\begin{array}{lll}
\ld\sE,A_{23}\rd=0,&\lf\sE,A_{13}\rf=0,&\lf\sE,A_{12}\rf=0,\\
\lf\sE,B_1\rf=0,&\ld\sE,B_2\rd=0,&\ld\sE,B_3\rd=0,
\end{array}\quad\cE_c\in\sE,\;\cE_a,\,\cE_b\not\in\sE.
\end{equation}

Let us consider now the matrix $\sE$ of the group $\sAut_+(\C_n)\simeq
Q_4/\dZ_2$. In this case $\sE=\cE_1\cE_2\cdots\cE_m$, $m\equiv 1\pmod{2}$.
At $\cE_a,\,\cE_b,\,\cE_c\in\sE$ from (\ref{R3}) it follows that
$2(a+b)\equiv 2(m+1)\pmod{2}$, therefore, in this case $\sE$ always commutes
with $A_{ik}$. In its turn, from (\ref{R4}) it follows that $\sE$ always
commutes with $B_i$, since $m\equiv 1\pmod{2}$. Thus, we have the relations
(\ref{C3}). At $\cE_a,\cE_b,\cE_c\not\in\sE$ (except the cases $n=2$ and
$n=4$) from (\ref{R5}) it follows that $\sE$ always commutes with $A_{ik}$,
and from (\ref{R6}) it follows that $\sE$ always anticommutes with $B_i$.
Therefore, for the case $\cE_a,\cE_b,\cE_c\not\in\sE$ we have the relations
(\ref{C2}). Analogously, at $\cE_a,\cE_b\in\sE$, $\cE_c\not\in\sE$ in
accordance with (\ref{R3}) and (\ref{R4}) $\sE$ commutes with $A_{23}$ and
$B_1,B_3$, and in accordance with (\ref{R7}) and (\ref{R6}) anticommutes
with $A_{13},A_{12}$ and $B_1$. Therefore, for this case we have the
relations (\ref{C9}). At $\cE_a\in\sE$, $\cE_b,\cE_c\not\in\sE$ from
(\ref{R8}) it follows that $\sE$ anticommutes with $A_{23}$. In virtue of
(\ref{R5}) $\sE$ commutes with $A_{13}$ and anticommutes with $A_{12}$ in
virtue of (\ref{R7}). Correspondingly, from (\ref{R6}) and (\ref{R4}) it
follows that $\sE$ anticommutes with $B_1,B_3$ and commutes with $B_2$.
Therefore, for the case $\cE_a\in\sE$, $\cE_b,\cE_c\not\in\sE$ we have the
relations (\ref{C7}). Further, cyclic permutations of the indices in
$\cE_i,\cE_j\in\sE$, $\cE_k\not\in\sE$ and $\cE_i\in\sE$, 
$\cE_j,\cE_k\not\in\sE$, $i,j,k=\{a,b,c\}$, give for $\cE_a,\cE_c\in\sE$,
$\cE_b\not\in\sE$ the relations (\ref{C8}), for $\cE_b,\cE_c\in\sE$,
$\cE_a\not\in\sE$ the relations (\ref{C5}), for $\cE_b\in\sE$,
$\cE_a,\cE_c\not\in\sE$ the relations (\ref{C6}) and for $\cE_c\in\sE$,
$\cE_a,\cE_b\not\in\sE$ the relations (\ref{C4}).

Let us consider now the permutation conditions of the operators
(\ref{O1})--(\ref{O2}) with the matrix $\sC$ of the antiautomorphism
$\cA\rightarrow\widetilde{\cA^\star}$ (full reflection). Over the field
$\F=\C$ the matrix $\sC$ has two different forms (Theorem 9 in \cite{Var99}):
1) $\sC=\cE_1\cE_2\cdots\cE_m$ at $m\equiv 0\pmod{2}$, the group
$\sAut_-(\C_n)\simeq\dZ_2\otimes\dZ_2$; 2) $\sC=\cE_{m+1}\cE_{m+2}\cdots\cE_n$
at $m\equiv 1\pmod{2}$, the group $\sAut_+(\C_n)\simeq Q_4/\dZ_2$. So, let
$\sC=\cE_1\cE_2\cdots\cE_m$ be a matrix of 
$\cA\rightarrow\widetilde{\cA^\star}$, $m\equiv 1\pmod{2}$, then by analogy
with the matrix $\sE=\cE_{m+1}\cE_{m+2}\cdots\cE_n$ of $\cA\rightarrow
\widetilde{\cA}$, $m\equiv 0\pmod{2}$, we have for $\sC$ the relations of
the form (\ref{C2})--(\ref{C9}). In its turn, the matrix 
$\sC=\cE_{m+1}\cE_{m+2}\cdots\cE_n$ is analogous to 
$\sE=\cE_1\cE_2\cdots\cE_m$, $m\equiv 1\pmod{2}$, therefore, in this case
we have also the relations (\ref{C2})--(\ref{C9}) for $\sC$.

Now we have all possible combinations of permutation relations between the
matrices of infinitesimal operators (\ref{O1})--(\ref{O2}) of the proper
Lorentz group $\fG_+$ and matrices of the fundamental automorphisms of the
complex Clifford algebra $\C_n$ associated with the complex representation
$\fC^{l_0+l_1-1,0}$ of $\fG_+$. It is obvious that the relations (\ref{C1})
take place for any representation $\fC^{l_0+l_1-1,0}$ of the group
$\fG_+$. Further, if $\fC^{l_0+l_1-1,0}$ with $2(l_0+l_1-1)\equiv 0\pmod{4}$
and if $\sAut_-(\C_n)\simeq\dZ_2\otimes\dZ_2$ with $\sE=\cE_{m+1}\cE_{m+2}
\cdots\cE_n$, $\sC=\cE_1\cE_2\cdots\cE_m$, $m\equiv 0\pmod{2}$, then at
$\cE_a,\cE_b,\cE_c\in\sE$ ($\cE_a,\cE_b,\cE_c\not\in\sC$) we have the
relations (\ref{C2}) for $\sE$ and the relations of the form (\ref{C3}) for
$\sC$ and, therefore, we have the relations (\ref{T1}) and (\ref{T2}) of the
present Theorem. Correspondingly, at $\cE_a,\cE_b,\cE_c\not\in\sE$
($\cE_a,\cE_b,\cE_c\in\sC$) we have the relations (\ref{C3}) for $\sE$ and
the relations of the form (\ref{C2}) for $\sC$, that is, the relations 
(\ref{T5}) and (\ref{T6}) of Theorem. At $\cE_a,\cE_b\in\sE$, 
$\cE_c\not\in\sE$ ($\cE_a,\cE_b\not\in\sC$, $\cE_c\in\sC$) we obtain the
relations (\ref{C4}) for $\sE$ and the relations of the form (\ref{C9}) for
$\sC$ ((\ref{T9})--(\ref{T10}) in Theorem) and so on. Analogously, if
$\fC^{l_0+l_1-1,0}$ with $2(l_0+l_1-1)\equiv 2\pmod{4}$ and if
$\sAut_+(\C_n)\simeq Q_4/\dZ_2$ with $\sE=\cE_1\cE_2\cdots\cE_m$,
$\sC=\cE_{m+1}\cE_{m+2}\cdots\cE_n$, $m\equiv 1\pmod{2}$, then at
$\cE_a,\cE_b,\cE_c\in\sE$ ($\cE_a,\cE_b,\cE_c\not\in\sC$) we have the
relations (\ref{C3}) for $\sE$ and the relations of the form (\ref{C2})
for $\sC$ (relations (\ref{T2}) and (\ref{T1}) in Theorem) and so on.

Further, let us consider the following combinations of $A_{ik}$ and $B_i$
(rising and lowering operators):
\begin{alignat}{3}
H_+ &=iA_{23}-A_{13}, &\quad
H_- &=iA_{23}+A_{13}, &\quad
H_3 &=iA_{13},\nonumber\\
F_+ &=iB_1-B_2, &\quad
F_- &=iB_1+B_2, &\quad
F_3 &=iB_3,\label{I}
\end{alignat}
satisfying in virtue of (\ref{commut}) the relations
\begin{gather}
\ld H_+,H_3\rd=-H_+,\;\;\ld H_-,H_3\rd=H_-,\;\;\ld H_+,H_-\rd=2H_3,\nonumber\\
\ld H_+,F_+\rd=\ld H_-,F_-\rd=\ld H_3,F_3\rd=0,\nonumber\\
\ld F_+,F_3\rd=-H_+,\;\;\ld F_-,F_3\rd=H_-,\;\;\ld F_+,F_-\rd=-2H_3,\nonumber\\
\ld H_+,F_3\rd=F_+,\;\;\ld H_-,F_3\rd=-F_-,\nonumber\\
\ld H_-,F_+\rd=-\ld H_+,F_-\rd=2F_3,\nonumber\\
\ld F_+,H_3\rd=-F_+,\;\;\ld F_-,H_3\rd=F_-.\label{commut2}
\end{gather}
It is easy to see that for $\sW$ from (\ref{C1}) and (\ref{I}) we have always
the relations (\ref{T0}). Further, for $\sAut_-(\C_n)\simeq\dZ_2\otimes\dZ_2$
at $\cE_a,\cE_b,\cE_c\in\sE$ from (\ref{C2}) and (\ref{C3}) in virtue of
(\ref{I}) it follow the relations (\ref{T3}) and (\ref{T4}). Analogously,
at $\cE_a,\cE_b,\cE_c\not\in\sE$ from (\ref{C3}) and (\ref{C2}) we obtain
the relations (\ref{T7}) and (\ref{T8}). In contrast with this, at
$\cE_a,\cE_b\in\sE$, $\cE_c\not\in\sE$ the combinations (\ref{C4}) and
(\ref{C9}) do not form permutation relations with operators $H_{+,-,3}$ and
$F_{+,-,3}$, since $\sE$ and $\sC$ commute with $A_{23}$ and anticommute
with $A_{13}$, and $B_1$ commutes with $\sC$ and anticommutes with $\sE$
(inverse relations take place for $B_2$). Other two relations
(\ref{T21})--(\ref{T24}) and (\ref{T13})--(\ref{T16}) for
$\sAut_-(\C_n)\simeq\dZ_2\otimes\dZ_2$ correspond to $\cE_a,\cE_c\in\sE$,
$\cE_b\not\in\sE$ and $\cE_b\in\sE$, $\cE_a,\cE_c\not\in\sE$. It is easy
to see that in such a way for the group $\sAut_+(\C_n)\simeq Q_4/\dZ_2$ we
obtain the relations (\ref{T7})--(\ref{T8}), (\ref{T3})--(\ref{T4}),
(\ref{T21})--(\ref{T24}) and (\ref{T13})--(\ref{T16}) correspondingly
for $\cE_a,\cE_b,\cE_c\in\sE$, $\cE_a,\cE_b,\cE_c\not\in\sE$,
$\cE_a,\cE_c\in\sE$, $\cE_b\not\in\sE$ and $\cE_b\in\sE$, 
$\cE_a,\cE_c\not\in\sE$.

In accordance with \cite{GMS} a representation conjugated to
$\fC^{l_0+l_1-1,0}$ is defined by a pair
\[
(l_0,l_1)=\left(-\frac{r}{2},\,\frac{r}{2}+1\right),
\]
that is, this representation has a form $\fC^{0,l_0-l_1+1}$. In its turn,
a representation conjugated to fundamental representation $\fC^{1,0}$ is
$\fC^{0,-1}$. Let us find infinitesimal operators of the representation
$\fC^{0,-1}$. At $l_0=-1/2$ and $l_1=3/2$ from (\ref{I1})--(\ref{I6'}) and
(\ref{A1})--(\ref{B3}) we obtain
\begin{alignat}{3}
A^{-1/2}_{23} &=-\frac{i}{2}\ar\begin{bmatrix}
0 & 1\\
1 & 0
\end{bmatrix}, &\quad
A^{-1/2}_{13} &=\frac{1}{2}\ar\begin{bmatrix}
0 & 1\\
-1 & 0
\end{bmatrix}, &\quad
A^{-1/2}_{12} &=\frac{1}{2}\ar\begin{bmatrix}
i & 0\\
0 &-i
\end{bmatrix},\nonumber\\[0.2cm]
B^{-1/2}_1 &=\frac{1}{2}\ar\begin{bmatrix}
0 & 1\\
1 & 0
\end{bmatrix}, &\quad
B^{-1/2}_2 &=\frac{1}{2}\ar\begin{bmatrix}
0 & i\\
-i & 0
\end{bmatrix}, &\quad
B^{-1/2}_3 &=\frac{i}{2}\ar\begin{bmatrix}
-i & 0\\
0 & i
\end{bmatrix}.\nonumber
\end{alignat}
Or
\begin{alignat}{3}
A^{-1/2}_{23} &=-\frac{1}{2}\sigma_2\sigma_3, &\quad
A^{-1/2}_{13} &=-\frac{1}{2}\sigma_1\sigma_3, &\quad
A^{-1/2}_{12} &=\frac{1}{2}\sigma_1\sigma_2,\nonumber\\
B^{-1/2}_1 &=\frac{1}{2}\sigma_1, &\quad
B^{-1/2}_2 &=-\frac{1}{2}\sigma_2, &\quad
B^{-1/2}_3 &=\frac{1}{2}\sigma_3.\label{Op}
\end{alignat}
It is easy to see that operators (\ref{Op}) differ from 
(\ref{IF1})--(\ref{IF2}) only in the sign at the operators $B_i$ of the
hyperbolic rotations. This result is a direct consequence of the
well--known definition of the group $SL(2;\C)$ as a complexification of the
special unimodular group $SU(2)$ (see \cite{Vil68}). Indeed, the group
$SL(2;\C)$ has six parameters $a_1,a_2,a_3,ia_1,ia_2,ia_3$, where
$a_1,a_2,a_3\in SU(2)$. It is easy to verify that operators (\ref{Op})
satisfy the relations (\ref{commut}). Therefore, as in case of the
the representation $\fC^{l_0+l_1-1,0}$ infinitesimal operators of the
conjugated representation $\fC^{0,l_0-l_1+1}$ are defined as follows
\begin{alignat}{3}
A^{j^\prime}_{23} &\sim-\frac{1}{2}\cE_a\cE_b, &\quad
A^{j^\prime}_{13} &\sim-\frac{1}{2}\cE_c\cE_b, &\quad
A^{j^\prime}_{12} &\sim\frac{1}{2}\cE_c\cE_a,\nonumber\\
B^{j^\prime}_1 &\sim\frac{1}{2}\cE_c, &\quad
B^{j^\prime}_2 &\sim-\frac{1}{2}\cE_a, &\quad
B^{j^\prime}_3 &\sim\frac{1}{2}\cE_b,\label{Op2}
\end{alignat}
where $\cE_a,\cE_b,\cE_c$ are tensor products of the form (\ref{6.6}).
It is not difficult to verify that operators (\ref{Op2}) satisfy the
relations (\ref{commut}), and their linear combinations $H_{+,-,3}$,
$F_{+,-,3}$ satisfy the relations (\ref{commut2}). Since the structure of
the operators (\ref{Op2}) is analogous to the structure of the operators
(\ref{O1})--(\ref{O2}), then all the permutation conditions between the
operators of discrete symmetries and operators (\ref{O1})--(\ref{O2})
of the representation $\fC^{l_0+l_1-1,0}$ are valid also for the conjugated
representation $\fC^{0,l_0-l_1+1}$ and, obviously, for a representation
$\fC^{l_0+l_1-1,l_0-l_1+1}$.\\[0.2cm]
2) Real representations.\\
As known \cite{GMS}, if an irreducible representation of the proper Lorentz
group $\fG_+$ is defined by the pair $(l_0,l_1)$, then a conjugated
representation is also irreducible and defined by a pair $\pm(l_0,-l_1)$.
Hence it follows that the irreducible representation is equivalent to its
conjugated representation only in case when this representation is defined
by a pair $(0,l_1)$ or $(l_0,0)$, that is, either of the two numbers
$l_0$ and $l_1$ is equal to zero. We assume that $l_1=0$. In its turn,
for the complex Clifford algebra $\C_n$
($\overset{\ast}{\C}_n$) associated
with the representation $\fC^{l_0+l_1-1,0}\,(\fC^{0,l_0-l_1+1})$ of the
group $\fG_+$ the equivalence of the representation to its conjugated
representation induces a relation 
$\C_n=\overset{\ast}{\C}_n$, which,
obviously, is fulfilled only in case when the algebra $\C_n=\C\otimes\cl_{p,q}$
is reduced into its real subalgebra $\cl_{p,q}$ ($p+q=n$). Thus, a
restriction of the complex representation $\fC^{l_0+l_1-1,0}$
(or $\fC^{0,l_0-l_1+1}$) of the group $\fG_+$ onto a real representation,
$(l_0,l_1)\rightarrow(l_0,0)$, induces a restriction $\C_n\rightarrow\cl_{p,q}$.
Further, over the field $\F=\R$ at $p+q\equiv 0\pmod{2}$ there are four
types of real algebras $\cl_{p,q}$: two types $p-q\equiv 0,2\pmod{8}$ with
a real division ring $\K\simeq\R$ and two types $p-q\equiv 4,6\pmod{8}$ with a
quaternionic division ring $\K\simeq\BH$. Thus, we have four classes of the
real representations of the group $\fG_+$:
\begin{eqnarray}
\fR^{l_0}_0&\leftrightarrow&\cl_{p,q},\;p-q\equiv 0\pmod{8},\;\K\simeq\R,
\nonumber\\
\fR^{l_0}_2&\leftrightarrow&\cl_{p,q},\;p-q\equiv 2\pmod{8},\;\K\simeq\R,
\nonumber\\
\fH^{l_0}_4&\leftrightarrow&\cl_{p,q},\;p-q\equiv 4\pmod{8},\;\K\simeq\BH,
\nonumber\\
\fH^{l_0}_6&\leftrightarrow&\cl_{p,q},\;p-q\equiv 6\pmod{8},\;\K\simeq\BH,
\label{Ident}
\end{eqnarray}
We will call the representations $\fH^{l_0}_4$ and $\fH^{l_0}_6$ are
{\it quaternionic representations} of the group $\fG_+$. It is not difficult
to see that for the real representations with the pair $(l_0,0)$ all the
coefficients $A_l=il_0l_1/l(l+1)$ are equal to zero, since $l_1=0$.
Therefore, all the infinitesimal operators $B^j_1,B^j_2,B^j_3$ (see
formulas (\ref{B1})--(\ref{B3})) of hyperbolic rotations are
also equal to zero. Hence it follows that the restriction
$(l_0,l_1)\rightarrow(l_0,0)$ induces a restriction of the group $\fG_+$
onto its subgroup $SO(3)$ of three--dimensional rotations. Thus, real
representations with the pair $(l_0,0)$ are representations of the subgroup
$SO(3)$. This result directly follows from the complexification of $SU(2)$
which is equivalent to $SL(2;\C)$. Indeed, the parameters $a_1,a_2,a_3$
compose a real part of $SL(2;\C)$ which under complex conjugation remain
unaltered, whereas the parameters $\pm ia_1,\pm ia_2,\pm ia_3$ of the complex
part of $SL(2;\C)$ under complex conjugation are mutually annihilate for
the representations with the pairs $(l_0,l_1)$ and $\pm(l_0,-l_1)$.

Let us find a relation of the number $l_0$ with dimension of the real algebra
$\cl_{p,q}$. If $p+q\equiv 0\pmod{2}$ and $\omega^2=\e^2_{12\ldots p+q}=1$,
then $\cl_{p,q}$ is called {\it positive} ($\cl_{p,q}>0$ at 
$p-q\equiv 0,4\pmod{8}$) and correspondingly {\it negative} if $\omega^2=-1$
($\cl_{p,q}<0$ at $p-q\equiv 2,6\pmod{8}$). Further, in accordance with
Karoubi Theorem \cite[Prop.~3.16]{Kar79} it follows that if $\cl(V,Q)>0$,
and $\dim V$ is even, then $\cl(V\oplus V^\prime,Q\oplus Q^\prime)\simeq
\cl(V,Q)\otimes\cl(V^\prime,Q^\prime)$, and also if $\cl(V,Q)<0$, and
$\dim V$ is even, then $\cl(V\oplus V^\prime, Q\oplus Q^\prime)\simeq
\cl(V,Q)\otimes\cl(V^\prime,-Q^\prime)$, where $V$ is a vector space
associated with $\cl_{p,q}$, $Q$ is a quadratic form of $V$. Using the
Karoubi Theorem we obtain for the algebra $\cl_{p,q}$ a following
factorization
\begin{equation}\label{Ten}
\cl_{p,q}\simeq\underbrace{\cl_{s_i,t_j}\otimes\cl_{s_i,t_j}\otimes\cdots
\otimes\cl_{s_i,t_j}}_{r\;\text{times}}
\end{equation}
where $s_i,t_j\in\{0,1,2\}$. For example, there are two different
factorizations $\cl_{1,1}\otimes\cl_{0,2}$ and $\cl_{1,1}\otimes\cl_{2,0}$
for the spacetime algebra $\cl_{1,3}$ and Majorana algebra $\cl_{3,1}$.
It is obvious that $l_0=r/2$ and $n=2r=p+q=4l_0$, therefore, $l_0=(p+q)/4$.

So, we begin with the representation of the class $\fR^{l_0}_0$. In accordance
with Theorem 4 in \cite{Var00} for the algebra $\cl_{p,q}$ of the type
$p-q\equiv 0\pmod{8}$ a matrix of the automorphism $\cA\rightarrow\cA^\star$
has a form $\sW=\cE_1\cE_2\cdots\cE_{p+q}$ and $\sW^2=\sI$. It is obvious
that for $\cE^2_a=\cE^2_b=\cE^2_c=\sI$ permutation conditions of the matrix
$\sW$ with $A_{ik}$ are analogous to (\ref{C1}), that is, $\sW$ always
commutes with the operators $A_{ik}$ of the subgroup $SO(3)$. It is
sufficient to consider permutation conditions of $\sW$ with the operators
$A_{ik}$ of $SO(3)$ only, since $B_i=0$ for the real representations.
In this case the relations (\ref{commut}) take a form
\begin{equation}\label{commut3}
\ld A_{23},A_{13}\rd=A_{12}, \quad \ld A_{13},A_{12}\rd=A_{23}, \quad
\ld A_{12},A_{23}\rd=A_{13}.
\end{equation}
Assume now that $\cE^2_a=\cE^2_b=\cE^2_c=-\sI$, then it is easy to verify
that operators
\begin{equation}\label{O3}
A_{23}\sim\frac{1}{2}\cE_a\cE_b,\quad A_{13}\sim\frac{1}{2}\cE_c\cE_b,\quad
A_{12}\sim-\frac{1}{2}\cE_c\cE_a
\end{equation}
satisfy the relations (\ref{commut3}) and commute with the matrix $\sW$ of
$\cA\rightarrow\cA^\star$. It is easy to see that at $\cE^2_i=-\sI$,
$\cE^2_j=\cE^2_k=\sI$ and $\cE^2_i=\cE^2_j=-\sI$, $\cE^2_k=\sI$
($i,j,k=\{a,b,c\}$) the operators $A_{ik}$ do not satisfy the relations
(\ref{commut3}). Therefore, there exist only two possibilities
$\cE^2_a=\cE^2_b=\cE^2_c=\sI$ and $\cE^2_a=\cE^2_b=\cE^2_c=-\sI$
corresponding to the operators (\ref{O1}) and (\ref{O3}), respectively.

Further, for the type $p-q\equiv 0\pmod{8}$ ($p=q=m$) at
$\sE=\cE_{p+1}\cE_{p+2}\cdots\cE_{p+q}$ and $\sC=\cE_1\cE_2\cdots\cE_p$
there exist Abelian groups $\sAut_-(\cl_{p,q})\simeq\dZ_2\otimes\dZ_2$
with the signature $(+,+,+)$ and $\sAut_-(\cl_{p,q})\simeq\dZ_4$ with
$(+,-,-)$ correspondingly at $p,q\equiv 0\pmod{4}$ and $p,q\equiv 2\pmod{4}$,
and also at $\sE=\cE_1\cE_2\cdots\cE_p$ and $\sC=\cE_{p+1}\cE_{p+2}\cdots
\cE_{p+q}$ there exist non--Abelian groups $\sAut_+(\cl_{p,q})\simeq
D_4/\dZ_2$ with $(+,-,+)$ and $\sAut_+(\cl_{p,q})\simeq D_4/\dZ_2$ with
$(+,+,-)$ correspondingly at $p,q\equiv 3\pmod{4}$ and $p,q\equiv 1\pmod{4}$
(Theorem 4 in \cite{Var00}). Besides, for the algebras $\cl_{8t,0}$ of the
type $p-q\equiv 0\pmod{8}$, $t=1,2,\ldots$, the matrices $\sE\sim\sI$,
$\sC\sim\cE_1\cE_2\cdots\cE_p$ and $\sW$ form an Abelian group
$\sAut_-(\cl_{p,0})\simeq\dZ_2\otimes\dZ_2$. Correspondingly, for the
algebras $\cl_{0,8t}$ of the type $p-q\equiv 0\pmod{8}$ the matrices
$\sE\sim\cE_1\cE_2\cdots\cE_q$, $\sC\sim\sI$ and $\sW$ also form the
group $\sAut_-(\cl_{0,q})\simeq\dZ_2\otimes\dZ_2$.

So, let $\sE=\cE_{m+1}\cE_{m+2}\cdots\cE_{2m}$ and $\sC=\cE_1\cE_2\cdots\cE_m$
be matrices of $\cA\rightarrow\widetilde{\cA}$ and
$\cA\rightarrow\widetilde{\cA^\star}$, $p=q=m$, $m\equiv 0\pmod{2}$.
Then for the operators (\ref{O1}) we obtain
\begin{equation}\label{C34}
A_{ik}\sE=(-1)^{2m}\sE A_{ik},
\end{equation}
\begin{eqnarray}
A_{ik}\sC&=&\pm(-1)^{i+j-2}\frac{1}{2}\sigma(i)\sigma(j)\cE_1\cE_2\cdots
\cE_{i-1}\cE_{i+1}\cdots\cE_{j-1}\cE_{j+1}\cdots\cE_m,\nonumber\\
\sC A_{ik}&=&\pm(-1)^{2m-i-j}\frac{1}{2}\sigma(i)\sigma(j)\cE_1\cE_2\cdots
\cE_{i-1}\cE_{i+1}\cdots\cE_{j-1}\cE_{j+1}\cdots\cE_m,\label{C35}
\end{eqnarray}
since $\cE_a,\cE_b,\cE_c\not\in\sE$ and a function $\sigma(n)=\sigma(m-n)$
has a form
\[
\sigma(n)=\begin{cases}
-1 & \text{if $n\leq 0$}\\
+1 & \text{if $n> 0$}
\end{cases}
\]
Analogously, for the operators (\ref{O3}) we find
\begin{eqnarray}
A_{ik}\sE&=&\pm(-1)^{i+j-2}\frac{1}{2}\sigma(i)\sigma(j)\cE_{m+1}\cE_{m+2}
\cdots\cE_{i-1}\cE_{i+1}\cdots\cE_{j-1}\cE_{j+1}\cdots\cE_{2m},\nonumber\\
\sE A_{ik}&=&\pm(-1)^{2m-i-j}\frac{1}{2}\sigma(i)\sigma(j)\cE_{m+1}\cE_{m+2}
\cdots\cE_{i-1}\cE_{i+1}\cdots\cE_{j-1}\cE_{j+1}\cdots\cE_{2m},\label{C36}
\end{eqnarray}
\begin{equation}\label{C37}
A_{ik}\sC=(-1)^{2m}\sC A_{ik}.
\end{equation}
It is easy to see that in both cases the matrices $\sE$ and $\sC$ always
commute with the operators $A_{ik}$. Therefore, for the type 
$p-q\equiv 0\pmod{8}$ at $p,q\equiv 0\pmod{4}$ and $p,q\equiv 2\pmod{4}$ the
elements of the Abelian groups $\sAut_-(\cl_{p,q})\simeq\dZ_2\otimes\dZ_2$
and $\sAut_-(\cl_{p,q})\simeq\dZ_4$ with $(+,-,-)$ are always commute with
infinitesimal operators $A_{ik}$ of $SO(3)$. Correspondingly, for the algebra
$\cl_{8t,0}$ the elements of $\sAut_-(\cl_{8t,0})\simeq\dZ_2\otimes\dZ_2$
are also commute with $A_{ik}$. The analogous statement takes place for
other degenerate case $\sAut_-(\cl_{0,8t})\simeq\dZ_2\otimes\dZ_2$. In the
case of non--Abelian groups $\sAut_+(\cl_{p,q})\simeq D_4/\dZ_2$
(signatures $(+,-,+)$ and $(+,+,-)$) we obtain the same permutation conditions
as (\ref{C34})--(\ref{C37}), that is, in this case the matrices of the
fundamental automorphisms always commute with $A_{ik}$. Thus, for the real
representation of the class $\fR^{l_0}_0$ operators of the discrete subgroup
always commute with all the infinitesimal operators of $SO(3)$.

Further, for the real representation of the class $\fR^{l_0}_2$, type
$p-q\equiv 2\pmod{8}$, at $\sE=\cE_{p+1}\cE_{p+2}\cdots\cE_{p+q}$ and
$\sC=\cE_1\cE_2\cdots\cE_p$ there exist Abelian groups $\sAut_-(\cl_{p,q})
\simeq\dZ_4$ with $(-,-,+)$ and $\sAut_-(\cl_{p,q})\simeq\dZ_4$ with
$(-,+,-)$ correspondingly at $p\equiv 0\pmod{4}$, $q\equiv 2\pmod{4}$ and
$p\equiv 2\pmod{4}$, $q\equiv 0\pmod{4}$, and also at $\sE=\cE_1\cE_2\cdots
\cE_p$ and $\sC=\cE_{p+1}\cE_{p+2}\cdots\cE_{p+q}$ there exist
non--Abelian groups $\sAut_+(\cl_{p,q})\simeq Q_4/\dZ_2$ with $(-,-,-)$ and
$\sAut_+(\cl_{p,q})\simeq D_4/\dZ_2$ with $(-,+,+)$ correspondingly at
$p\equiv 3\pmod{4}$, $q\equiv 1\pmod{4}$ and $p\equiv 1\pmod{4}$,
$q\equiv 3\pmod{4}$ (Theorem 4 in \cite{Var00}). So, for the Abelian groups
at $\cE_a,\cE_b,\cE_c\not\in\sE$ (operators (\ref{O1})) we obtain
\begin{equation}\label{C38}
A_{ik}\sE=(-1)^{2q}\sE A_{ik},
\end{equation}
\begin{eqnarray}
A_{ik}\sC&=&\pm(-1)^{i+j-2}\frac{1}{2}\sigma(i)\sigma(j)\cE_1\cE_2\cdots
\cE_{i-1}\cE_{i+1}\cdots\cE_{j-1}\cE_{j+1}\cdots\cE_p,\nonumber\\
\sC A_{ik}&=&\pm(-1)^{2p-i-j}\frac{1}{2}\sigma(i)\sigma(j)\cE_1\cE_2\cdots
\cE_{i-1}\cE_{i+1}\cdots\cE_{j-1}\cE_{j+1}\cdots\cE_p.\label{C39}
\end{eqnarray}
Correspondingly, for the operators (\ref{O3}) ($\cE_a,\cE_b,\cE_c\in\sE$)
we have
\begin{eqnarray}
A_{ik}\sE&=&\pm(-1)^{i+j-2}\frac{1}{2}\sigma(i)\sigma(j)\cE_{p+1}\cE_{p+2}
\cdots\cE_{i-1}\cE_{i+1}\cdots\cE_{j-1}\cE_{j+1}\cdots\cE_{p+q},\nonumber\\
\sE A_{ik}&=&\pm(-1)^{2q-i-j}\frac{1}{2}\sigma(i)\sigma(j)\cE_{p+1}\cE_{p+2}
\cdots\cE_{i-1}\cE_{i+1}\cdots\cE_{j-1}\cE_{j+1}\cdots\cE_{p+q},\label{C40}
\end{eqnarray}
\begin{equation}\label{C41}
A_{ik}\sC=(-1)^{2p}\sC A_{ik}.
\end{equation}
The analogous relations take place for the non--Abelian groups. From
(\ref{C38})--(\ref{C41}) it is easy to see that the matrices $\sE$ and
$\sC$ always commute with $A_{ik}$. Therefore, for the real representation
of the class $\fR^{l_0}_2$ operators of the discrete subgroup always commute
with all the infinitesimal operators of $SO(3)$.

Let us consider now quaternionic representations. Quaternionic representations
of the classes $\fH^{l_0}_4$ and $\fH^{l_0}_6$, types $p-q\equiv 4,6\pmod{8}$,
in virtue of the more wide ring $\K\simeq\BH$ have a more complicated
structure of the reflection groups than in the case of $\K\simeq\R$.
Indeed, if $\sE=\cE_{j_1}\cE_{j_2}\cdots\cE_{j_k}$ is a product of $k$
skewsymmetric matrices (among which $l$ matrices have `$+$'-square and
$t$ matrices have `$-$'-square) and 
$\sC=\cE_{i_1}\cE_{i_2}\cdots\cE_{i_{p+q-k}}$ is a product of $p+q-k$
symmetric matrices (among which there are $h$ `$+$'-squares and $g$ 
`$-$-squares), then at $k\equiv 0\pmod{2}$ for the type $p-q\equiv 4\pmod{8}$
there exist Abelian groups $\sAut_-(\cl_{p,q})\simeq\dZ_2\otimes\dZ_2$ and
$\sAut_-(\cl_{p,q})\simeq\dZ_4$ with $(+,-,-)$ if correspondingly
$l-t,h-g\equiv 0,1,4,5\pmod{8}$ and $l-t,h-g\equiv 2,3,6,7\pmod{8}$, and
also at $k\equiv 1\pmod{2}$ for the type $p-q\equiv 6\pmod{8}$ there exist
$\sAut_-(\cl_{p,q})\simeq\dZ_4$ with $(-,+,-)$ and $\sAut_-(\cl_{p,q})\simeq
\dZ_4$ with $(-,-,+)$ if correspondingly $l-t\equiv 0,1,4,5\pmod{8}$,
$h-g\equiv 2,3,6,7\pmod{8}$ and $l-t\equiv 2,3,6,7\pmod{8}$,
$h-g\equiv 0,1,4,5\pmod{8}$. Inversely, if $\sE=\cE_{i_1}\cE_{i_2}\cdots
\cE_{i_{p+q-k}}$ and $\sC=\cE_{j_1}\cE_{j_2}\cdots\cE_{j_k}$, then
at $k\equiv 1\pmod{8}$ for the type $p-q\equiv 4\pmod{8}$ there exist
non--Abelian groups $\sAut_+(\cl_{p,q})\simeq D_4/\dZ_2$ with $(+,-,+)$ and
$\sAut_+(\cl_{p,q})\simeq D_4/\dZ_2$ with $(+,+,-)$ if correspondingly
$h-g\equiv 2,3,6,7\pmod{8}$, $l-t\equiv 0,1,4,5\pmod{8}$ and
$h-g\equiv 0,1,4,5\pmod{8}$, $l-t\equiv 2,3,6,7\pmod{8}$, and also at
$k\equiv 1\pmod{2}$ for the type $p-q\equiv 6\pmod{8}$ there exist
$\sAut_+(\cl_{p,q})\simeq Q_4/\dZ_2$ with $(-,-,-)$ and
$\sAut_+(\cl_{p,q})\simeq D_4/\dZ_2$ with $(-,+,+)$ if correspondingly
$h-g,l-t\equiv 2,3,6,7\pmod{8}$ and $h-g,l-t\equiv 0,1,4,5\pmod{8}$
(see Theorem 4 in \cite{Var00}).

So, let $\sE=\cE_{j_1}\cE_{j_2}\cdots\cE_{j_k}$ and
$\sC=\cE_{i_1}\cE_{i_2}\cdots\cE_{i_{p+q-k}}$ be the matrices of
$\cA\rightarrow\widetilde{\cA}$ and $\cA\rightarrow\widetilde{\cA^\star}$,
$k\equiv 0\pmod{2}$. Assume that $\cE_a,\cE_b,\cE_c\in\sE$, that is, all the
matrices $\cE_i$ in the operators (\ref{O1}) or (\ref{O3}) are
skewsymmetric. Then for the operators (\ref{O1}) and (\ref{O3}) we obtain
\begin{eqnarray}
A_{23}\sE&=&-(-1)^{b+a-2}\frac{1}{2}\sigma(j_a)\sigma(j_b)\cE_{j_1}\cE_{j_2}
\cdots\cE_{j_{a-1}}\cE_{j_{a+1}}\cdots\cE_{j_{b-1}}\cE_{j_{b+1}}\cdots
\cE_{j_k},\nonumber\\
\sE A_{23}&=&-(-1)^{2k-a-b}\frac{1}{2}\sigma(j_a)\sigma(j_b)\cE_{j_1}\cE_{j_2}
\cdots\cE_{j_{a-1}}\cE_{j_{a+1}}\cdots\cE_{j_{b-1}}\cE_{j_{b+1}}\cdots
\cE_{j_k},\label{C42}
\end{eqnarray}
\begin{equation}\label{C43}
A_{23}\sC=(-1)^{2(p+q-k)}\sC A_{23},
\end{equation}
that is, $\sE$ and $\sC$ commute with $A_{23}$ (correspondingly with
$A_{13}$, $A_{12}$). It is easy to see that relations (\ref{C42}) and
(\ref{C43}) are analogous to the relations (\ref{R3}) and (\ref{R5}) for the
field $\F=\C$. Therefore, from (\ref{C42}) and (\ref{C43}) we obtain the
relations (\ref{TR11}) of Theorem. Further, assume that $\cE_a,\cE_b,\cE_c
\not\in\sE$, that is, all the matrices $\cE_i$ in the operators
(\ref{O1}) and (\ref{O3}) are symmetric. Then
\begin{equation}\label{C44}
A_{23}\sE=(-1)^{2k}\sE A_{23},
\end{equation}
\begin{eqnarray}
A_{23}\sC&=&-(-1)^{b+a-2}\frac{1}{2}\sigma(i_a)\sigma(i_b)\cE_{i_1}\cE_{i_2}
\cdots\cE_{i_{a-1}}\cE_{i_{a+1}}\cdots\cE_{i_{b-1}}\cE_{i_{b+1}}\cdots
\cE_{i_{p+q-k}},\nonumber\\
\sC A_{23}&=&-(-1)^{2(p+q-k)-a-b}\frac{1}{2}\sigma(i_a)\sigma(i_b)\cE_{i_1}
\cE_{i_2}\cdots\cE_{i_{a-1}}\cE_{i_{a+1}}\cdots\cE_{i_{b-1}}\cE_{i_{b+1}}
\cdots\cE_{i_{p+q-k}}\label{C45}
\end{eqnarray}
and analogous relations take place for $A_{13},A_{12}$. It is easy to verify
that from (\ref{C44}) and (\ref{C45}) we obtain the same relations
(\ref{TR11}), since (\ref{TR11}) are relations (\ref{T1}) or (\ref{T5})
at $B_i=0$. Therefore, over the ring $\K\simeq\BH$ (quasicomplex case)
the elements of Abelian reflection groups of the quaternionic
representations $\fH^{l_0}_{4,6}$ satisfy the relations
(\ref{T1})--(\ref{T24}) over the field $\F=\C$ at $B_i=0$. Indeed, at
$\cE_a,\cE_b\in\sE$, $\cE_c\not\in\sE$ and $\cE_c\in\sE$,
$\cE_a,\cE_b\not\in\sE$ we have relations (\ref{TR12}) which are particular
cases of (\ref{T9}) at $B_i=0$ and so on. It is easy to verify that the
same relations take place for non--Abelian reflection groups at
$\sE=\cE_{i_1}\cE_{i_2}\cdots\cE_{i_{p+q-k}}$ and $\sC=\cE_{j_1}\cE_{j_2}
\cdots\cE_{j_k}$, $k\equiv 1\pmod{2}$.
\end{proof}
{\bf Remark}. Theorem exhausts all possible permutation relations between
transformations $P,\,T,\,PT$ and infinitesimal operators of the group
$\fG_+$. The relations (\ref{T0})--(\ref{T0'}) take place always, that is,
at any $n\equiv 0\pmod{2}$ (except the case $n=2$). In turn, the relations
(\ref{T1})--(\ref{T24}) are divided into two classes. The first class
contains relations with operators $H_{+,-,3}$ and $F_{+,-,3}$ (the relations
(\ref{T1})--(\ref{T4}), (\ref{T5})--(\ref{T8}), (\ref{T13})--(\ref{T16}),
(\ref{T21})--(\ref{T24})). The second class does not contain the relations
with $H_{+,-,3}$, $F_{+,-,3}$ (the relations (\ref{T9})--(\ref{T10}),
(\ref{T11})--(\ref{T12}), (\ref{T17})--(\ref{T18}), (\ref{T19})--(\ref{T20}).
Besides, in accordance with \cite{GMS} for the transformation $T$ there
are only two possibilities $T=P$ and $T=-P$ (both these cases correspond
to relation (\ref{T1})). However, from other relations it follows that
$T\neq\pm P$, as it should be take place in general case. The exceptional
case $n=2$  corresponds to neutrino field and further it will be explored
in the following sections within quotient representations of the group
$\fG_+$. Permutations relations with respect to symmetric subspaces
$\Sym_{(k,r)}$ can be obtained by similar manner.
\section{Atiyah--Bott--Shapiro periodicity on the Lorentz group}
In accordance with the section 2 the finite--dimensional representations
$\fC$, $\overset{\ast}{\fC}$, 
$\fC\oplus\overset{\ast}{\fC}$ related with
the algebras $\C_{2k}$, $\overset{\ast}{\C}_{2r}$, 
$\C_{2k}\oplus\overset{\ast}{\C}_{2k}$
of the type $n\equiv 0\pmod{2}$ and the quotient
representations ${}^\chi\fC$, ${}^\chi\overset{\ast}{\fC}$,
${}^\chi\fC\cup{}^\chi\fC$
(${}^\chi\fC\oplus{}^\chi\overset{\ast}{\fC}$) 
related with the quotient
algebras ${}^\epsilon\C_{2k}$, ${}^\epsilon\overset{\ast}{\C}_{2r}$,
${}^\epsilon\C_{2k}\cup{}^\epsilon\overset{\ast}{\C}_{2k}$ corresponding to the type
$n\equiv 1\pmod{2}$ form a full system $\fM=\fM^0\oplus\fM^1$ of the
finite--dimensional representations of the proper Lorentz group $\fG_+$.
This extension of the Lorentz group allows to apply Atiyah--Bott--Shapiro
periodicity \cite{AtBSh} on the system $\fM$, and also it allows to define on
$\fM$ some cyclic relations that give rise to interesting applications in
particle physics (in the spirit of Gell-Mann--Ne'eman eightfold way 
\cite{GN64}).

As known, ABS--periodicity based on the
$\dZ_2$--graded structure of the Clifford algebra $\cl$. Indeed, let
$\cl^+$ (correspondingly $\cl^-$) be a set consisting of all even
(correspondingly odd) 
elements of the algebra $\cl$. The set $\cl^+$ is a subalgebra of
$\cl$. It is obvious that
$\cl=\cl^+\oplus\cl^-$, and also $\cl^+\cl^+
\subset\cl^+,\,\cl^+\cl^-\subset\cl^-,\,
\cl^-\cl^+\subset\cl^-,\,\cl^-\cl^-\subset
\cl^+$. A degree $\deg a$ of the even (correspondingly odd) 
element $a\in\cl$ is equal
to 0 (correspondingly 1). 
Let $\mathfrak{A}$ and $\mathfrak{B}$ be the two associative
$\dZ_2$--graded algebras over the field $\F$; then a multiplication of
homogeneous elements
$\mathfrak{a}^\prime\in\mathfrak{A}$ and $\mathfrak{b}\in\mathfrak{B}$ in a
graded tensor product
$\mathfrak{A}\hat{\otimes}\mathfrak{B}$ is defined as follows: 
$(\mathfrak{a}\otimes \mathfrak{b})(\mathfrak{a}^\prime
\otimes \mathfrak{b}^\prime)=(-1)^{\deg\mathfrak{b}\deg\mathfrak{a}^\prime}
\mathfrak{a}\mathfrak{a}^\prime\otimes\mathfrak{b}\mathfrak{b}^\prime$.
The graded tensor product of the two graded central simple algebras is also
graded central simple 
\cite[Theorem 2]{Wal64}. Over the field $\F=\C$ the Clifford algebra
$\C_n$ is central simple if $n\equiv 0\pmod{2}$. It is known that for the
Clifford algebras with odd dimensionality there is a following isomorphism
$\C^+_{n+1}\simeq\C_n$ \cite{Rash}. Thus, $\C^+_{n+1}$ is central simple
algebra. Further, in accordance with Chevalley Theorem \cite{Che55} for the
graded tensor product there is an isomorphism 
$\C_{n^\prime}\hat{\otimes}\C_{n^{\prime\prime}}\simeq\C_{n^\prime+
n^{\prime\prime}}$. Two algebras $\C_{n^\p}$ and $\C_{n^{\p\p}}$ are said to
be of the same class if $n^\p\equiv n^{\p\p}\pmod{2}$. The graded central
simple Clifford algebras over the field $\F=\C$ form two similarity classes,
which, as it is easy to see, coincide with the two types of the algebras
$\C_n$: $n\equiv 0,1\pmod{2}$. The set of these 2 types (classes) forms
a Brauer--Wall group $BW_{\C}$ \cite{Wal64,Lou81} that is isomorphic to a
cyclic group $\dZ_2$. Thus, the algebra $\C_n$ is an element of the
Brauer--Wall group $BW_{\C}\simeq\dZ_2$, and a group operation is the graded
tensor product $\hat{\otimes}$. Coming back to representations of the group
$\fG_+$ we see that in virtue of identifications $\C_n\leftrightarrow\fC$
($n\equiv 0\pmod{2}$) and $\C_n\leftrightarrow\fC\cup\fC$ ($n\equiv 1\pmod{2}$)
a group action of $BW_{\C}\simeq\dZ_2$ can be transferred onto the system
$\fM=\fM^0\oplus\fM^1$. Indeed, a cyclic structure of the group
$BW_{\C}\simeq\dZ_2$ is defined by a transition 
$\C^+_n\overset{h}{\longrightarrow}\C_n$, where the type of the algebra
$\C_n$ is defined by a formula $n=h+2r$, here $h\in\{0,1\}$, $r\in\dZ$
\cite{BTr87,BT88}. Therefore, the action of $BW_{\C}\simeq\dZ_2$ on $\fM$
is defined by a transition 
$\overset{+}{\fC}\overset{h}{\longrightarrow}
\fC$, where $\overset{+}{\fC}\!{}^{l_0+l_1-1,0}\simeq\fC^{l_0+l_1-2,0}$ when
$\fC\in\fM^0$ ($h=1$) and $\overset{+}{\fC}=\left(\fC^{l_0+l_1-1,0}\cup
\fC^{l_0+l_1-1,0}\right)^+\sim{}^\chi\fC^{l_0+l_1-1,0}$ when $\fC\in\fM^1$
($h=0$), $\dim\fC=h+2r$ ($\dim\fC=l_0+l_1-1$ if $\fC\in\fM^0$ and
$\dim\fC=2(l_0+l_1-1)$ if $\fC\in\fM^1$). For example, in virtue of
$\C^+_2\simeq\C_1$ a transition $\C^+_2\rightarrow\C_2$ ($\C_1\rightarrow
\C_2$) induces on the system $\fM$ a transition 
$\overset{+}{\fC}\!{}^{1,0}
\rightarrow\fC^{1,0}$ that in virtue of $\overset{+}{\fC}\!{}^{1,0}\simeq
\fC^{0,0}$ is equivalent to $\fC^{0,0}\rightarrow\fC^{1,0}$ ($\fC^{0,0}$ is
one--dimensional representation of $\fG_+$) and, therefore, $h=1$.
In its turn, a transition $\C^+_3\rightarrow\C_3$ induces on the system
$\fM$ a transition $\left(\fC^{1,0}\cup\fC^{1,0}\right)^+\rightarrow
\fC^{1,0}\cup\fC^{1,0}$ or ${}^\chi\fC^{1,0}\rightarrow\fC^{1,0}\cup
\fC^{1,0}$ and, therefore, $h=0$. In such a way, we see that a cyclic
structure of the group $BW_{\C}\simeq\dZ_2$ induces on the system $\fM$
modulo 2 periodic relations which can be explicitly showed on the
Trautmann--like diagram (spinorial clock \cite{BTr87,BT88}, see also
\cite{Var00}):
\medskip
\[
\unitlength=0.5mm
%\linethickness{0.4pt}
\begin{picture}(50.00,50.00)(0,0)
\put(5,25){0}
\put(42,25){1}
\put(22,-4){$\fC$}
\put(17,55){$\fC\cup\fC$}
\put(3,-13){$n\equiv 0\!\!\!\!\pmod{2}$}
\put(3,64){$n\equiv 1\!\!\!\!\pmod{2}$}
%1
\put(20,49.49){$\cdot$}
\put(19.5,49.39){$\cdot$}
\put(19,49.27){$\cdot$}
\put(18.5,49.14){$\cdot$}
\put(18,49){$\cdot$}
\put(17.5,48.85){$\cdot$}
\put(17,48.68){$\cdot$}
\put(16.5,48.51){$\cdot$}
\put(16,48.32){$\cdot$}
\put(15.5,48.12){$\cdot$}
\put(15,47.91){$\cdot$}
\put(14.5,47.69){$\cdot$}
\put(14,47.45){$\cdot$}
\put(13.5,47.2){$\cdot$}
\put(13,46.93){$\cdot$}
\put(12.5,46.65){$\cdot$}
\put(12,46.35){$\cdot$}
\put(11.5,46.04){$\cdot$}
\put(11,45.71){$\cdot$}
\put(10.5,45.36){$\cdot$}
\put(10,45){$\cdot$}
\put(9.5,44.61){$\cdot$}
\put(9,44.21){$\cdot$}
\put(8.5,43.78){$\cdot$}
\put(8,43.33){$\cdot$}
\put(7.5,42.85){$\cdot$}
\put(7,42.35){$\cdot$}
\put(6.5,41.81){$\cdot$}
\put(6,41.25){$\cdot$}
\put(5.5,40.64){$\cdot$}
\put(5,40){$\cdot$}
\put(4.5,39.3){$\cdot$}
\put(4,38.56){$\cdot$}
\put(3.5,37.76){$\cdot$}
\put(3,36.87){$\cdot$}
\put(2.5,35.89){$\cdot$}
\put(2,34.79){$\cdot$}
\put(1.5,33.53){$\cdot$}
\put(1,32){$\cdot$}
\put(0.5,29.97){$\cdot$}
%2
\put(30,49.49){$\cdot$}
\put(30.5,49.39){$\cdot$}
\put(31,49.27){$\cdot$}
\put(31.5,49.14){$\cdot$}
\put(32,49){$\cdot$}
\put(32.5,48.85){$\cdot$}
\put(33,48.68){$\cdot$}
\put(33.5,48.51){$\cdot$}
\put(34,48.32){$\cdot$}
\put(34.5,48.12){$\cdot$}
\put(35,47.91){$\cdot$}
\put(35.5,47.69){$\cdot$}
\put(36,47.45){$\cdot$}
\put(36.5,47.2){$\cdot$}
\put(37,46.93){$\cdot$}
\put(37.5,46.65){$\cdot$}
\put(38,46.35){$\cdot$}
\put(38.5,46.04){$\cdot$}
\put(39,45.71){$\cdot$}
\put(39.5,45.36){$\cdot$}
\put(40,45){$\cdot$}
\put(40.5,44.61){$\cdot$}
\put(41,44.21){$\cdot$}
\put(41.5,43.78){$\cdot$}
\put(42,43.33){$\cdot$}
\put(42.5,42.85){$\cdot$}
\put(43,42.35){$\cdot$}
\put(43.5,41.81){$\cdot$}
\put(44,41.25){$\cdot$}
\put(44.5,40.64){$\cdot$}
\put(45,40){$\cdot$}
\put(45.5,39.3){$\cdot$}
\put(46,38.56){$\cdot$}
\put(46.5,37.76){$\cdot$}
\put(47,36.87){$\cdot$}
\put(47.5,35.89){$\cdot$}
\put(48,34.79){$\cdot$}
\put(48.5,33.53){$\cdot$}
\put(49,32){$\cdot$}
\put(49.5,29.97){$\cdot$}
%1'
\put(0,25){$\cdot$}
\put(0,24.5){$\cdot$}
\put(0.02,24){$\cdot$}
\put(0.04,23.5){$\cdot$}
\put(0.08,23){$\cdot$}
\put(0.12,22.5){$\cdot$}
\put(0.18,22){$\cdot$}
\put(0.25,21.5){$\cdot$}
\put(0.32,21){$\cdot$}
\put(0.4,20.5){$\cdot$}
\put(0.5,20){$\cdot$}
\put(0.61,19.5){$\cdot$}
\put(0.73,19){$\cdot$}
\put(0.85,18.5){$\cdot$}
\put(1,18){$\cdot$}
\put(1.15,17.5){$\cdot$}
\put(1.31,17){$\cdot$}
\put(1.49,16.5){$\cdot$}
\put(1.68,16){$\cdot$}
\put(1.88,15.5){$\cdot$}
\put(2.09,15){$\cdot$}
\put(2.31,14.5){$\cdot$}
\put(2.55,14){$\cdot$}
\put(2.8,13.5){$\cdot$}
\put(3.06,13){$\cdot$}
\put(0,25.5){$\cdot$}
\put(0.02,26){$\cdot$}
\put(0.04,26.5){$\cdot$}
\put(0.08,27){$\cdot$}
\put(0.12,27.5){$\cdot$}
\put(0.18,28){$\cdot$}
\put(0.25,28.5){$\cdot$}
\put(0.32,29){$\cdot$}
\put(0.4,29.5){$\cdot$}
\put(0.5,30){$\cdot$}
\put(0.61,30.5){$\cdot$}
\put(0.73,31){$\cdot$}
\put(0.85,31.5){$\cdot$}
\put(1,32){$\cdot$}
\put(1.15,32.5){$\cdot$}
\put(1.31,33){$\cdot$}
\put(1.49,33.5){$\cdot$}
\put(1.68,34){$\cdot$}
\put(1.88,34.5){$\cdot$}
\put(2.09,35){$\cdot$}
\put(2.31,35.5){$\cdot$}
\put(2.55,36){$\cdot$}
\put(2.8,36.5){$\cdot$}
\put(3.06,37){$\cdot$}
%2'
\put(50,25){$\cdot$}
\put(49.99,24.5){$\cdot$}
\put(49.98,24){$\cdot$}
\put(49.95,23.5){$\cdot$}
\put(49.92,23){$\cdot$}
\put(49.87,22.5){$\cdot$}
\put(49.82,22){$\cdot$}
\put(49.75,21.5){$\cdot$}
\put(49.68,21){$\cdot$}
\put(49.51,20.5){$\cdot$}
\put(49.49,20){$\cdot$}
\put(49.39,19.5){$\cdot$}
\put(49.27,19){$\cdot$}
\put(49.14,18.5){$\cdot$}
\put(49,18){$\cdot$}
\put(48.85,17.5){$\cdot$}
\put(48.69,17){$\cdot$}
\put(48.51,16.5){$\cdot$}
\put(48.32,16){$\cdot$}
\put(48.12,15.5){$\cdot$}
\put(47.91,15){$\cdot$}
\put(47.69,14.5){$\cdot$}
\put(47.45,14){$\cdot$}
\put(47.2,13.5){$\cdot$}
\put(46.93,13){$\cdot$}
\put(50,25){$\cdot$}
\put(49.99,25.5){$\cdot$}
\put(49.98,26){$\cdot$}
\put(49.95,26.5){$\cdot$}
\put(49.92,27){$\cdot$}
\put(49.87,27.5){$\cdot$}
\put(49.82,28){$\cdot$}
\put(49.75,28.5){$\cdot$}
\put(49.68,29){$\cdot$}
\put(49.51,29.5){$\cdot$}
\put(49.49,30){$\cdot$}
\put(49.39,30.5){$\cdot$}
\put(49.27,31){$\cdot$}
\put(49.14,31.5){$\cdot$}
\put(49,32){$\cdot$}
\put(48.85,32.5){$\cdot$}
\put(48.69,33){$\cdot$}
\put(48.51,33.5){$\cdot$}
\put(48.32,34){$\cdot$}
\put(48.12,34.5){$\cdot$}
\put(47.91,35){$\cdot$}
\put(47.69,35.5){$\cdot$}
\put(47.45,36){$\cdot$}
\put(47.2,36.5){$\cdot$}
\put(46.93,37){$\cdot$}
%4
\put(20,0.5){$\cdot$}
\put(19.5,0.61){$\cdot$}
\put(19,0.73){$\cdot$}
\put(18.5,0.86){$\cdot$}
\put(18,1){$\cdot$}
\put(17.5,1.15){$\cdot$}
\put(17,1.31){$\cdot$}
\put(16.5,1.49){$\cdot$}
\put(16,1.68){$\cdot$}
\put(15.5,1.87){$\cdot$}
\put(15,2.09){$\cdot$}
\put(14.5,2.31){$\cdot$}
\put(14,2.55){$\cdot$}
\put(13.5,2.8){$\cdot$}
\put(13,3.06){$\cdot$}
\put(12.5,3.35){$\cdot$}
\put(12,3.64){$\cdot$}
\put(11.5,3.96){$\cdot$}
\put(11,4.29){$\cdot$}
\put(10.5,4.63){$\cdot$}
\put(10,5){$\cdot$}
\put(9.5,5.38){$\cdot$}
\put(9,5.79){$\cdot$}
\put(8.5,6.22){$\cdot$}
\put(8,6.67){$\cdot$}
\put(7.5,7.15){$\cdot$}
\put(7,7.65){$\cdot$}
\put(6.5,8.18){$\cdot$}
\put(6,8.75){$\cdot$}
\put(5.5,9.35){$\cdot$}
\put(5,10){$\cdot$}
\put(4.5,10.69){$\cdot$}
\put(4,11.43){$\cdot$}
\put(3.5,12.24){$\cdot$}
\put(3,13.12){$\cdot$}
\put(2.5,14.10){$\cdot$}
\put(2,15.20){$\cdot$}
\put(1.5,16.47){$\cdot$}
\put(1,18){$\cdot$}
\put(0.5,20.02){$\cdot$}
%3'
%\put(30,0.5){\vector(0,0){1}}
\put(30,0.5){$\cdot$}
\put(30.5,0.61){$\cdot$}
\put(31,0.73){$\cdot$}
\put(31.5,0.86){$\cdot$}
\put(32,1){$\cdot$}
\put(32.5,1.15){$\cdot$}
\put(33,1.31){$\cdot$}
\put(33.5,1.49){$\cdot$}
\put(34,1.68){$\cdot$}
\put(34.5,1.87){$\cdot$}
\put(35,2.09){$\cdot$}
\put(35.5,2.31){$\cdot$}
\put(36,2.55){$\cdot$}
\put(36.5,2.8){$\cdot$}
\put(37,3.06){$\cdot$}
\put(37.5,3.35){$\cdot$}
\put(38,3.64){$\cdot$}
\put(38.5,3.96){$\cdot$}
\put(39,4.29){$\cdot$}
\put(39.5,4.63){$\cdot$}
\put(40,5){$\cdot$}
\put(40.5,5.38){$\cdot$}
\put(41,5.79){$\cdot$}
\put(41.5,6.22){$\cdot$}
\put(42,6.67){$\cdot$}
\put(42.5,7.15){$\cdot$}
\put(43,7.65){$\cdot$}
\put(43.5,8.18){$\cdot$}
\put(44,8.75){$\cdot$}
\put(44.5,9.35){$\cdot$}
\put(45,10){$\cdot$}
\put(45.5,10.69){$\cdot$}
\put(46,11.43){$\cdot$}
\put(46.5,12.24){$\cdot$}
\put(47,13.12){$\cdot$}
\put(47.5,14.10){$\cdot$}
\put(48,15.20){$\cdot$}
\put(48.5,16.47){$\cdot$}
\put(49,18){$\cdot$}
\put(49.5,20.02){$\cdot$}

\end{picture}
\]
\vspace{1ex}
\begin{center}
\begin{minipage}{25pc}{\small
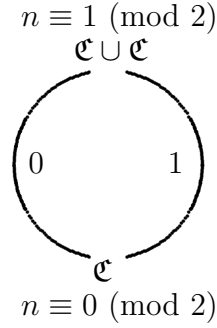
{\bf Fig.1} The action of the Brauer--Wall group 
$BW_{\C}\simeq\dZ_2$ on the full system $\fM=\fM^0\oplus\fM^1$ of complex
finite--dimensional representations $\fC$ of the proper Lorentz group $\fG_+$.} 
\end{minipage}
\end{center}
\bigskip
It is obvious that a group structure over $\C_n$, defined by the group
$BW_{\C}\simeq\dZ_2$, immediately relates with a modulo 2 periodicity of the
complex Clifford algebras \cite{AtBSh,Kar79}: $\C_{n+2}\simeq\C_n\otimes\C_2$.
Therefore, we have the following relations for $\fC$ and
$\overset{\ast}{\fC}$:
\begin{eqnarray}
\fC^{l_0+l_1,0}&\simeq&\fC^{l_0+l_1-1,0}\otimes\fC^{1,0},\nonumber\\
\fC^{0,l_0-l_1}&\simeq&\fC^{0,l_0-l_1+1}\otimes\fC^{0,-1}\nonumber
\end{eqnarray}
and correspondingly for 
$\fC\otimes\overset{\ast}{\fC}$
\[
\fC^{l_0+l_1,l_0-l_1}\simeq\fC^{l_0+l_1-1,l_0-l_1+1}\otimes\fC^{1,-1}.
\]
Thus, the action of $BW_{\C}\simeq\dZ_2$ form a cycle of the period 2 on
the system $\fM$, where the basic 2-period factor is the fundamental
representation $\fC^{1,0}$ ($\fC^{0,-1}$) of the group $\fG_+$. This
cyclic structure is intimately related with de Broglie--Jordan neutrino
theory of light and, moreover, this structure is a natural generalization of
BJ--theory. Indeed, in the simplest case we obtain two relations
$\fC^{2,0}\simeq\fC^{1,0}\otimes\fC^{1,0}$ and $\fC^{0,-2}\simeq
\fC^{0,-1}\otimes\fC^{0,-1}$, which, as it is easy to see, correspond two
helicity states of the photon 
(left-- and right--handed polarization)\footnote{Such a description
corresponds to Helmholtz--Silberstein representation of the electromagnetic
field as the complex linear combinations $\bF=\bE+i\bH$,
$\overset{\ast}{\bF}=\bE-i\bH$ that form a basis of the 
Majorana--Oppenheimer quantum electrodynamics 
\cite{Opp31,RF68,Esp98,Dvo97b} (see also recent development on this
subject based on the Joos--Weinberg and Bargmann--Wigner formalisms
\cite{Dvo97}).}. In such a way, subsequent rotations of the
representation 2-cycle give all other higher spin physical fields and
follows to de Broglie and Jordan this structure should be called as
`{\it neutrino theory of everything}'.

Further, when we consider restriction of the complex representation $\fC$
onto real representations $\fR$ and $\fH$, that corresponds to restriction
of the group $\fG_+$ onto its subgroup $SO(3)$\footnote{Physical feilds
defined within such representations describe neutral particles, or particles
at rest such as atomic nuclei.}, we come to a more high--graded modulo 8
periodicity over the field $\F=\R$. Indeed,
the Clifford algebra $\cl_{p,q}$ is central simple
if $p-q\not\equiv 1,5\pmod{8}$. It is known that for the Clifford algebra
with odd dimensionality, the isomorphisms are as follows:
$\cl^+_{p,q+1}\simeq\cl_{p,q}$ and $\cl^+_{p+1,q}\simeq\cl_{q,p}$ 
\cite{Rash,Port}. Thus, $\cl^+_{p,q+1}$ and $\cl^+_{p+1,q}$
are central simple algebras. Further, in accordance with Chevalley Theorem
\cite{Che55} for the graded tensor product there is an isomorphism
$\cl_{p,q}\hat{\otimes}\cl_{p^{\p},q^{\p}}\simeq
\cl_{p+p^{\p},q+q^{\p}}$. Two algebras $\cl_{p,q}$ and $\cl_{p^{\p},q^{\p}}$
are said to be of the same class if $p+q^{\p}\equiv p^{\p}+q\pmod{8}$.
The graded central simple Clifford algebras over the field $\F=\R$
form eight similarity classes, which, as it is easy to see, coincide
with the eight types of the algebras $\cl_{p,q}$.
The set of these 8 types (classes) forms a Brauer--Wall group $BW_{\R}$
\cite{Wal64,Lou81} that is isomorphic to a cyclic group $\dZ_8$. 
Therefore, in virtue of identifications (\ref{Ident}) a group action of
$BW_{\R}\simeq\dZ_8$ can be transferred onto the system $\fM=\fM^+\oplus\fM^-$.
In its turn, a cyclic structure of the group $BW_{\R}\simeq\dZ_8$ is defined by
a transition $\cl^+_{p,q}\overset{h}{\longrightarrow}\cl_{p,q}$, where the
type of the algebra $\cl_{p,q}$ is defined by a formula
$q-p=h+8r$, here $h\in\{0,\ldots,7\}$, $r\in\dZ$ \cite{BTr87,BT88}. Thus,
the action of $BW_{\R}\simeq\dZ_8$ on $\fM$ is defined by a transition
$\overset{+}{\fD}\!{}^{l_0}\overset{h}{\longrightarrow}\fD^{l_0}$, where
$\fD^{\l_0}=\left\{\fR^{l_0}_{0,2},\fH^{l_0}_{4,6},\fC^{l_0}_{3,7},
\fR^{l_0}_{0,2}\cup\fR^{l_0}_{0,2},\fH^{l_0}_{4,6}\cup\fH^{l_0}_{4,6}\right\}$,
and $\overset{+}{\fD}\!{}^{r/2}\simeq\fD^{\frac{r-1}{2}}$ when $\fD\in\fM^+$ and
$\left(\fD^{r/2}\cup\fD^{r/2}\right)^+\simeq{}^\chi\fD^{r/2}$ when
$\fD\in\fM^-$, $r$ is a number of tensor products in (\ref{Ten}). Therefore,
a cyclic structure of the group $BW_{\R}\simeq\dZ_8$ induces on the system
$\fM$ modulo 8 periodic relations which can be explicitly showed on the
following diagram (the round on the diagram is realized by an hour--hand):
\bigskip
\[
\unitlength=0.5mm
%\linethickness{0.4pt}
\begin{picture}(100.00,110.00)
%\put(50,50){\circle{10}[tr]}
%\circle{20}
%\put(75,93){\line(1,-1){1}}
%\put(50.5,50.3){\vector(-1,-1){25}}
%\put(50,50){\vector(1,1){25}}
%\put(50,50){\vector(1,-1){25}}
%\put(50,50){\vector(-1,1){25}}
%\put(50,50){\vector(0,-1){35}}
%\put(50,50){\vector(0,1){35}}
%\put(50,50){\vector(1,0){35}}
%\put(50,50){\vector(-1,0){35}}
%\put(50,85){\vector(1,-1){15}}
%\put(50,85){\line(2,-3){15}}
%\put(50,85){\line(-2,-3){15}}
%\framebox(50,100)[c]{ALGEBRA}
%\oval(0,0)[tr]
%\put(0,0){$\lambda$}
%\put(50,50){$\cdot$}
%\put(50,100){$\cdot$}
%\put(100,50){$\cdot$}
%\put(100,100){$\cdot$}
%\put(100,0){$\cdot$}
%\put(50,0){$\cdot$}
%\put(0,50){$\cdot$}

%\put(0,100){$\bullet$}
% II quadrant
\put(97,67){$\fC^{l_0}_7$}\put(108,64){$p-q\equiv 7\!\!\!\!\pmod{8}$}
\put(80,80){1}
\put(75,93.3){$\cdot$}
\put(75.5,93){$\cdot$}
\put(76,92.7){$\cdot$}
\put(76.5,92.4){$\cdot$}
\put(77,92.08){$\cdot$}
\put(77.5,91.76){$\cdot$}
\put(78,91.42){$\cdot$}
\put(78.5,91.08){$\cdot$}
\put(79,90.73){$\cdot$}
\put(79.5,90.37){$\cdot$}
\put(80,90.0){$\cdot$}
\put(80.5,89.62){$\cdot$}
\put(81,89.23){$\cdot$}
\put(81.5,88.83){$\cdot$}
\put(82,88.42){$\cdot$}
\put(82.5,87.99){$\cdot$}
\put(83,87.56){$\cdot$}
\put(83.5,87.12){$\cdot$}
\put(84,86.66){$\cdot$}
\put(84.5,86.19){$\cdot$}
\put(85,85.70){$\cdot$}
\put(85.5,85.21){$\cdot$}
\put(86,84.69){$\cdot$}
\put(86.5,84.17){$\cdot$}
\put(87,83.63){$\cdot$}
\put(87.5,83.07){$\cdot$}
\put(88,82.49){$\cdot$}
\put(88.5,81.9){$\cdot$}
\put(89,81.29){$\cdot$}
\put(89.5,80.65){$\cdot$}
\put(90,80){$\cdot$}
\put(90.5,79.32){$\cdot$}
\put(91,78.62){$\cdot$}
\put(91.5,77.89){$\cdot$}
\put(92,77.13){$\cdot$}
\put(92.5,76.34){$\cdot$}
\put(93,75.51){$\cdot$}
\put(93.5,74.65){$\cdot$}
\put(94,73.74){$\cdot$}
\put(94.5,72.79){$\cdot$}
\put(96.5,73.74){\vector(1,-2){1}}
% IV quadrant
\put(80,20){3}
\put(97,31){$\fH^{l_0}_6$}\put(108,28){$p-q\equiv 6\!\!\!\!\pmod{8}$}
\put(75,6.7){$\cdot$}
\put(75.5,7){$\cdot$}
\put(76,7.29){$\cdot$}
\put(76.5,7.6){$\cdot$}
\put(77,7.91){$\cdot$}
\put(77.5,8.24){$\cdot$}
\put(78,8.57){$\cdot$}
\put(78.5,8.91){$\cdot$}
\put(79,9.27){$\cdot$}
\put(79.5,9.63){$\cdot$}
\put(80,10){$\cdot$}
\put(80.5,10.38){$\cdot$}
\put(81,10.77){$\cdot$}
\put(81.5,11.17){$\cdot$}
\put(82,11.58){$\cdot$}
\put(82.5,12.00){$\cdot$}
\put(83,12.44){$\cdot$}
\put(83.5,12.88){$\cdot$}
\put(84,13.34){$\cdot$}
\put(84.5,13.8){$\cdot$}
\put(85,14.29){$\cdot$}
\put(85.5,14.79){$\cdot$}
\put(86,15.3){$\cdot$}
\put(86.5,15.82){$\cdot$}
\put(87,16.37){$\cdot$}
\put(87.5,16.92){$\cdot$}
\put(88,17.5){$\cdot$}
\put(88.5,18.09){$\cdot$}
\put(89,18.71){$\cdot$}
\put(89.5,19.34){$\cdot$}
\put(90,20){$\cdot$}
\put(90.5,20.68){$\cdot$}
\put(91,21.38){$\cdot$}
\put(91.5,22.11){$\cdot$}
\put(92,22.87){$\cdot$}
\put(92.5,23.66){$\cdot$}
\put(93,24.48){$\cdot$}
\put(93.5,25.34){$\cdot$}
\put(94,26.25){$\cdot$}
\put(94.5,27.20){$\cdot$}
\put(95,28.20){$\cdot$}
% I quadrant
\put(20,80){7}
\put(25,93.3){$\cdot$}
\put(24.5,93){$\cdot$}
\put(24,92.7){$\cdot$}
\put(23.5,92.49){$\cdot$}
\put(23,92.08){$\cdot$}
\put(22.5,91.75){$\cdot$}
\put(22,91.42){$\cdot$}
\put(21.5,91.08){$\cdot$}
\put(21,90.73){$\cdot$}
\put(20.5,90.37){$\cdot$}
\put(20,90){$\cdot$}
\put(19.5,89.62){$\cdot$}
\put(19,89.23){$\cdot$}
\put(18.5,88.83){$\cdot$}
\put(18,88.42){$\cdot$}
\put(17.5,87.99){$\cdot$}
\put(17,87.56){$\cdot$}
\put(16.5,87.12){$\cdot$}
\put(16,86.66){$\cdot$}
\put(15.5,86.19){$\cdot$}
\put(15,85.70){$\cdot$}
\put(14.5,85.21){$\cdot$}
\put(14,84.69){$\cdot$}
\put(13.5,84.17){$\cdot$}
\put(13,83.63){$\cdot$}
\put(12.5,83.07){$\cdot$}
\put(12,82.49){$\cdot$}
\put(11.5,81.9){$\cdot$}
\put(11,81.29){$\cdot$}
\put(10.5,80.65){$\cdot$}
\put(10,80){$\cdot$}
\put(9.5,79.32){$\cdot$}
\put(9,78.62){$\cdot$}
\put(8.5,77.89){$\cdot$}
\put(8,77.13){$\cdot$}
\put(7.5,76.34){$\cdot$}
\put(7,75.51){$\cdot$}
\put(6.5,74.65){$\cdot$}
\put(6,73.79){$\cdot$}
\put(5.5,72.79){$\cdot$}
\put(5,71.79){$\cdot$}
% III quadrant
\put(20,20){5}
\put(25,6.7){$\cdot$}
\put(24.5,7){$\cdot$}
\put(24,7.29){$\cdot$}
\put(23.5,7.6){$\cdot$}
\put(23,7.91){$\cdot$}
\put(22.5,8.24){$\cdot$}
\put(22,8.57){$\cdot$}
\put(21.5,8.91){$\cdot$}
\put(21,9.27){$\cdot$}
\put(20.5,9.63){$\cdot$}
\put(20,10){$\cdot$}
\put(19.5,10.38){$\cdot$}
\put(19,10.77){$\cdot$}
\put(18.5,11.17){$\cdot$}
\put(18,11.58){$\cdot$}
\put(17.5,12){$\cdot$}
\put(17,12.44){$\cdot$}
\put(16.5,12.88){$\cdot$}
\put(16,13.34){$\cdot$}
\put(15.5,13.8){$\cdot$}
\put(15,14.29){$\cdot$}
\put(14.5,14.79){$\cdot$}
\put(14,15.3){$\cdot$}
\put(13.5,15.82){$\cdot$}
\put(13,16.37){$\cdot$}
\put(12.5,16.92){$\cdot$}
\put(12,17.5){$\cdot$}
\put(11.5,18.09){$\cdot$}
\put(11,18.71){$\cdot$}
\put(10.5,19.34){$\cdot$}
\put(10,20){$\cdot$}
\put(9.5,20.68){$\cdot$}
\put(9,21.38){$\cdot$}
\put(8.5,22.11){$\cdot$}
\put(8,22.87){$\cdot$}
\put(7.5,23.66){$\cdot$}
\put(7,24.48){$\cdot$}
\put(6.5,25.34){$\cdot$}
\put(6,26.25){$\cdot$}
\put(5.5,27.20){$\cdot$}
\put(5,28.20){$\cdot$}
%quadrants I-II
\put(-1,97){$\fR^{l_0}_{0,2}\cup\fR^{l_0}_{0,2}$}
\put(-55,107){$p-q\equiv 1\!\!\!\!\pmod{8}$}
\put(50,93){0}
\put(50,100){$\cdot$}
\put(49.5,99.99){$\cdot$}
\put(49,99.98){$\cdot$}
\put(48.5,99.97){$\cdot$}
\put(48,99.96){$\cdot$}
\put(47.5,99.94){$\cdot$}
\put(47,99.91){$\cdot$}
\put(46.5,99.86){$\cdot$}
\put(46,99.84){$\cdot$}
\put(45.5,99.8){$\cdot$}
\put(45,99.75){$\cdot$}
\put(44.5,99.7){$\cdot$}
\put(44,99.64){$\cdot$}
\put(43.5,99.57){$\cdot$}
\put(43,99.51){$\cdot$}
\put(42.5,99.43){$\cdot$}
\put(42,99.35){$\cdot$}
\put(41.5,99.27){$\cdot$}
\put(41,99.18){$\cdot$}
\put(40.5,99.09){$\cdot$}
\put(40,98.99){$\cdot$}
\put(39.5,98.88){$\cdot$}
\put(39,98.77){$\cdot$}
\put(38.5,98.66){$\cdot$}
\put(38,98.54){$\cdot$}
\put(37.5,98.41){$\cdot$}
\put(37,98.28){$\cdot$}
\put(50.5,99.99){$\cdot$}
\put(51,99.98){$\cdot$}
\put(51.5,99.97){$\cdot$}
\put(52,99.96){$\cdot$}
\put(52.5,99.94){$\cdot$}
\put(53,99.91){$\cdot$}
\put(53.5,99.86){$\cdot$}
\put(54,99.84){$\cdot$}
\put(54.5,99.8){$\cdot$}
\put(55,99.75){$\cdot$}
\put(55.5,99.7){$\cdot$}
\put(56,99.64){$\cdot$}
\put(56.5,99.57){$\cdot$}
\put(57,99.51){$\cdot$}
\put(57.5,99.43){$\cdot$}
\put(58,99.35){$\cdot$}
\put(58.5,99.27){$\cdot$}
\put(59,99.18){$\cdot$}
\put(59.5,99.09){$\cdot$}
\put(60,98.99){$\cdot$}
\put(60.5,98.88){$\cdot$}
\put(61,98.77){$\cdot$}
\put(61.5,98.66){$\cdot$}
\put(62,98.54){$\cdot$}
\put(62.5,98.41){$\cdot$}
\put(63,98.28){$\cdot$}
\put(65,97){$\fR^{l_0}_0$}\put(73,108){$p-q\equiv 0\!\!\!\!\pmod{8}$}
%quadrants III-IV
\put(50,7){4}
\put(67,2){$\fH^{l_0}_{4,6}\cup\fH^{l_0}_{4,6}$}
\put(90,-6){$p-q\equiv 5\!\!\!\!\pmod{8}$}
\put(50,0){$\cdot$}
\put(50.5,0){$\cdot$}
\put(51,0.01){$\cdot$}
\put(51.5,0.02){$\cdot$}
\put(52,0.04){$\cdot$}
\put(52.5,0.06){$\cdot$}
\put(53,0.09){$\cdot$}
\put(53.5,0.12){$\cdot$}
\put(54,0.16){$\cdot$}
\put(54.5,0.2){$\cdot$}
\put(55,0.25){$\cdot$}
\put(55.5,0.3){$\cdot$}
\put(56,0.36){$\cdot$}
\put(56.5,0.42){$\cdot$}
\put(57,0.49){$\cdot$}
\put(57.5,0.56){$\cdot$}
\put(58,0.64){$\cdot$}
\put(58.5,0.73){$\cdot$}
\put(59,0.82){$\cdot$}
\put(59.5,0.91){$\cdot$}
\put(60,1.01){$\cdot$}
\put(60.5,1.11){$\cdot$}
\put(61,1.22){$\cdot$}
\put(61.5,1.34){$\cdot$}
\put(62,1.46){$\cdot$}
\put(62.5,1.59){$\cdot$}
\put(63,1.72){$\cdot$}
\put(49.5,0){$\cdot$}
\put(49,0.01){$\cdot$}
\put(48.5,0.02){$\cdot$}
\put(48,0.04){$\cdot$}
\put(47.5,0.06){$\cdot$}
\put(47,0.09){$\cdot$}
\put(46.5,0.12){$\cdot$}
\put(46,0.16){$\cdot$}
\put(45.5,0.2){$\cdot$}
\put(45,0.25){$\cdot$}
\put(44.5,0.3){$\cdot$}
\put(44,0.36){$\cdot$}
\put(43.5,0.42){$\cdot$}
\put(43,0.49){$\cdot$}
\put(42.5,0.56){$\cdot$}
\put(42,0.64){$\cdot$}
\put(41.5,0.73){$\cdot$}
\put(41,0.82){$\cdot$}
\put(40.5,0.91){$\cdot$}
\put(40,1.01){$\cdot$}
\put(39.5,1.11){$\cdot$}
\put(39,1.22){$\cdot$}
\put(38.5,1.34){$\cdot$}
\put(38,1.46){$\cdot$}
\put(37.5,1.59){$\cdot$}
\put(37,1.72){$\cdot$}
\put(28,3){$\fH^{l_0}_4$}\put(-40,-4){$p-q\equiv 4\!\!\!\!\pmod{8}$}
%quadrants II-IV
\put(93,50){2}
\put(98.28,63){$\cdot$}
\put(98.41,62.5){$\cdot$}
\put(98.54,62){$\cdot$}
\put(98.66,61.5){$\cdot$}
\put(98.77,61){$\cdot$}
\put(98.88,60.5){$\cdot$}
\put(98.99,60){$\cdot$}
\put(99.09,59.5){$\cdot$}
\put(99.18,59){$\cdot$}
\put(99.27,58.5){$\cdot$}
\put(99.35,58){$\cdot$}
\put(99.43,57.5){$\cdot$}
\put(99.51,57){$\cdot$}
\put(99.57,56.5){$\cdot$}
\put(99.64,56){$\cdot$}
\put(99.7,55.5){$\cdot$}
\put(99.75,55){$\cdot$}
\put(99.8,54.5){$\cdot$}
\put(99.84,54){$\cdot$}
\put(99.86,53.5){$\cdot$}
\put(99.91,53){$\cdot$}
\put(99.94,52.5){$\cdot$}
\put(99.96,52){$\cdot$}
\put(99.97,51.5){$\cdot$}
\put(99.98,51){$\cdot$}
\put(99.99,50.5){$\cdot$}
\put(100,50){$\cdot$}
\put(98.28,37){$\cdot$}
\put(98.41,37.5){$\cdot$}
\put(98.54,38){$\cdot$}
\put(98.66,38.5){$\cdot$}
\put(98.77,39){$\cdot$}
\put(98.88,39.5){$\cdot$}
\put(98.99,40){$\cdot$}
\put(99.09,40.5){$\cdot$}
\put(99.18,41){$\cdot$}
\put(99.27,41.5){$\cdot$}
\put(99.35,42){$\cdot$}
\put(99.43,42.5){$\cdot$}
\put(99.51,43){$\cdot$}
\put(99.57,43.5){$\cdot$}
\put(99.64,44){$\cdot$}
\put(99.7,44.5){$\cdot$}
\put(99.75,45){$\cdot$}
\put(99.8,45.5){$\cdot$}
\put(99.84,46){$\cdot$}
\put(99.86,46.5){$\cdot$}
\put(99.91,47){$\cdot$}
\put(99.94,47.5){$\cdot$}
\put(99.96,48){$\cdot$}
\put(99.97,48.5){$\cdot$}
\put(99.98,49){$\cdot$}
\put(99.99,49.5){$\cdot$}
%quadrants III-I
\put(7,50){6}
\put(1,32){$\fC^{l_0}_3$}\put(-65,29){$p-q\equiv 3\!\!\!\!\pmod{8}$}
\put(1.72,63){$\cdot$}
\put(1.59,62.5){$\cdot$}
\put(1.46,62){$\cdot$}
\put(1.34,61.5){$\cdot$}
\put(1.22,61){$\cdot$}
\put(1.11,60.5){$\cdot$}
\put(1.01,60){$\cdot$}
\put(0.99,59.5){$\cdot$}
\put(0.82,59){$\cdot$}
\put(0.73,58.5){$\cdot$}
\put(0.64,58){$\cdot$}
\put(0.56,57.5){$\cdot$}
\put(0.49,57){$\cdot$}
\put(0.42,56.5){$\cdot$}
\put(0.36,56){$\cdot$}
\put(0.3,55.5){$\cdot$}
\put(0.25,55){$\cdot$}
\put(0.2,54.5){$\cdot$}
\put(0.16,54){$\cdot$}
\put(0.12,53.5){$\cdot$}
\put(0.09,53){$\cdot$}
\put(0.06,52.5){$\cdot$}
\put(0.04,52){$\cdot$}
\put(0.02,51.5){$\cdot$}
\put(0.01,51){$\cdot$}
\put(0,50.5){$\cdot$}
\put(0,50){$\cdot$}
\put(1.72,37){$\cdot$}
\put(1.59,37.5){$\cdot$}
\put(1.46,38){$\cdot$}
\put(1.34,38.5){$\cdot$}
\put(1.22,39){$\cdot$}
\put(1.11,39.5){$\cdot$}
\put(1.01,40){$\cdot$}
\put(0.99,40.5){$\cdot$}
\put(0.82,41){$\cdot$}
\put(0.73,41.5){$\cdot$}
\put(0.64,42){$\cdot$}
\put(0.56,42.5){$\cdot$}
\put(0.49,43){$\cdot$}
\put(0.42,43.5){$\cdot$}
\put(0.36,44){$\cdot$}
\put(0.3,44.5){$\cdot$}
\put(0.25,45){$\cdot$}
\put(0.2,45.5){$\cdot$}
\put(0.16,46){$\cdot$}
\put(0.12,46.5){$\cdot$}
\put(0.09,47){$\cdot$}
\put(0.06,47.5){$\cdot$}
\put(0.04,48){$\cdot$}
\put(0.02,48.5){$\cdot$}
\put(0.01,49){$\cdot$}
\put(0,49.5){$\cdot$}
\put(0.5,67){$\fR^{l_0}_2$}\put(-65,75){$p-q\equiv 2\!\!\!\!\pmod{8}$}
\end{picture}
\]

\vspace{2ex}
\begin{center}
\begin{minipage}{25pc}{\small
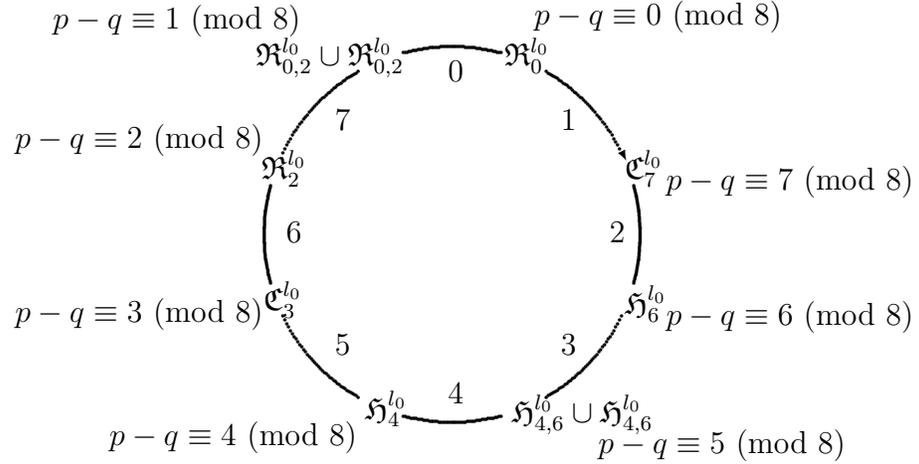
{\bf Fig.2} The action of the Brauer--Wall group $BW_{\R}\simeq\dZ_8$ on
the full system $\fM=\fM^+\oplus\fM^-$ of real representations $\fD$ of the
proper Lorentz group $\fG_+$, $l_0=\frac{p+q}{4}$.}
\end{minipage}
\end{center}
\medskip
Further, it is well--known that a group structure over $\cl_{p,q}$, defined by
$BW_{\R}\simeq\dZ_8$, immediately relates with the Atiyah--Bott--Shapiro
periodicity \cite{AtBSh}. In accordance with \cite{AtBSh}, the Clifford
algebra over the field $\F=\R$ is modulo 8 periodic:
$\cl_{p+8,q}\simeq\cl_{p,q}\otimes\cl_{8,0}\,(\cl_{p,q+8}\simeq\cl_{p,q}
\otimes\cl_{0,8})$. Therefore, we have a following relation
\[
\fD^{l_0+2}\simeq\fD^{l_0}\otimes\fR^2_0,
\]
since $\fR^2_0\leftrightarrow\cl_{8,0}\;(\cl_{0,8})$ and in virtue of
Karoubi Theorem from (\ref{Ten}) it follows that
$\cl_{8,0}\simeq\cl_{2,0}\otimes\cl_{0,2}\otimes\cl_{0,2}\otimes\cl_{2,0}$
($\cl_{0,8}\simeq\cl_{0,2}\otimes\cl_{2,0}\otimes\cl_{2,0}
\otimes\cl_{0,2}$)\footnote{The minimal left ideal of $\cl_{8,0}$ is equal
to $\dS_{16}$ and in virtue of the real ring $\K\simeq\R$ is defined within
the full matrix algebra $\M_{16}(\R)$. At first glance, from the
factorization of $\cl_{8,0}$ it follows that
$\M_2(\R)\otimes\BH\otimes\BH\otimes\M_2(\R)\not\simeq\M_{16}(\R)$, but it
is wrong, since there is an isomorphism $\BH\otimes\BH\simeq\M_4(\R)$
(see Appendix B in \cite{BDGK}).}, therefore, $r=4$, $l_0=r/2=2$.
On the other hand, in terms of minimal left ideal the modulo 8
periodicity looks like
\[
\dS_{n+8}\simeq\dS_n\otimes\dS_{16}.
\]
In virtue of the mapping $\gamma_{8,0}:\,\cl_{8,0}\rightarrow\M_2(\dO)$
\cite{MS96} (see also excellent review \cite{Bae01}) the latter relation
can be written in the form
\[
\dS_{n+8}\simeq\dS_n\otimes\dO^2,
\]
where $\dO$ is {\it an octonion algebra}. Since the algebra $\cl_{8,0}\simeq
\cl_{0,8}$ admits an octonionic representation, then in virtue of the
modulo 8 periodicity the octonionic representations can be defined for all
high dimensions and, therefore, on the system $\fM=\fM^+\oplus\fM^-$ we
have a relation
\[
\fD^{l_0+2}\simeq\fD^{l_0}\otimes\fO,
\]
where $\fO$ is {\it an octonionic representation} of the group $\fG_+$
($\fO\sim\fR^2_0$). Thus, the action of $BW_{\R}\simeq\dZ_8$ form a cycle
of the period 8 on the system $\fM$. This is intimately related with an
octonionic structure. In 1973, G\"{u}naydin and G\"{u}rsey showed that an
automorphism group of the algebra $\dO$ is isomorphic to an exceptional Lie
group $G_2$ that contains $SU(3)$ as a subgroup \cite{GG73}. The
G\"{u}naydin--G\"{u}rsey construction allows to incorporate the quark
phenomenology into a general algebraic framework. Moreover, this
construction allows to define the quark structure on the system $\fM$
within octonionic representations of the proper Lorentz group $\fG_+$.
It is obvious that within such a framework the quark structure cannot be
considered as a fundamental physical structure underlieing of the world
(as it suggested by QCD). This is fully derivative structure firstly
appearred in 8-dimension and further reproduced into high dimensions by the
round of 8-cycle generated by the group $BW_{\R}\simeq\dZ_8$ from 8 ad
infinitum (growth of quark's flavors with increase of energy). One can say
that such a description, included very powerful algebraic tools, opens an
another way of understanding of the Gell-Mann--Ne'emann eightfold way in
particle physics.\section{Pseudoautomorphism $\cA\longrightarrow\overline{\cA}$
and charge conjugation}
As noted previously, an extraction of the minimal left ideal of the complex
algebra $\C_n\simeq\C_2\otimes\C_2\otimes\cdots\otimes\C_2$ induces a space
of the finite--dimensional spintensor representation of the group
$\fG_+$. Besides, the algebra $\C_n$ is associated with a complex vector
space $\C^n$. Let $n=p+q$, then an extraction operation of the real subspace
$\R^{p,q}$ in $\C^n$  forms the foundation of definition of the discrete
transformation known in physics as
{\it a charge conjugation} $C$. Indeed, let
$\{\e_1,\ldots,\e_n\}$ be an orthobasis in the space $\C^n$, $\e^2_i=1$.
Let us remain the first $p$ vectors of this basis unchanged, and other $q$
vectors multiply by the factor $i$. Then the basis
\begin{equation}\label{6.23}
\left\{\e_1,\ldots,\e_p,i\e_{p+1},\ldots,i\e_{p+q}\right\}
\end{equation}
allows to extract the subspace $\R^{p,q}$ in $\C^n$. Namely,
for the vectors $\R^{p,q}$ we take the vectors of
$\C^n$ which decompose on the basis
(\ref{6.23}) with real coefficients. In such a way we obtain a real vector
space $\R^{p,q}$ endowed (in general case) with a non--degenerate
quadratic form
\[
Q(x)=x^2_1+x^2_2+\ldots+x^2_p-x^2_{p+1}-x^2_{p+2}-\ldots-x^2_{p+q},
\]
where $x_1,\ldots,x_{p+q}$ are coordinates of the vector $\bx$ 
in the basis (\ref{6.23}).
It is easy to see that the extraction of
$\R^{p,q}$ in $\C^n$ induces an extraction of
{\it a real subalgebra} $\cl_{p,q}$ in $\C_n$. Therefore, any element
$\cA\in\C_n$ can be unambiguously represented in the form
\[
\cA=\cA_1+i\cA_2,
\]
where $\cA_1,\,\cA_2\in\cl_{p,q}$. The one--to--one mapping
\begin{equation}\label{6.24}
\cA\longrightarrow\overline{\cA}=\cA_1-i\cA_2
\end{equation}
transforms the algebra $\C_n$ into itself with preservation of addition
and multiplication operations for the elements $\cA$; the operation of
multiplication of the element $\cA$ by the number transforms to an operation
of multiplication by the complex conjugate number.
Any mapping of $\C_n$ satisfying these conditions is called
{\it a pseudoautomorphism}. Thus, the extraction of the subspace
$\R^{p,q}$ in the space $\C^n$ induces in the algebra $\C_n$ 
a pseudoautomorphism $\cA\rightarrow\overline{\cA}$ \cite{Rash}.

Let us consider a spinor representation of the pseudoautomorphism
$\cA\rightarrow\overline{\cA}$ of the algebra $\C_n$ when $n\equiv 0\s\pmod{2}$.
In the spinor representation the every element $\cA\in\C_n$ should be
represented by some matrix $\sA$, and the pseudoautomorphism (\ref{6.24})
takes a form of the pseudoautomorphism of the full matrix algebra 
$\M_{2^{n/2}}$:
\[
\sA\longrightarrow\overline{\sA}.
\]\begin{sloppypar}\noindent
On the other hand, a transformation replacing the matrix $\sA$ by the
complex conjugate matrix, $\sA\rightarrow\dot{\sA}$, is also some
pseudoautomorphism of the algebra $\M_{2^{n/2}}$. The composition of the two
pseudoautomorpisms $\dot{\sA}\rightarrow\sA$ and
$\sA\rightarrow\overline{\sA}$, $\dot{\sA}\rightarrow\sA\rightarrow
\overline{\sA}$, is an internal automorphism
$\dot{\sA}\rightarrow\overline{\sA}$ of the full matrix algebra $\M_{2^{n/2}}$:
\end{sloppypar}
\begin{equation}\label{6.25}
\overline{\sA}=\Pi\dot{\sA}\Pi^{-1},
\end{equation}
where $\Pi$ is a matrix of the pseudoautomorphism 
$\cA\rightarrow\overline{\cA}$ in the spinor representation.
The sufficient condition for definition of the pseudoautomorphism
$\cA\rightarrow\overline{\cA}$ is a choice of the matrix
$\Pi$ in such a way that the transformation 
$\sA\rightarrow\Pi\dot{\sA}\Pi^{-1}$ transfers into itself the matrices
$\cE_1,\ldots,\cE_p,i\cE_{p+1},\ldots,i\cE_{p+q}$
(the matrices of the spinbasis of $\cl_{p,q}$), that is,
\begin{equation}\label{6.26}
\cE_i\longrightarrow\cE_i=\Pi\dot{\cE}_i\Pi^{-1}\quad
(i=1,\ldots,p+q).
\end{equation}
\begin{theorem}\label{tpseudo}
Let $\C_n$ be a complex Clifford algebra when $n\equiv 0\s\pmod{2}$
and let $\cl_{p,q}\subset\C_n$ be its subalgebra with a real division ring
$\K\simeq\R$ when $p-q\equiv 0,2\s\pmod{8}$ and quaternionic division ring
$\K\simeq\BH$ when $p-q\equiv 4,6\s\pmod{8}$, $n=p+q$. Then in dependence
on the division ring structure of the real subalgebra $\cl_{p,q}$ the matrix
$\Pi$ of the pseudoautomorphism $\cA\rightarrow\overline{\cA}$ 
has the following form:\\[0.2cm]
1) $\K\simeq\R$, $p-q\equiv 0,2\s\pmod{8}$.\\[0.1cm]
The matrix $\Pi$ for any spinor representation over the ring $\K\simeq\R$
is proportional to the unit matrix.\\[0.2cm]
2) $\K\simeq\BH$, $p-q\equiv 4,6\s\pmod{8}$.\\[0.1cm]
$\Pi=\cE_{\alpha_1}\cE_{\alpha_2}\cdots\cE_{\alpha_a}$ when 
$a\equiv 0\s\pmod{2}$ and
$\Pi=\cE_{\beta_1}\cE_{\beta_2}\cdots\cE_{\beta_b}$ when $b\equiv 1\s\pmod{2}$,
where $a$ complex matrices $\cE_{\alpha_t}$ 
and $b$ real matrices $\cE_{\beta_s}$ form a basis of the spinor
representation of the algebra $\cl_{p,q}$ over the ring $\K\simeq\BH$,
$a+b=p+q,\,0<t\leq a,\,0<s\leq b$. At this point
\begin{eqnarray}
\Pi\dot{\Pi}&=&\phantom{-}\sI\quad\text{if $a,b\equiv 0,1\s\pmod{4}$},
\nonumber\\
\Pi\dot{\Pi}&=&-\sI\quad\text{if $a,b\equiv 2,3\s\pmod{4}$},\nonumber
\end{eqnarray}
where $\sI$ is the unit matrix.
\end{theorem}
\begin{proof}
The algebra $\C_n$ ($n\equiv 0\s\pmod{2}$, $n=p+q$) in virtue of
$\C_n=\C\otimes\cl_{p,q}$ and definition of the division ring
$\K\simeq f\cl_{p,q}f$ 
($f$ is a primitive idempotent of the algebra $\cl_{p,q}$)
has four different real subalgebras: $p-q\equiv 0,2\s\pmod{8}$
for the real division ring $\K\simeq\R$ and $p-q\equiv 4,6\s\pmod{8}$ for
the quaternionic division ring $\K\simeq\BH$.\\[0.2cm]
1) $\K\simeq\R$.\\[0.1cm]
Since for the types $p-q\equiv 0,2\s\pmod{8}$ there is an isomorphism
$\cl_{p,q}\simeq\M_{2^{\frac{p+q}{2}}}(\R)$ (Wedderburn--Artin Theorem), then
all the matrices $\cE_i$ of the spinbasis of $\cl_{p,q}$ are real and
$\dot{\cE}_i=\cE_i$. Therefore, in this case the condition (\ref{6.26})
can be written as follows
\[
\cE_i\longrightarrow\cE_i=\Pi\cE_i\Pi^{-1},
\]
whence $\cE_i\Pi=\Pi\cE_i$. Thus, for the algebras $\cl_{p,q}$ of the types
$p-q\equiv 0,2\s\pmod{8}$ the matrix $\Pi$ of the pseudoautomorphism
$\cA\rightarrow\overline{\cA}$ commutes with all the matrices $\cE_i$.
It is easy to see that
$\Pi\sim\sI$.\\[0.2cm]
2) $\K\simeq\BH$.\\[0.1cm]
In turn, for the quaternionic types $p-q\equiv 4,6\s\pmod{8}$ there is an
isomorphism $\cl_{p,q}\simeq\M_{2^{\frac{p+q}{2}}}(\BH)$. Therefore, among
the matrices of the spinbasis of the algebra $\cl_{p,q}$ there are matrices
$\cE_\alpha$ satisfying the condition $\dot{\cE}_\alpha=-\cE_\alpha$. 
Let $a$ be a quantity of the complex matrices, then the spinbasis of $\cl_{p,q}$
is divided into two subsets. The first subset
$\{\dot{\cE}_{\alpha_t}=-\cE_{\alpha_t}\}$ contains complex matrices,
$0<t\leq a$, and the second subset
$\{\dot{\cE}_{\beta_s}=\cE_{\beta_s}\}$ contains real matrices,
$0<s\leq p+q-a$. In accordance with a spinbasis structure of the algebra
$\cl_{p,q}\simeq\M_{2^{\frac{p+q}{2}}}(\BH)$ the condition (\ref{6.26})
can be written as follows
\[
\cE_{\alpha_t}\longrightarrow-\cE_{\alpha_t}=\Pi\cE_{\alpha_t}\Pi^{-1},\quad
\cE_{\beta_s}\longrightarrow\cE_{\beta_s}=\Pi\cE_{\beta_s}\Pi^{-1}.
\]
Whence
\begin{equation}\label{6.27}
\cE_{\alpha_t}\Pi=-\Pi\cE_{\alpha_t},\quad
\cE_{\beta_s}\Pi=\Pi\cE_{\beta_s}.
\end{equation}
Thus, for the quaternionic types $p-q\equiv 4,6\s\pmod{8}$ the matrix
$\Pi$ of the pseudoautomorphism $\cA\rightarrow\overline{\cA}$ anticommutes
with a complex part of the spinbasis of $\cl_{p,q}$ and commutes with
a real part of the same spinbasis. From (\ref{6.27}) it follows that a
structure of the matrix $\Pi$ is analogous to the structure of
the matrices $\sE$ and $\sC$ of the antiautomorphisms
$\cA\rightarrow\widetilde{\cA}$ and
$\cA\rightarrow\widetilde{\cA^\star}$, correspondingly 
(see Theorem 4 in \cite{Var00}), that is, the matrix
$\Pi$ of the pseudoautomorphism $\cA\rightarrow\overline{\cA}$ of the algebra
$\C_n$ is a product of only complex matrices, or only real matrices
of the spinbasis of the subalgebra $\cl_{p,q}$.

So, let $0<a<p+q$ and let $\Pi=\cE_{\alpha_1}\cE_{\alpha_2}\cdots
\cE_{\alpha_a}$ be a matrix of $\cA\rightarrow\overline{\cA}$,
then permutation conditions of the matrix $\Pi$ 
with the matrices $\cE_{\beta_s}$
of the real part ($0<s\leq p+q-a$) and with the matrices
$\cE_{\alpha_t}$ of the complex part ($0<t\leq a$) have the form
\begin{equation}\label{6.28}
\Pi\cE_{\beta_s}=(-1)^a\cE_{\beta_s}\Pi,
\end{equation}
\begin{eqnarray}
\Pi\cE_{\alpha_t}&=&(-1)^{a-t}\sigma(\alpha_t)\cE_{\alpha_1}\cE_{\alpha_2}
\cdots\cE_{\alpha_{t-1}}\cE_{\alpha_{t+1}}\cdots\cE_{\alpha_a},\nonumber\\
\cE_{\alpha_t}\Pi&=&(-1)^{t-1}\sigma(\alpha_t)\cE_{\alpha_1}\cE_{\alpha_2}
\cdots\cE_{\alpha_{t-1}}\cE_{\alpha_{t+1}}\cdots\cE_{\alpha_a},\label{6.29}
\end{eqnarray}
that is, when $a\equiv 0\s\pmod{2}$ the matrix $\Pi$ commutes with the real
part and anticommutes with the complex part of the spinbasis of $\cl_{p,q}$.
Correspondingly, when $a\equiv 1\s\pmod{2}$ the matrix $\Pi$ anticommutes
with the real part and commutes with the complex part. Further, let
$\Pi=\cE_{\beta_1}\cE_{\beta_2}\cdots\cE_{\beta_{p+q-a}}$ be a product of the
real matrices, then
\begin{eqnarray}
\Pi\cE_{\beta_s}&=&(-1)^{p+q-a-s}\sigma(\beta_s)\cE_{\beta_1}\cE_{\beta_2}
\cdots\cE_{\beta_{s-1}}\cE_{\beta_{s+1}}\cdots\cE_{\beta_{p+q-a}},\nonumber\\
\cE_{\beta_s}\Pi&=&(-1)^{s-1}\sigma(\beta_s)\cE_{\beta_1}\cE_{\beta_2}
\cdots\cE_{\beta_{s-1}}\cE_{\beta_{s+1}}\cdots\cE_{\beta_{p+q-a}},\label{6.30}
\end{eqnarray}
\begin{equation}\label{6.31}
\Pi\cE_{\alpha_t}=(-1)^{p+q-a}\cE_{\alpha_t}\Pi,
\end{equation}
that is, when $p+q-a\equiv 0\s\pmod{2}$ the matrix $\Pi$ anticommutes with
the real part and commutes with the complex part 
of the spinbasis of $\cl_{p,q}$. Correspondingly, when
$p+q-a\equiv 1\s\pmod{2}$ the matrix $\Pi$ commutes with the real part and
anticommutes with the complex part.

The comparison of the conditions (\ref{6.28})--(\ref{6.29}) 
with the condition (\ref{6.27}) shows that the matrix
$\Pi=\cE_{\alpha_1}\cE_{\alpha_2}\cdots\cE_{\alpha_a}$ exists only at
$a\equiv 0\s\pmod{2}$, that is, $\Pi$ is a product of the complex matrices
$\cE_{\alpha_t}$ of the even number. In its turn, a comparison of
(\ref{6.30})--(\ref{6.31}) with
(\ref{6.27}) shows that the matrix $\Pi=\cE_{\beta_1}\cE_{\beta_2}\cdots
\cE_{\beta_{p+q-a}}$ exists only at $p+q-a\equiv 1\s\pmod{2}$, that is,
$\Pi$ is a product of the real matrices $\cE_{\beta_s}$ of the odd number.

Let us calculate now the product $\Pi\dot{\Pi}$. 
Let $\Pi=\cE_{\beta_1}\cE_{\beta_2}\cdots
\cE_{\beta_{p+q-a}}$ be a product of the $p+q-a$ real matrices.
Since $\dot{\cE}_{\beta_s}=\cE_{\beta_s}$, then
$\dot{\Pi}=\Pi$ and $\Pi\dot{\Pi}=\Pi^2$. Therefore,
\begin{equation}\label{6.32}
\Pi\dot{\Pi}=(\cE_{\beta_1}\cE_{\beta_2}\cdots\cE_{\beta_{p+q-a}})^2=
(-1)^{\frac{(p+q-a)(p+q-a-1)}{2}}\cdot\sI.
\end{equation}
Further, let $\Pi=\cE_{\alpha_1}\cE_{\alpha_2}\cdots\cE_{\alpha_a}$ be a
product of the $a$ complex matrices. Then
$\dot{\cE}_{\alpha_t}=-\cE_{\alpha_t}$ and $\dot{\Pi}=(-1)^a\Pi=\Pi$, since
$a\equiv 0\s\pmod{2}$. Therefore,
\begin{equation}\label{6.33}
\Pi\dot{\Pi}=(\cE_{\alpha_1}\cE_{\alpha_2}\cdots\cE_{\alpha_a})^2=
(-1)^{\frac{a(a-1)}{2}}\cdot\sI.
\end{equation}
Let $p+q-a=b$ be a quantity of the real matrices $\cE_{\beta_s}$ of the
spinbasis of $\cl_{p,q}$, then $p+q=a+b$. Since $p+q$ is always even number
for the quaternionic types $p-q\equiv 4,6\s\pmod{8}$, then $a$ and $b$ 
are simultaneously even or odd numbers. Thus, from (\ref{6.32}) and (\ref{6.33})
it follows
\[
\Pi\dot{\Pi}=\begin{cases}
\phantom{-}\sI,& \text{if $a,b\equiv 0,1\s\!\!\pmod{4}$},\\
-\sI,& \text{if $a,b\equiv 2,3\s\!\!\pmod{4}$},
\end{cases}
\]
which required to be proved.
\end{proof}

In the present form of quantum field theory complex fields correspond
to charged particles. Thus, the extraction of the subalgebra $\cl_{p,q}$ with
the real ring $\K\simeq\R$ in $\C_n$, $p-q\equiv 0,2\s\pmod{8}$,
corresponds to physical fields describing {\it truly neutral particles}
such as photon and neutral mesons ($\pi^0,\,\eta^0,\,\rho^0,\,
\omega^0,\,\varphi^0,\,K^0$). In turn, the subalgebras $\cl_{p,q}$ with the
ring $\K\simeq\BH$, $p-q\equiv 4,6\s\pmod{8}$ correspond to charged or
neutral fields.

As known \cite{GY48b}, the charge conjugation $C$ should be satisfied the
following requirement
\begin{equation}\label{Com}
CI^{ik}=I^{ik}C,
\end{equation}
where $I^{ik}$ are infinitesimal operators of the group $\fG_+$\footnote{The
requirement $CP=PC$ presented also in the Gel'fand--Yaglom work \cite{GY48b}
is superfluous, since the inverse relation $CP=-PC$ is valid in
BWW--type quantum field theories \cite{AJG93}.}. This requirement is necessary
for the definition of the operation $C$ on the representation spaces of
$\fC$. Let us find permutation conditions of the matrix $\Pi$ with $I^{ik}$
defined by the relations (\ref{O1})--(\ref{O2}). It is obvious that in the
case of $\K\simeq\R$ the matrix $\Pi\simeq\sI$ commutes with all the
operators $I^{ik}$ and, therefore, the relations (\ref{Com}) hold. In the
case of $\K\simeq\BH$ and $\Pi=\cE_{\alpha_1}\cE_{\alpha_2}\cdots\cE_{\alpha_a}$
it is easy to verify that at $\cE_a,\cE_b,\cE_c\in\Pi$ (all 
$\cE_a,\cE_b,\cE_c$ are complex matrices) the matrix $\Pi$ commutes with
$A_{ik}$ and anticommutes with $B_i$ (permutation conditions in this case
are analogous to (\ref{R3}) and (\ref{R4})). In its turn, when
$\cE_a,\cE_b,\cE_c\not\in\Pi$ (all $\cE_a,\cE_b,\cE_c$ are real matrices) the
matrix $\Pi$ commutes with all the operators $I^{ik}$. It is easy to see
that all other cases given by the cyclic permutations $\cE_i,\cE_j\in\Pi$,
$\cE_k\not\in\Pi$ and $\cE_i\in\Pi$, $\cE_j,\cE_k\not\in\Pi$
($i,j,k\in\{a,b,c\}$) do not satisfy the relations (\ref{Com}). For example,
at $\cE_a,\cE_b\in\Pi$ and $\cE_c\not\in\Pi$ the matrix $\Pi$ commutes with
$A_{23}$ and $B_1$ and anticommutes with $A_{13},A_{12},B_2,B_3$. Further,
in the case of $\Pi=\cE_{\beta_1}\cE_{\beta_2}\cdots\cE_{\beta_b}$ it is not
difficult to see that only at $\cE_a,\cE_b,\cE_c\in\Pi$ (all
$\cE_a,\cE_b,\cE_c$ are real matrices) the matrix $\Pi$ commutes with all
$I^{ik}$. Therefore, in both cases $\Pi=\cE_{\alpha_1}\cE_{\alpha_2}\cdots
\cE_{\alpha_a}$ and $\Pi=\cE_{\beta_1}\cE_{\beta_2}\cdots\cE_{\beta_b}$ the
relations (\ref{Com}) hold when all the matrices $\cE_a,\cE_b,\cE_c$
belonging to (\ref{O1})--(\ref{O2}) are real.

When we restrict the complex representation $\fC$ (charged particles)
of $\fG_+$ to real representation $\fR$ (truly neutral particles) and
$\fH$ (neutral particles) we see that in this case the charge conjugation
is reduced to an identical transformation $\sI$ for $\fR$ and to a
particle--antiparticle conjugation $C^\prime$ 
for $\fH$. Moreover, as follows from
Theorem \ref{tinf} for the real representations $B_i=0$ and, therefore,
the relations (\ref{Com}) take a form
\begin{equation}\label{Com2}
C^\prime A_{ik}=A_{ik}C^\prime.
\end{equation}
Over the ring $\K\simeq\R$ the relations (\ref{Com2}) hold identically.
It is easy to verify that over the ring $\K\simeq\BH$ for the matrix
$\Pi=\cE_{\alpha_1}\cE_{\alpha_2}\cdots\cE_{\alpha_a}$ the relations
(\ref{Com2}) hold at $\cE_a,\cE_b,\cE_c\in\Pi$ and 
$\cE_a,\cE_b,\cE_c\not\in\Pi$. The same result takes place for the matrix
$\Pi=\cE_{\beta_1}\cE_{\beta_2}\cdots\cE_{\beta_b}$. All other cases given by
the cyclic permutations do not satisfy the relations (\ref{Com2}).
Therefore, in both cases the relations (\ref{Com2}) hold when all the
matrices $\cE_a,\cE_b,\cE_c$ in (\ref{O1}) (correspondingly in (\ref{O3}))
are complex or real.
Let us consider the action of the pseudoautomorphism 
$\cA\rightarrow\overline{\cA}$ on the spinors
(\ref{6.5}) (`vectors' of the fundamental representation of the group
$\fG_+$). The matrix $\Pi$ allows to compare to each spinor
$\xi^\alpha$ its conjugated spinor $\overline{\xi}^\alpha$ by the following
rule
\begin{equation}\label{6.33'}
\overline{\xi}^\alpha=\Pi^\alpha_{\dot{\alpha}}\xi^{\dot{\alpha}},
\end{equation}
here $\xi^{\dot{\alpha}}=(\xi^\alpha)^\cdot$. In accordance with Theorem
\ref{tpseudo} for the matrix $\Pi^\alpha_{\dot{\beta}}$ 
we have $\dot{\Pi}=\Pi^{-1}$ or
$\dot{\Pi}=-\Pi^{-1}$, where $\Pi^{-1}=\Pi^{\dot{\alpha}}_\beta$.
Then a twice conjugated spinor looks like
\[
\overline{\xi}^\alpha=\overline{\Pi^\alpha_{\dot{\beta}}\xi^{\dot{\beta}}}=
\Pi^\alpha_{\dot{\alpha}}(\Pi^\alpha_{\dot{\beta}}\xi^{\dot{\beta}})^\cdot=
\Pi^\alpha_{\dot{\alpha}}(\pm\Pi^{\dot{\alpha}}_\beta)\xi^\beta=
\pm\xi^\alpha.
\]
Therefore, the twice conjugated spinor coincides with the initial spinor
in the case of the real subalgebra of $\C_2$ with the ring
$\K\simeq\R$ (the algebras $\cl_{1,1}$ and $\cl_{2,0}$), and also in the case
of $\K\simeq\BH$ (the algebra $\cl_{0,2}\simeq\BH$) at
$a-b\equiv 0,1\pmod{4}$. Since for the algebra $\cl_{0,2}\simeq\BH$ we have
always $a-b\equiv 0\pmod{4}$, then a property of the reciprocal conjugacy
of the spinors
$\xi^\alpha$ ($\alpha=1,2$) is an invariant fact for the fundamental
representation of the group $\fG_+$ (this property is very important in
physics, since this is an algebraic expression of the requirement
$C^2=1$). Further, since the `vector' (spintensor) of the 
finite--dimensional representation of the group
$\fG_+$ is defined by the tensor product
$\xi^{\alpha_1\alpha_2\cdots\alpha_k}=\sum\xi^{\alpha_1}\otimes
\xi^{\alpha_2}\otimes\cdots\otimes\xi^{\alpha_k}$, then its conjugated
spintensor takes a form
\begin{equation}\label{6.33''}
\overline{\xi}^{\alpha_1\alpha_2\cdots\alpha_k}=
\sum\Pi^{\alpha_1}_{\dot{\alpha}_1}\Pi^{\alpha_2}_{\dot{\alpha}_2}\cdots
\Pi^{\alpha_k}_{\dot{\alpha}_k}\xi^{\dot{\alpha}_1\dot{\alpha}_2\cdots
\dot{\alpha}_k},
\end{equation}\begin{sloppypar}\noindent
It is obvious that the condition of reciprocal conjugacy 
$\overline{\overline{\xi}}\!{}^{\alpha_1\alpha_2\cdots\alpha_k}=
\xi^{\alpha_1\alpha_2
\cdots\alpha_k}$ is also fulfilled for (\ref{6.33''}), since for each matrix
$\Pi^{\alpha_i}_{\dot{\alpha}_i}$ in (\ref{6.33''}) we have
$\dot{\Pi}=\Pi^{-1}$ (all the matrices $\Pi^{\alpha_i}_{\dot{\alpha}_i}$
are defined for the algebra $\C_2$). \end{sloppypar}

Let us define now permutation conditions of the matrix $\Pi$ of the
pseudoautomorphism
$\cA\rightarrow\overline{\cA}$ (charge conjugation) with the matrix $\sW$ 
of the automorphism $\cA\rightarrow\cA^\star$ (space inversion).
First of all, in accordance with Theorem \ref{tpseudo} in the case of
$\cl_{p,q}$ with the real ring $\K\simeq\R$ (types $p-q\equiv 0,2\pmod{8}$)
the matrix $\Pi$ is proportional to the unit matrix and, therefore,
commutes with the matrix $\sW$. In the case of
$\K\simeq\BH$ (types $p-q\equiv 4,6\pmod{8}$) from Theorem
\ref{tpseudo} it follows two possibilities:
$\Pi=\cE_{\alpha_1}\cE_{\alpha_2}\cdots\cE_{\alpha_a}$ is a product of
$a$ complex matrices at $a\equiv 0\pmod{2}$ and
$\Pi=\cE_{\beta_1}\cE_{\beta_2}\cdots\cE_{\beta_b}$ is a product of
$b$ real matrices at $b\equiv 1\pmod{2}$. 
Since $a+b=p+q$, then the matrix $\sW$ can be represented by the product
$\cE_{\alpha_1}\cE_{\alpha_2}\cdots\cE_{\alpha_a}\cE_{\beta_1}\cE_{\beta_2}
\cdots\cE_{\beta_b}$. Then for $\Pi=\cE_{\alpha_1}\cE_{\alpha_2}\cdots
\cE_{\alpha_a}$ we have
\begin{eqnarray}
\Pi\sW&=&(-1)^{\frac{a(a-1)}{2}}\sigma(\alpha_1)\sigma(\alpha_2)\cdots
\sigma(\alpha_a)\cE_{\beta_1}\cE_{\beta_2}\cdots\cE_{\beta_b},\nonumber\\
\sW\Pi&=&(-1)^{\frac{a(a-1)}{2}+ba}\sigma(\alpha_1)\sigma(\alpha_2)\cdots
\sigma(\alpha_a)\cE_{\beta_1}\cE_{\beta_2}\cdots\cE_{\beta_b}.\nonumber
\end{eqnarray}
Hence it follows that at $ab\equiv 0\pmod{2}$ the matrices $\Pi$ and $\sW$
always commute, since $a\equiv 0\pmod{2}$. Taking
$\Pi=\cE_{\beta_1}\cE_{\beta_2}\cdots\cE_{\beta_b}$ we obtain following
conditions:
\begin{eqnarray}
\Pi\sW&=&(-1)^{\frac{b(b-1)}{2}+ab}\sigma(\beta_1)\sigma(\beta_2)\cdots
\sigma(\beta_b)\cE_{\alpha_1}\cE_{\alpha_2}\cdots\cE_{\alpha_a},\nonumber\\
\sW\Pi&=&(-1)^{\frac{b(b-1)}{2}}\sigma(\beta_1)\sigma(\beta_2)\cdots
\sigma(\beta_b)\cE_{\alpha_1}\cE_{\alpha_2}\cdots\cE_{\alpha_a}.\nonumber
\end{eqnarray}
Hence it follows that $ab\equiv 1\pmod{2}$, since in this case 
$b\equiv 1\pmod{2}$,
and $p+q=a+b$ is even number, $a$ is odd number. Therefore, at
$ab\equiv 1\pmod{2}$ the matrices $\Pi$ and $\sW$ always anticommute.

It should be noted one important feature related with the anticommutation of
the matrices
$\Pi$ and $\sW$, $\Pi\sW=-\sW\Pi$, that corresponds to relation
$CP=-PC$. The latter relation holds for Bargmann--Wightmann--Wigner type
quantum field theories in which bosons and antibosons have mutually
opposite intrinsic parities \cite{AJG93}. Thus, in this case
the matrix of the operator $C$ is a product of real matrices of odd number.
\section{Quotient representations of the Lorentz group}
\begin{theorem}\label{tfactor}1) $\F=\C$.
Let $\cA\rightarrow\overline{\cA}$, $\cA\rightarrow\cA^\star$,
$\cA\rightarrow\widetilde{\cA}$ be the automorphisms of the odd--dimensional
complex Clifford algebra $\C_{n+1}$ ($n+1\equiv 1,3\s\pmod{4}$) corresponding
the discrete transformations $C,\,P,\,T$ (charge conjugation, space inversion,
time reversal) and let ${}^\epsilon\C_n$ be a quotient algebra obtained in the
result of the homomorphic mapping $\epsilon:\;\C_{n+1}
\rightarrow\C_n$. Then over the field $\F=\C$ 
in dependence on the structure of
${}^\epsilon\C_n$ all the quotient representations of the
Lorentz group are divided in the following six classes:
\begin{eqnarray}
1)\;{}^\chi\fC^{l_0+l_1-1,0}_{a_1}&:&\{T,\,C\sim I\},\nonumber\\
2)\;{}^\chi\fC^{l_0+l_1-1,0}_{a_2}&:&\{T,\,C\},\nonumber\\
3)\;{}^\chi\fC^{l_0+l_1-1,0}_{b}&:&\{T,\,CP,\,CPT\},\nonumber\\
%\end{array}\nonumber\\
%\text{$j$ is half--integer ($n\equiv 2\s\pmod{4}$)}\quad\begin{array}{lcl}
4)\;{}^\chi\fC^{l_0+l_1-1,0}_{c}&:&\{PT,\,C,\,CPT\},\nonumber\\
5)\;{}^\chi\fC^{l_0+l_1-1,0}_{d_1}&:&\{PT,\,CP\sim IP,\,CT\sim IT\},\nonumber\\
6)\;{}^\chi\fC^{l_0+l_1-1,0}_{d_2}&:&\{PT,\,CP,\,CT\}.\nonumber
\end{eqnarray}
2) $\F=\R$. Real quotient representations are divided into four different
classes:
\begin{eqnarray}
7)\;{}^\chi\fR^{l_0}_{e_1}&:&\{T,\,C\sim I,\,CT\sim IT\},\nonumber\\
8)\;{}^\chi\fR^{l_0}_{e_2}&:&\{T,\,CP\sim IP,\,CPT\sim IPT\},\nonumber\\
9)\;{}^\chi\fH^{l_0}_{f_1}&:&\{T,\,C\sim C^\prime,\,CT\sim C^\prime T \},
\nonumber\\
10)\;{}^\chi\fH^{l_0}_{f_2}&:&\{T,\,CP\sim C^\prime P,\,CPT\sim C^\prime PT\}.
\nonumber
\end{eqnarray}
%Correspondingly, for the fields of type $(j,0)\oplus(0,j)$ there exist the
%following six classes:
%\[
%{}^\chi\fD^{(j,0}_{class}(\sigma)\oplus{}^\chi\fD^{(0,j)}_{class}(\sigma),
%\]
%where $class=\{a_1,a_2,b,c,d_1,d_2\}$.
\end{theorem}
\begin{proof}
1) Complex representations.\\
Before we proceed to find an explicit form of the quotient representations
${}^\chi\fC$ it is necessary to consider in details a
structure of the quotient algebras
${}^\epsilon\C_n$ obtaining in the result of the homomorphic mapping
$\epsilon:\,\C_{n+1}\rightarrow\C_n$. The structure of the quotient algebra
${}^\epsilon\C_n$ depends on the transfer of the automorphisms
$\cA\rightarrow\cA^\star,
\;\cA\rightarrow\widetilde{\cA},\;\cA\rightarrow\widetilde{\cA^\star},\;
\cA\rightarrow\overline{\cA}$ of the algebra $\C_{n+1}$ under action of the
homomorphism $\epsilon$ onto its subalgebra $\C_n$. As noted previously
(see conclusion of Theorem \ref{tprod}), the homomorphisms $\epsilon$ and
$\chi$ have an analogous texture. The action of the homomorphism
$\epsilon$ is defined as follows
\[
\epsilon:\;\cA^1+\varepsilon\omega\cA^2\longrightarrow\cA^1+\cA^2,
\]
where $\cA^1,\,\cA^2\in\C_n$, $\omega=\e_{12\cdots n+1}$, and
\[
\varepsilon=\begin{cases}
1,& \text{if $n+1\equiv 1\s\!\!\pmod{4}$},\\
i,& \text{if $n+1\equiv 3\s\!\!\pmod{4}$};
\end{cases}
\]
so that $(\varepsilon\omega)^2=1$. At this point $\varepsilon\omega\rightarrow 1$
and the quotient algebra has a form
\[
{}^\epsilon\C_n\simeq\C_{n+1}/\Ker\epsilon,
\]
where $\Ker\epsilon=\left\{\cA^1-\varepsilon\omega\cA^1\right\}$ is a kernel
of the homomorphism $\epsilon$. 

For the transfer of the antiautomorphism $\cA\rightarrow\widetilde{\cA}$ 
from $\C_{n+1}$ into $\C_n$ it is necessary that
\begin{equation}\label{6.34}
\widetilde{\varepsilon\omega}=\varepsilon\omega.
\end{equation}
Indeed, since under action of $\epsilon$ the elements
1 and $\varepsilon\omega$ are equally mapped into the unit, then transformed
elements $\widetilde{1}$ and $\widetilde{\varepsilon\omega}$ are also
should be mapped into 1, but $\widetilde{1}=1\rightarrow 1$, and
$\widetilde{\varepsilon\omega}=\pm\varepsilon\omega\rightarrow\pm 1$ in
virtue of $\widetilde{\omega}=(-1)^{\frac{n(n-1)}{2}}\omega$, whence
\begin{equation}\label{6.34'}
\widetilde{\omega}=\begin{cases}
\omega,& \text{if $n+1\equiv 1\s\!\!\pmod{4}$};\\
-\omega,& \text{if $n+1\equiv 3\s\!\!\pmod{4}$}.
\end{cases}
\end{equation}
Therefore, {\it under action of the homomorphism $\epsilon$ the antiautomorphism
$\cA\rightarrow\widetilde{\cA}$ is transferred from $\C_{n+1}$ 
into $\C_n$ only at $n\equiv 0\s\pmod{4}$}.

In its turn, for the transfer of the automorphism $\cA\rightarrow\cA^\star$
it is necessary that $(\varepsilon\omega)^\star=\varepsilon\omega$. However,
since the element $\omega$ is odd and $\omega^\star=(-1)^{n+1}
\omega$, then we have always
\begin{equation}\label{6.35}
\omega^\star=-\omega.
\end{equation}
Thus, {\it the automorphism $\cA\rightarrow\cA^\star$ is never transferred
from $\C_{n+1}$ into $\C_n$}.

Further, for the transfer of the antiautomorphism 
$\cA\rightarrow\widetilde{\cA^\star}$
from $\C_{n+1}$ into $\C_n$ it is necessary that
\begin{equation}\label{6.36}
\widetilde{(\varepsilon\omega)^\star}=\varepsilon\omega.
\end{equation}
It is easy to see that the condition (\ref{6.36}) is satisfied only at
$n+1\equiv 3\s\pmod{4}$, since in this case from the second equality of
(\ref{6.34'}) and (\ref{6.35}) it follows
\begin{equation}\label{6.36'}
\widetilde{(\varepsilon\omega)^\star}=\varepsilon\widetilde{\omega^\star}=
-\varepsilon\omega^\star=\varepsilon\omega.
\end{equation}
Therefore, {\it under action of the homomorphism $\epsilon$ the 
antiautomorphism
$\cA\rightarrow\widetilde{\cA^\star}$ is transferred from $\C_{n+1}$ into $\C_n$
only at $n\equiv 2\s\pmod{4}$}.

Let $n+1=p+q$. Defining in $\C_{n+1}$ the basis $\{\e_1,\ldots,\e_p,i\e_{p+1},
\ldots, i\e_{p+q}\}$ we extract the real subalgebra $\cl_{p,q}$, where at
$p-q\equiv 3,7\s\pmod{8}$ we have a complex division ring
$\K\simeq\C$, and at $p-q\equiv 1\s\pmod{8}$ and $p-q\equiv 5\s\pmod{8}$
correspondingly a double real division ring $\K\simeq\R\oplus\R$ and a double
quaternionic division ring $\K\simeq\BH\oplus\BH$. The product
$\e_1\e_2\cdots\e_pi\e_{p+1}\cdots i\e_{p+q}=i^q\omega\in\C_{n+1}$ sets
a volume element of the real subalgebra $\cl_{p,q}$. At this point we have a
condition $\overline{(i^q\omega)}=i^q\omega$, that is,
$(-i)^q\overline{\omega}=i^q\omega$, whence
\begin{equation}\label{6.37}
\overline{\omega}=(-1)^q\omega.
\end{equation}
When $q$ is even, from (\ref{6.37}) it follows $\overline{\omega}=\omega$ and,
therefore, the pseudoautomorphism $\cA\rightarrow\overline{\cA}$ is transferred
at $q\equiv 0\s\pmod{2}$, and since $p+q$ is odd number, then we have always
$p\equiv 1\s\pmod{2}$. In more detail, at $n+1\equiv 3\s\pmod{4}$ the
pseudoautomorphism $\cA\rightarrow\overline{\cA}$ is transferred from $\C_{n+1}$
into $\C_n$ if the real subalgebra $\cl_{p,q}$ possesses the complex ring
$\K\simeq\C$, $p-q\equiv 3,7\s\pmod{8}$, and is not transferred
($\overline{\omega}=-\omega,\;q\equiv 1\s\pmod{2},\,p\equiv 0\s\pmod{2}$)
in the case of $\cl_{p,q}$ with double rings $\K\simeq\R\oplus\R$ and
$\K\simeq\BH\oplus\BH$, $p-q\equiv 1,5\s\pmod{8}$. In its turn, at
$n+1\equiv 1\s\pmod{4}$ the pseudoautomorphism $\cA\rightarrow\overline{\cA}$
is transferred from $\C_{n+1}$ into $\C_n$ if the subalgebra $\cl_{p,q}$ has
the type $p-q\equiv 1,5\s\pmod{8}$ and is not transferred in the case
of $\cl_{p,q}$ with $p-q\equiv 3,7\s\pmod{8}$. 
Besides, in virtue of (\ref{6.35}) at
$n+1\equiv 3\s\pmod{4}$ with $p-q\equiv 1,5\s\pmod{8}$ 
and at $n+1\equiv 1\s\pmod{4}$ with 
$p-q\equiv 3,7\s\pmod{8}$ a pseudoautomorphism
$\cA\rightarrow\overline{\cA^\star}$ (a composition of the pseudoautomorphism
$\cA\rightarrow\overline{\cA}$ with the automorphism $\cA\rightarrow\cA^\star$)
is transferred from $\C_{n+1}$ into $\C_n$, since
\[
\overline{\varepsilon\omega^\star}=\varepsilon\omega.
\]
Further, in virtue of the second equality of (\ref{6.34'}) 
at $n+1\equiv 3\s\pmod{4}$ with
$p-q\equiv 1,5\s\pmod{8}$  
a pseudoantiautomorphism $\cA\rightarrow\overline{\widetilde{\cA}}$
(a composition of the pseudoautomorphism $\cA\rightarrow\overline{\cA}$ with
the antiautomorphism $\cA\rightarrow\widetilde{\cA}$) is transferred from
$\C_{n+1}$ into $\C_n$, since
\[
\overline{\widetilde{\varepsilon\omega}}=\varepsilon\omega.
\]
Finally, a pseudoantiautomorphism 
$\cA\rightarrow\overline{\widetilde{\cA^\star}}$ (a composition of the
pseudoautomorphism $\cA\rightarrow\overline{\cA}$ with the antiautomorphism
$\cA\rightarrow\widetilde{\cA^\star}$), corresponded to $CPT$--transformation,
is transferred from $\C_{n+1}$ into $\C_n$ at $n+1\equiv 3\pmod{4}$ and
$\cl_{p,q}$ with $p-q\equiv 3,7\pmod{8}$, since in this case in virtue of
(\ref{6.36'}) and (\ref{6.37}) we have
\[
\overline{\widetilde{(\varepsilon\omega^\star)}}=\varepsilon\omega.
\]
Also at $n+1\equiv 1\pmod{4}$ and $q\equiv 1\pmod{2}$ we obtain
\[
\overline{\widetilde{(\varepsilon\omega^\star)}}=-
\widetilde{(\varepsilon\omega^\star)}=-(\varepsilon\omega)^\star=
\varepsilon\omega,
\]
therefore, the transformation $\cA\rightarrow\overline{\widetilde{\cA^\star}}$
is transferred at $n+1\equiv 1\pmod{4}$ and $\cl_{p,q}$ with
$p-q\equiv 3,7\pmod{8}$.

The conditions for the transfer of the fundamental automorphisms of the algebra
$\C_{n+1}$ into its subalgebra $\C_n$ under action of the homomorphism
$\epsilon$ allow to define in evident way an explicit form of the quotient
algebras ${}^\epsilon\C_n$.\\[0.2cm]
1) The quotient algebra ${}^\epsilon\C_n$, $n\equiv 0\s\pmod{4}$.\\[0.1cm]
As noted previously, in the case $n+1\equiv 1\s\pmod{4}$ 
the antiautomorphism $\cA\rightarrow\widetilde{\cA}$ and
pseudoautomorphism $\cA\rightarrow\overline{\cA}$ are transferred from
$\C_{n+1}$ into $\C_n$ if the subalgebra
$\cl_{p,q}\subset\C_{n+1}$ possesses the double rings $\K\simeq\R\oplus\R$,
$\K\simeq\BH\oplus\BH$ ($p-q\equiv 1,5\s\pmod{8}$), and also the
pseudoautomorphism $\cA\rightarrow\overline{\cA^\star}$ and
pseudoantiautomorpism $\cA\rightarrow\overline{\widetilde{\cA^\star}}$
are transferred if
$\cl_{p,q}$ has the complex ring $\K\simeq\C$ ($p-q\equiv 3,7\s\pmod{8}$).
It is easy to see that in dependence on the type of $\cl_{p,q}$ the structure
of the quotient algebras ${}^\epsilon\C_n$ of this type is divided into two
different classes:\\[0.1cm]
{\bf a}) The class of quotient algebras ${}^\epsilon\C_n$ containing the
antiautomorphism
$\cA\rightarrow\widetilde{\cA}$ and pseudoautomorphism $\cA\rightarrow
\overline{\cA}$. It is obvious that in dependence on a division ring
structure of the subalgebra
$\cl_{p,q}\subset\C_{n+1}$ this class is divided into two subclasses:
\begin{description}
\item[$a_1$)] ${}^\epsilon\C_n$ with $\cA\rightarrow\widetilde{\cA}$,
$\cA\rightarrow\overline{\cA}$ at $\cl_{p,q}$ with the ring
$\K\simeq\R\oplus\R$, $p-q\equiv 1\s\pmod{8}$.
\item[$a_2$)] ${}^\epsilon\C_n$ with $\cA\rightarrow\widetilde{\cA}$,
$\cA\rightarrow\overline{\cA}$ at $\cl_{p,q}$ with the ring
$\K\simeq\BH\oplus\BH$, $p-q\equiv 5\s\pmod{8}$.
\end{description}
{\bf b}) The class of quotient algebras ${}^\epsilon\C_n$ containing the
transformations
$\cA\rightarrow\widetilde{\cA}$, $\cA\rightarrow
\overline{\cA^\star}$, $\cA\rightarrow\overline{\widetilde{\cA^\star}}$ 
if the subalgebra $\cl_{p,q}\subset\C_{n+1}$ has the
complex ring $\K\simeq\C$, $p-q\equiv 3,7\s\pmod{8}$.\\[0.2cm]
2) The quotient algebra ${}^\epsilon\C_n$, $n\equiv 2\s\pmod{4}$.\\[0.1cm]
In the case $n+1\equiv 3\s\pmod{4}$ the antiautomorphism
$\cA\rightarrow\widetilde{\cA^\star}$, pseudoautomorphism
$\cA\rightarrow\overline{\cA}$ and pseudoantiautomorphism
$\cA\rightarrow\overline{\widetilde{\cA^\star}}$
are transferred from $\C_{n+1}$ into $\C_n$
if the subalgebra $\cl_{p,q}\subset\C_{n+1}$
possesses the complex ring $\K\simeq\C$ ($p-q\equiv 3,7\s\pmod{8}$), and also
the pseudoautomorphism $\cA\rightarrow\overline{\cA^\star}$ and
pseudoantiautomorphism $\cA\rightarrow\overline{\widetilde{\cA}}$ are
transferred if $\cl_{p,q}$ has the double rings $\K\simeq\R\oplus\R$,
$\K\simeq\BH\oplus\BH$ ($p-q\equiv 1,5\s\pmod{8}$). In dependence on the type of
$\cl_{p,q}\subset\C_{n+1}$ all the quotient algebras ${}^\epsilon\C_n$ of this
type are divided into following two classes:\\[0.1cm]
{\bf c}) The class of quotient algebras ${}^\epsilon\C_n$ containing the
transformations
$\cA\rightarrow\widetilde{\cA^\star}$,
$\cA\rightarrow\overline{\cA}$, 
$\cA\rightarrow\overline{\widetilde{\cA^\star}}$
if the subalgebra $\cl_{p,q}$ has the ring
$\K\simeq\C$, $p-q\equiv 3,7\s\pmod{8}$.\\[0.1cm]
{\bf d}) The class of quotient algebras ${}^\epsilon\C_n$ containing the
antiautomorphism
$\cA\rightarrow\widetilde{\cA^\star}$, pseudoautomorphism
$\cA\rightarrow\overline{\cA^\star}$ and pseudoautomorphism
$\cA\rightarrow\overline{\widetilde{\cA}}$. At this point, in dependence on the
division ring structure of $\cl_{p,q}$ we have two subclasses
\begin{description}
\item[$d_1$)] ${}^\epsilon\C_n$ with $\cA\rightarrow\widetilde{\cA^\star}$,
$\cA\rightarrow\overline{\cA^\star}$ and $\cA\rightarrow
\overline{\widetilde{\cA}}$ at $\cl_{p,q}$ with the ring $\K\simeq\R\oplus\R$,
$p-q\equiv 1\s\pmod{8}$.
\item[$d_2$)] ${}^\epsilon\C_n$ with $\cA\rightarrow\widetilde{\cA^\star}$,
$\cA\rightarrow\overline{\cA^\star}$ and $\cA\rightarrow
\overline{\widetilde{\cA}}$ at $\cl_{p,q}$ with the ring $\K\simeq\BH\oplus\BH$,
$p-q\equiv 5\s\pmod{8}$.
\end{description}
Thus, we have 6 different classes of the quotient algebras ${}^\epsilon\C_n$.
Further, in accordance with \cite{Var99} 
the automorphism $\cA\rightarrow\cA^\star$ corresponds to space inversion
$P$, the antiautomorphisms
$\cA\rightarrow\widetilde{\cA}$ and $\cA\rightarrow\widetilde{\cA^\star}$
set correspondingly time reversal $T$ and full reflection
$PT$, and the pseudoautomorphism $\cA\rightarrow\overline{\cA}$ corresponds to
charge conjugation $C$. Taking into account this relation and Theorem
\ref{tprod} we come to classification presented in Theorem for complex
quotient representations.\\[0.2cm]
2) Real representations.\\
Let us define real quotient representations of the group $\fG_+$. First
of all, in the case of types $p-q\equiv 3,7\pmod{8}$ we have the
isomorphism (\ref{Iso}) and, therefore, these representations are equivalent
to complex representations considered in the section 3. Further, when
$p-q\equiv 1,5\pmod{8}$ we have the real algebras $\cl_{p,q}$ with the
rings $\K\simeq\R\oplus\R$, $\K\simeq\BH\oplus\BH$ and, therefore, there
exist homomorphic mappings $\epsilon:\,\cl_{p,q}\rightarrow\cl_{p,q-1}$,
$\epsilon:\,\cl_{p,q}\rightarrow\cl_{q,p-1}$. In this case the quotient
algebra has a form
\[
{}^\epsilon\cl_{p,q-1}\simeq\cl_{p,q}/\Ker\epsilon
\]
or
\[
{}^\epsilon\cl_{q,p-1}\simeq\cl_{p,q}/\Ker\epsilon,
\]
where $\Ker\epsilon=\left\{\cA^1-\omega\cA^1\right\}$ is a kernel of $\epsilon$, 
since in accordance with
\[
\omega^2=\begin{cases}
-1& \text{if $p-q\equiv 2,3,6,7\pmod{8}$},\\
+1& \text{if $p-q\equiv 0,1,4,5\pmod{8}$}
\end{cases}
\]
at $p-q\equiv 1,5\pmod{8}$ we have always $\omega^2=1$ and, therefore,
$\varepsilon=1$. Thus, for the transfer of the antiautomorphism
$\cA\rightarrow\widetilde{\cA}$ from $\cl_{p,q}$ into $\cl_{p,q-1}$
($\cl_{q,p-1}$) it is necessary that
\[
\widetilde{\omega}=\omega
\]
In virtue of the relation $\widetilde{\omega}=(-1)^{\frac{(p+q)(p+q+1)}{2}}
\omega$ we obtain
\begin{equation}\label{Real1}
\widetilde{\omega}=\begin{cases}
+\omega& \text{if $p-q\equiv 1,5\pmod{8}$},\\
-\omega& \text{if $p-q\equiv 3,7\pmod{8}$}.
\end{cases}
\end{equation}
Therefore, for the algebras over the field $\F=\R$ the antiautomorphism
$\cA\rightarrow\widetilde{\cA}$ is transferred at the mappings
$\cl_{p,q}\rightarrow\cl_{p,q-1}$, $\cl_{p,q}\rightarrow\cl_{q,p-1}$, where
$p-q\equiv 1,5\pmod{8}$.

In its turn, for the transfer of the automorphism $\cA\rightarrow\cA^\star$
it is necessary that $\omega^\star=\omega$. However, since the element
$\omega$ is odd and $\omega^\star=(-1)^{p+q}\omega$, then we have always
\begin{equation}\label{Real2}
\omega^\star=-\omega.
\end{equation}
Thus, the automorphism $\cA\rightarrow\cA^\star$ is never transferred from
$\cl_{p,q}$ into $\cl_{p,q-1}$ ($\cl_{q,p-1}$)

Further, for the transfer of the antiautomorphism 
$\cA\rightarrow\widetilde{\cA^\star}$ it is necessary that
\[
\widetilde{\omega^\star}=\omega.
\]
From (\ref{Real1}) and (\ref{Real2}) for the types $p-q\equiv 1,5\pmod{8}$
we obtain
\begin{equation}\label{Real3}
\widetilde{\omega^\star}=\omega^\star=-\omega.
\end{equation}
Therefore, under action of the homomorphism $\epsilon$ the antiautomorphism
$\cA\rightarrow\widetilde{\cA^\star}$ is never transferred from $\cl_{p,q}$
into $\cl_{p,q-1}$ ($\cl_{q,p-1}$).

As noted previously, for the real representations of $\fG_+$ the
pseudoautomorphism $\cA\rightarrow\overline{\cA}$ is reduced into identical
transformation $\sI$ for $\fR^{l_0}_{0,2}$ and to particle--antiparticle
conjugation $C^\prime$ for $\fH^{l_0}_{4,6}$. The volume element $\omega$
of $\cl_{p,q}$ (types $p-q\equiv 1,5\pmod{8}$) can be represented by the
product $\e_1\e_2\cdots\e_p\e^{\p}_{p+1}\e^{\p}_{p+2}\cdots\e^{\p}_{p+q}$,
where $\e^{\p}_{p+j}=i\e_{p+j}$, $\e^2_j=1$, $(\e^{\p}_{p+j})^2=-1$.
Therefore, for the transfer of $\cA\rightarrow\overline{\cA}$ from $\cl_{p,q}$
into $\cl_{p,q-1}$ ($\cl_{q,p-1}$) we have a condition
\[
\overline{\omega}=\omega,
\]
and in accordance with (\ref{6.37}) it follows that the pseudoautomorphism
$\cA\rightarrow\overline{\cA}$ is transferred at $q\equiv 0\pmod{2}$. 
Further, in virtue of the relation (\ref{Real2}) the pseudoautomorphism
$\cA\rightarrow\overline{\cA^\star}$ is transferred at $q\equiv 1\pmod{2}$,
since in this case we have
\[
\overline{\omega^\star}=\omega.
\]
Also from (\ref{Real1}) it follows that the pseudoantiautomorphism
$\cA\rightarrow\overline{\widetilde{\cA}}$ is transferred at
$p-q\equiv 1,5\pmod{8}$ and $q\equiv 0\pmod{2}$, since
\[
\overline{\widetilde{\omega}}=\omega.
\]
Finally, the pseudoantiautomorphism 
$\cA\rightarrow\overline{\widetilde{\cA^\star}}$ ($CPT$--transformation)
in virtue of (\ref{Real3}) and (\ref{6.37}) is transferred from $\cl_{p,q}$
into $\cl_{p,q-1}$ ($\cl_{q,p-1}$) at $p-q\equiv1,5\pmod{8}$ and
$q\equiv 1\pmod{2}$.

Now we are in a position that allows to classify the real quotient algebras
${}^\epsilon\cl_{p,q-1}$ (${}^\epsilon\cl_{q,p-1}$).\\
1) The quotient algebra ${}^\epsilon\cl_{p,q-1}$ (${}^\epsilon\cl_{q,p-1}$,
$p-q\equiv 1\pmod{8}$.\\
In this case the initial algebra $\cl_{p,q}$ has the double real division
ring $\K\simeq\R\oplus\R$ and its subalgebras $\cl_{p,q-1}$ and
$\cl_{q,p-1}$ are of the type $p-q\equiv 0\pmod{8}$ or $p-q\equiv 2\pmod{8}$
with the ring $\K\simeq\R$. Therefore, in accordance with Theorem 
\ref{tpseudo} for all such quotient algebras the pseudoautomorphism
$\cA\rightarrow\overline{\cA}$ is equivalent to the identical transformation
$\sI$. The antiautomorphism $\cA\rightarrow\widetilde{\cA}$ in this case
is transferred into $\cl_{p,q-1}$ ($\cl_{q,p-1}$) at any $p-q\equiv 1\pmod{8}$.
Further, in dependence on the number $q$ we have two different classes of
the quotient algebras of this type:
\begin{description}
\item[$e_1$)] ${}^\epsilon\cl_{p,q-1}$ (${}^\epsilon\cl_{q,p-1}$) with
$\cA\rightarrow\widetilde{\cA}$, $\cA\rightarrow\overline{\cA}$,
$\cA\rightarrow\overline{\widetilde{\cA}}$, $p-q\equiv 1\pmod{8}$,
$q\equiv 0\pmod{2}$.
\item[$e_2$)] ${}^\epsilon\cl_{p,q-1}$ (${}^\epsilon\cl_{q,p-1}$) with
$\cA\rightarrow\widetilde{\cA}$, $\cA\rightarrow\overline{\cA^\star}$,
$\cA\rightarrow\overline{\widetilde{\cA^\star}}$, $p-q\equiv 1\pmod{8}$,
$q\equiv 1\pmod{2}$.
\end{description}
2) The quotient algebras ${}^\epsilon\cl_{p,q-1}$ (${}^\epsilon\cl_{q,p-1}$),
$p-q\equiv 5\pmod{8}$.\\
In this case the initial algebra $\cl_{p,q}$ has the double quaternionic
division ring $\K\simeq\BH\oplus\BH$ and its subalgebras $\cl_{p,q-1}$ and
$\cl_{q,p-1}$ are of the type $p-q\equiv 4\pmod{8}$ or $p-q\equiv 6\pmod{8}$
with the ring $\K\simeq\BH$. Therefore, in this case the pseudoautomorphism
$\cA\rightarrow\overline{\cA}$ is equivalent to the particle--antiparticle
conjugation $C^\prime$. As in the previous case the antiautomorphism
$\cA\rightarrow\widetilde{\cA}$ is transferred at any $p-q\equiv 5\pmod{8}$.
For this type in dependence on the number $q$ there are two different classes:
\begin{description}
\item[$f_1$)] ${}^\epsilon\cl_{p,q-1}$ (${}^\epsilon\cl_{q,p-1}$) with
$\cA\rightarrow\widetilde{\cA}$, $\cA\rightarrow\overline{\cA}$,
$\cA\rightarrow\overline{\widetilde{\cA}}$, $p-q\equiv 5\pmod{8}$,
$q\equiv 0\pmod{2}$.
\item[$f_2$)] ${}^\epsilon\cl_{p,q-1}$ (${}^\epsilon\cl_{q,p-1}$) with
$\cA\rightarrow\widetilde{\cA}$, $\cA\rightarrow\overline{\cA^\star}$,
$\cA\rightarrow\overline{\widetilde{\cA^\star}}$, $p-q\equiv 5\pmod{8}$,
$q\equiv 1\pmod{2}$.
\end{description}
\end{proof}

\section{Quotient representation ${}^\chi\fC^{0,-1}_c$ and neutrino field}
Analysing the quotient representations of the group $\fG_+$ presented in
Theorem \ref{tfactor}, we see that only a repsesentation of the class
$c$ at $j=(l_0+l_1-1)/2=1/2$ is adequate for description of the neutrino
field. This representation admits full reflection
$PT$, charge conjugation $C$ and $CPT$--transformation
(space inversion $P$ is not defined). In contrast with this, the first three
classes $a_1,a_2$ and $b$ are unsuitable for description of neutrino,
since in this case $j$ is an integer number, $n\equiv 0\pmod{4}$ (bosonic
fields). In turn, the classes $d_1$ and $d_2$ admit 
$CT$--transformation that in accordance with $CPT$--Theorem is equivalent to
space inversion $P$, which, as known, is a forbidden operation for the
neutrino field. So, we have an homomorphic mapping
$\epsilon:\,\C_3\rightarrow\C_2$, where $\C_3$ is a simplest Clifford algebra
of the type $n+1\equiv 3\s\pmod{4}$. In accordance with Theorem
\ref{tfactor} under action of the homomorphism $\epsilon:\,\C_3
\rightarrow\C_2$ the transformations
$\cA\rightarrow\widetilde{\cA^\star}$, $\cA\rightarrow\overline{\cA}$ and
$\cA\rightarrow\overline{\widetilde{\cA^\star}}$ are transferred from $\C_3$
into $\C_2$. At this point, the real subalgebra $\cl_{3,0}\subset\C_3$ has
the complex ring $\K\simeq\C$, $p-q\equiv 3\s\pmod{8}$, and, therefore, the
matrix $\Pi$ of the pseudoautomorphism $\cA\rightarrow\overline{\cA}$ is not
unit, that according to Theorem \ref{tpseudo} corresponds to charged or
{\it neutral} fields.
In accordance with general scheme presented in the section 2 a decomposition
of the algebra $\C_3$ looks like
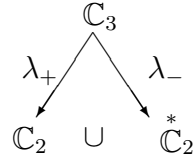
\[
\unitlength=0.5mm
\begin{picture}(70,50)
\put(35,40){\vector(2,-3){15}}
\put(35,40){\vector(-2,-3){15}}
\put(32.25,42){$\C_{3}$}
\put(16,28){$\lambda_{+}$}
\put(49.5,28){$\lambda_{-}$}
\put(13.5,9.20){$\C_{2}$}
\put(52.75,9){$\stackrel{\ast}{\C}_{2}$}
\put(32.5,10){$\cup$}
\end{picture}
\]
Here central idempotents
\begin{equation}\label{Cent}
\lambda_{-}=\frac{1-i\e_1\e_2\e_3}{2},\quad
\lambda_{+}=\frac{1+i\e_1\e_2\e_3}{2} 
\end{equation}
in accordance with \cite{CF97}
can be identified with helicity projection operators. In such a way, we have
two helicity states describing by the quotient algebras 
${}^\epsilon\C_2$ and ${}^\epsilon\overset{\ast}{\C}_2$, and a full
neutrino--antineutrino algebra is 
${}^\epsilon\C_2\cup{}^\epsilon\overset{\ast}{\C}_2$
(cf. electron--positron algebra
$\C_2\oplus\overset{\ast}{\C}_2$). 

Let $\varphi\in\C_3$ be an algebraic spinor of the form
\begin{equation}\label{Neut1}
\varphi=a^0+a^1\e_1+a^2\e_2+a^3\e_3+a^{12}\e_1\e_2+a^{13}\e_1\e_3+
a^{23}\e_2\e_3+a^{123}\e_1\e_2\e_3.
\end{equation}
Then it is easy to verify that spinors
\begin{equation}\label{Neut2}
\varphi^+=\lambda_+\varphi=\frac{1}{2}(1+i\e_1\e_2\e_3)\varphi,\quad
\varphi^-=\lambda_-\varphi=\frac{1}{2}(1-i\e_1\e_2\e_3)\varphi
\end{equation}
are mutually orthogonal, $\varphi^+\varphi^-=0$, since 
$\lambda_+\lambda_-=0$, and also $\varphi^+\in\C_2$,
$\varphi^-\in\overset{\ast}{\C}_2$. Further, it is obvious that a spinspace
of the algebra 
${}^\epsilon\C_2\cup{}^\epsilon\overset{\ast}{\C}_2$ is
$\dS_2\cup\dot{\dS}_2$. It should be noted here that structures of the
spinspaces $\dS_2\cup\dot{\dS}_2$ and $\dS_2\oplus\dot{\dS}_2$
are different. Indeed\footnote{See also \cite{Abl98}.},
\[
\dS_2\cup\dot{\dS}_2=\ar\begin{pmatrix}
\left[00,\dot{0}\dot{0}\right] & \left[01,\dot{0}\dot{1}\right]\\
\left[10,\dot{1}\dot{0}\right] & \left[11,\dot{1}\dot{1}\right]
\end{pmatrix},\quad
\dS_2\oplus\dot{\dS}_2=\begin{pmatrix}
00 & 01 & & \\
10 & 11 & & \\
   &    &\dot{0}\dot{0} & \dot{0}\dot{1}\\
   &    &\dot{1}\dot{0} & \dot{1}\dot{1}
\end{pmatrix}.
\]
Since spinor representations of the quotient algebras ${}^\epsilon\C_2$ and
${}^\epsilon\overset{\ast}{\C}_2$ 
are defined in terms of Pauli matrices
$\sigma_i$, then the algebraic spinors $\varphi^+\in{}^\epsilon\C_2$ and
$\varphi^-\in{}^\epsilon\overset{\ast}{\C}_2$ correspond to spinors
$\xi^{\alpha_i}\in\dS_2$ and $\xi^{\dot{\alpha}_i}\in\dot{\dS}_2$
($i=0,1$). Hence it immediately follow Weyl equations
\begin{equation}\label{Weyl}
\left(\frac{\partial}{\partial x^0}-\boldsymbol{\sigma}
\frac{\partial}{\partial\bx}\right)\xi^{\alpha}=0,\quad
\left(\frac{\partial}{\partial x^0}+\boldsymbol{\sigma}
\frac{\partial}{\partial\bx}\right)\xi^{\dot{\alpha}}=0.
\end{equation}
Therefore, two--component Weyl theory can be naturally formulated within
quotient representation ${}^\chi\fC^{1,0}_c\cup{}^\chi\fC^{0,-1}_c$ of the
group $\fG_+$. Further, in virtue of an isomorphism 
$\C_2\simeq\cl_{3,0}\simeq\cl^+_{1,3}$ ($\cl_{1,3}$ is the space--time
algebra) the spinor field of the quotient representation ${}^\chi\fC^{0,-1}_c$
(${}^\chi\fC^{1,0}_c$) can be expressed via the Dirac--Hestenes spinor
field $\phi(x)\in\cl_{3,0}$ \cite{Hest66,Hest67,Lou93}. 
Indeed, the Dirac--Hestenes spinor is represented by a following
biquaternion number
\begin{multline}
\phi=a^0+a^{01}\gamma_0\gamma_1+a^{02}\gamma_0\gamma_2+a^{03}\gamma_0\gamma_3+\\
a^{12}\gamma_1\gamma_2+a^{13}\gamma_1\gamma_3+a^{23}\gamma_2\gamma_3
+a^{0123}\gamma_0\gamma_1\gamma_2\gamma_3,
\label{173}
\end{multline}
or using $\gamma$--basis
\begin{equation}\label{173'}
\ar\gamma_0=\begin{pmatrix}
I & 0\\
0 & -I
\end{pmatrix},\;\;\Gamma_1=\begin{pmatrix}
0 & \sigma_1\\
-\sigma_1 & 0
\end{pmatrix},\;\;\Gamma_2=\begin{pmatrix}
0 & \sigma_2\\
-\sigma_2 & 0
\end{pmatrix},\;\;\Gamma_3=\begin{pmatrix}
0 & \sigma_3\\
-\sigma_3 & 0
\end{pmatrix},
\end{equation}
we can write (\ref{173}) in the matrix form
\begin{equation}\label{174}
\ar\phi=\begin{pmatrix}
\phi_1 & -\phi^\ast_2 & \phi_3 & \phi^\ast_4 \\
\phi_2 & \phi^\ast_1 & \phi_4 & -\phi^\ast_3\\
\phi_3 & \phi^\ast_4 & \phi_1 & -\phi^\ast_2\\
\phi_4 & -\phi^\ast_3 & \phi_2 & \phi^\ast_1
\end{pmatrix},
\end{equation}
where
\[
\phi_1=a^0-ia^{12},\quad
\phi_2=a^{13}-ia^{23},\quad
\phi_3=a^{03}-ia^{0123},\quad
\phi_4=a^{01}+ia^{02}.
\]
From (\ref{Neut1})--(\ref{Neut2}) and (\ref{173}) it is easy to see that
spinors $\varphi^+$ and $\varphi^-$ are algebraically equivalent to the
spinor $\phi\in\C_2\simeq\cl_{3,0}$. Further, since $\phi\in\cl^+_{1,3}$,
then actions of the antiautomorphisms $\cA\rightarrow\widetilde{\cA}$ and
$\cA\rightarrow\widetilde{\cA^\star}$ on the field $\phi$ are equivalent.
On the other hand, in accordance with Feynman--Stueckelberg interpretation,
time reversal for the chiral field is equivalent to charge conjugation
(particles reversed in time are antiparticles). Thus, for the field
$\phi\in{}^\chi\fC^{0,-1}_c$ we have $C\sim T$ and, therefore, this field
is $CP$--invariant.

The spinor (\ref{173}) (or (\ref{174})) satisfies the Dirac--Hestenes
equation
\begin{equation}\label{DH}
\partial\phi\gamma_2\gamma_1-\frac{mc}{\hbar}\phi\gamma_0=0,
\end{equation}
where $\partial=\gamma^\mu\frac{\partial}{\partial x^\mu}$ is the Dirac
operator. Let us show that a massless Dirac--Hestenes equation
\begin{equation}\label{DHM}
\partial\phi\gamma_2\gamma_1=0
\end{equation}
describes the neutrino field. Indeed, the matrix 
$\gamma_0\gamma_1\gamma_2\gamma_3=\gamma_5$ commutes with all the elements
of the biquaternion (\ref{173}) and, therefore, $\gamma_5$ is equivalent
to the volume element $\omega=\e_1\e_2\e_3$ of the biquaternion algebra
$\cl_{3,0}$. In such a way, we see that idempotents
\[
P_+=\frac{1+\gamma_5}{2},\quad P_-=\frac{1-\gamma_5}{2}
\]
cover the central idempotents (\ref{Cent}). Further, from (\ref{DHM}) we
obtain
\[
P_\pm\gamma^\mu\frac{\partial}{\partial x^\mu}\phi\gamma_2\gamma_1=
\gamma^\mu P_mp\frac{\partial}{\partial x^\mu}\phi\gamma_2\gamma_1=0,
\]
that is, there are two separated equations for 
$\phi^\pm=P_\pm\phi\gamma_2\gamma_1$:
\begin{equation}\label{DHM2}
\gamma^\mu\frac{\partial}{\partial x^\mu}\phi^\pm=0,
\end{equation}
where
\[
\phi^\pm=\frac{1}{2}(1\pm\gamma_5)\phi\gamma_2\gamma_1=\frac{i}{2}
\ar\begin{pmatrix}
\phi_1\mp\phi_3 & \phi^\ast_2\pm\phi^\ast_4 & \phi_3\mp\phi_1 &
-\phi^\ast_4\mp\phi^\ast_2\\
\phi_2\mp\phi_4 & -\phi^\ast_1\mp\phi^\ast_3 & \phi_4\mp\phi_2 &
\phi^\ast_3\pm\phi^\ast_1\\
\mp\phi_1+\phi_2 & \mp\phi^\ast_2-\phi^\ast_4 & \mp\phi_3+\phi_1 &
\pm\phi^\ast_4+\phi^\ast_2\\
\mp\phi_2+\phi_4 & \pm\phi^\ast_1+\phi^\ast_3 & \mp\phi_4+\phi_2 &
\mp\phi^\ast_3-\phi^\ast_1
\end{pmatrix}
\]
Therefore, each of the functions $\phi^+$ and $\phi^-$ contains only four
independent components and in the split form we have
\[
\phi^+=\ar\begin{pmatrix}
\psi_1 & \psi_2 & \psi_3 & \psi_4\\
-\psi_1 & -\psi_2 & -\psi_3 & -\psi_4
\end{pmatrix},\quad
\phi^-=\begin{pmatrix}
\psi_5 & \psi_6 & \psi_7 & \psi_8\\
\psi_5 & \psi_6 & \psi_7 & \psi_8
\end{pmatrix},
\]
where
\begin{gather}\ar
\psi_1=\frac{i}{2}\begin{pmatrix}
\phi_1-\phi_3\\
\phi_2-\phi_4
\end{pmatrix},\;\;
\psi_2=\frac{i}{2}\begin{pmatrix}
\phi^\ast_2+\phi^\ast_4\\
-\phi^\ast_1-\phi^\ast_3
\end{pmatrix},\;\;
\psi_3=\frac{i}{2}\begin{pmatrix}
\phi_3-\phi_1\\
\phi_4-\phi_2
\end{pmatrix},\;\;
\psi_4=\frac{i}{2}\begin{pmatrix}
-\phi^\ast_4-\phi^\ast_2\\
\phi^\ast_3+\phi^\ast_1
\end{pmatrix},\nonumber\\
\ar\psi_5=\frac{i}{2}\begin{pmatrix}
\phi_1+\phi_3\\
\phi_2+\phi_4
\end{pmatrix},\;\;
\psi_6=\frac{i}{2}\begin{pmatrix}
\phi^\ast_2-\phi^\ast_4\\
-\phi^\ast_1+\phi^\ast_3
\end{pmatrix},\;\;
\psi_7=\frac{i}{2}\begin{pmatrix}
\phi_3+\phi_1\\
\phi_4+\phi_2
\end{pmatrix},\;\;
\psi_8=\frac{i}{2}\begin{pmatrix}
-\phi^\ast_4+\phi^\ast_2\\
\phi^\ast_3-\phi^\ast_1
\end{pmatrix}.\nonumber
\end{gather}
Thus, in the $\gamma$--basis we obtain from (\ref{DHM2})
\[
\left(\frac{\partial}{\partial x^0}-\boldsymbol{\sigma}
\frac{\partial}{\partial\bx}\right)\psi_i=0,\quad
\left(\frac{\partial}{\partial x^0}+\boldsymbol{\sigma}
\frac{\partial}{\partial\bx}\right)\psi_{i+4}=0,\quad(i=1,2,3,4)
\]
These equations are equivalent to Weyl equations (\ref{Weyl}) and, therefore,
should be called {\it Weyl--Hestenes equations for neutrino field}.
\section*{Acknowledgments} I am grateful to Prof. J.S.R. Chisholm for
sending me his interesting works.

\end{document}